\documentclass[a4paper,12pt]{article}
\usepackage{amscd,amssymb,amsmath,amsthm}
\usepackage{rotating}
\usepackage{hyperref}
\usepackage{graphicx,caption,dsfont}
\usepackage{mathtools}

\usepackage{subcaption}
\usepackage{amsfonts,enumerate}
\usepackage{fullpage}
\usepackage[numbers,sort & compress]{natbib}
\usepackage{soul}
\usepackage{xfrac}
\usepackage{booktabs}
\numberwithin{equation}{section}
\usepackage{xcolor}

\newtheorem{thm}{Theorem}[section]

\newtheorem{nt}{Note}
\newtheorem{example}{{Example}}

\newtheoremstyle{case}{}{}{}{}{}{:}{ }{}
\theoremstyle{case}

\allowdisplaybreaks

\makeatletter
\def\and{%
  \end{tabular}%
  \begin{tabular}[t]{c}}%
\def\@fnsymbol#1{\ensuremath{\ifcase#1\or 1\or 2\or 3\or
   4\or 5\or 6\or 7\or 8\or 9\else\@ctrerr\fi}}
\makeatother
\vspace{-8ex}
\date{\today}
\begin{document}
\begin{center}
\begin{large}
\textbf{ Invariant subspace method to the initial and boundary value problem of the higher dimensional nonlinear time-fractional PDEs}
\end{large}
\end{center}
\vspace{-0.5cm}
\begin{center}
K.S. Priyendhu$^{1}$, P. Prakash$^{1*}$, M. Lakshmanan$^{2}$\\
%Department of Mathematics, Amrita School of Engineering, Coimbatore, Amrita Vishwa
%Vidyapeetham, India
$^{1}$Department of Mathematics,\\
Amrita School of Engineering, Coimbatore-641112,\\
  Amrita Vishwa Vidyapeetham, INDIA.\\
$^{2}$Department of Nonlinear Dynamics,\\
 Bharathidasan University,\\
Tiruchirappalli-620 024, INDIA.\\
E-mail ID's: vishnuindia89@gmail.com \&\ p$_{-}$prakash@cb.amrita.edu (P Prakash)\\
\hspace{1.6cm} : priyendhusasikumar@gmail.com (K.S. Priyendhu)\\
\hspace{1.6cm} : lakshman.cnld@gmail.com (M. Lakshmanan)\\
$*$-Corresponding Author
\end{center}
\begin{abstract}
	{This paper systematically explains how to apply the invariant subspace method using variable transformation for finding the exact solutions of the $(k+1)$-dimensional nonlinear time-fractional PDEs in detail. More precisely, we have shown how to transform the given $(k+1)$-dimensional nonlinear time-fractional PDEs into $(1+1)$-dimensional nonlinear time-fractional PDEs using the variable transformation procedure.  Also, we explain how to derive the exact solutions for the reduced equations using the invariant subspace method. Additionally, in this careful and systematic study, we will investigate how to find the various types of exact solutions of the $(3 + 1)$-dimensional nonlinear time-fractional convection-diffusion-reaction equation along with appropriate initial and boundary conditions for the first time. Moreover, the obtained exact solutions of the equation as mentioned above can be written in terms of polynomial, exponential, trigonometric, hyperbolic, and Mittag-Leffler functions. Finally, the discussed method is extended for the $(k+1)$-dimensional nonlinear time-fractional PDEs with several linear time delays, and the exact solution of the $(3 + 1)$-dimensional nonlinear time-fractional delay convection-diffusion-reaction equation is derived.}
\end{abstract}
\section{Introduction}
The theory and applications of fractional differential equations (FDEs) have gained much interest and importance both from the physical and mathematical points of view during the past few decades due to the exact description of diverse anomalous phenomena in science and engineering \cite{pi,kai,kts,vas5,mo11,mn,vas7,vas2}.
 Derivatives of non-integer order (fractional-order) have been successfully used to describe the memory effect in complex systems because non-integer order derivatives depend on all the past values of the relevant function and not only on the immediate past.
Hence, non-integer order models have been effectively used to investigate the anomalous behavior of systems. The relevant phenomena can be categorized by power-law long-term memory, fractal properties, and long-range interactions \cite{vas2,vas5,vas7}.
 Also, note that the non-integer order derivatives have a set of non-standard (unusual) properties such as violation of the standard form of the chain rule, Leibniz rule, and semi-group property, which are essential fundamental properties of non-integer order derivatives that allow us to add a memory effect in complex systems \cite{vas2,vas5,vas7,vet3}.
 Due to the unusual properties of non-integer order derivatives, finding the exact solutions of the non-integer order nonlinear PDEs is very complicated.

In the literature, there are no well-defined analytical methods for finding the exact solutions of non-integer order nonlinear PDEs due to the above-mentioned reasons. So, constructing the exact solutions for such non-integer order nonlinear PDEs is a challenging and difficult task.
Finding the exact solutions to nonlinear non-integer order PDE is important, which can help us to analyze and predict the qualitative and quantitative properties of  complex systems.
Many researchers in recent years have made remarkable contributions to the construction of solutions to nonlinear FDEs by developing analytical and numerical methods such as the differential transform method \cite{mo2}, Adomian decomposition method \cite{Ge,mo}, Lie symmetry analysis \cite{bb1,lk,CRD2,pra2,praf,kdv,ru4,pra1,nas,ppp,pra5},
new hybrid method \cite{mma}, invariant subspace method {\cite{ps1,ra,pra4,pra3,ps2,ru2,s1,pp1,pp2,ru3,pra2,gara1,gara2,IVSM,ppa,ppl,hashemi,abdel,chu2022}}, variable separation method \cite{rui,ruii} and so on.

Recent discussions show that due to the unusual properties of fractional derivatives, the invariant subspace method is a very effective and powerful analytical method to find the exact solutions for nonlinear fractional PDEs. The detailed study of the invariant subspace method was initially provided by Galaktionov and  Svirshchevskii \cite{gs} for deriving the exact solutions of the integer-order nonlinear evolution PDEs, which is commonly known as the generalized separation of the variable method. Since then, this method has been further studied by Ma and many others {\cite{wx,wy,wx1,yem,zq,liu,qu2009,shen}} for scalar and coupled nonlinear PDEs. In recent days, Gazizov and Kasatkin \cite{ru2}, and others \cite{ra,pra4,ru3,pra2,pp1,s1,pra3,pp2,ppa,ps2} have extended this method for finding the exact solutions of fractional scalar and coupled nonlinear higher-dimensional PDEs.

It is well-known that the following types of general separable and functional separable solutions are available for the $(1+1)$-dimensional nonlinear PDEs \cite{apaz} as
\begin{itemize}
\item[(i)] $u(x,t)=\sum\limits^{n}_{r=1}\varphi_r(t)\psi_r(x)$ and
\item[(ii)] $u(x,t)=U(z)$, $z=\sum\limits^{n}_{r=1}\varphi_r(t)\psi_r(x)$,
\end{itemize}
where the functions $\varphi_r(t)$ and $\psi_r(x)$ are to be determined which depend on the considered equation.
The above solutions (i) and (ii) can be derived from the generalized separable method and functional separable method, respectively.

Due to the unusual properties of non-integer order derivatives, the functional separable solution (ii) is impossible for the $(1+1)$-dimensional nonlinear time-fractional PDEs because of the time derivative involving the fractional order. However, the generalized separable solution (i) is possible for the $(1+1)$-dimensional nonlinear time-fractional PDEs for which the above type of solutions can be derived from the invariant subspace method.
{In \cite{zq}, Zhu and Qu have studied the exact solutions of the two-dimensional nonlinear evolution equations using the invariant subspace method.} In 2021, Abdel Kader et al. \cite{IVSM} have derived the exact solutions of the $(2+1)$-dimensional nonlinear time-fractional variable coefficient biological population model using the invariant subspace method along with the variable transformation.
 In the same year, Prakash et al. \cite{ppa} investigated the exact solutions for the $(2+1)$-dimensional nonlinear time-fractional PDEs using the direct approach of the invariant subspace method without any variable transformation.
 Very recently, Prakash et al. \cite{ppl} generalized the theory of the invariant subspace method for the $(k+1)$-dimensional nonlinear time-fractional PDEs. From this, we can expect the exact solution of the form
\begin{equation}\label{kda}
	u(x_1,\dots,x_k,t)=\sum\limits_{i_1,i_2,\dots,i_k}\psi_{i_1,i_2,\dots,i_k}(t)\left(\prod\limits_{r=1}^{k}\varphi_{i_r}^r(x_r)\right),\ i_r=1,2,\dots,n_r,\ r=1,2,\dots,k,
	\end{equation}
where the functions $\psi_{i_1,i_2,\dots,i_k}(t)$ and $\varphi_{i_r}^r(x_r)$ are to be determined which depend on the considered equation. The main aim of this work is to {apply} the invariant subspace method associated with variable transformation for finding the following type of exact solution of the $(k+1)$-dimensional nonlinear time-fractional PDE as
 \begin{equation}\label{ttt}
 u(x_1,x_2,\dots,x_k,t)=\sum\limits_{r=1}^{n}F_{r}(t)G_r(z),\ z=\lambda_1x_1+\lambda_2x_2+\dots+\lambda_kx_k,
 \end{equation}
 where $\lambda_i\in\mathbb{R},i=1,2,\dots,k,$ and the functions $F_{r}(t)$ and $G_r(z)$ are unknown and have to be determined which depend on the considered equation.
 %%%%%%%%%%%%%%%%%%%%%%%%%%%%%%%%%%%%%%%%%%%%%%%%%%%%%%%%%%%%%%%%%%%%%%%%%%%%%%%%%%%%%%%%%%%%%%%%%%%%%%%%%%%%%%%%%%%%%%%%%%%%%%%%%%%%%%%%%%%%%%%%%

The most popularly studied model is the anomalous diffusion-wave equation \cite{bb1,lk} generated with time-fractional derivative, which can be read as follows,
 \begin{equation}\label{1.1}
 	\dfrac{\partial^\alpha u}{\partial t^\alpha}=A\dfrac{\partial^2u}{\partial x^2}, \quad \alpha\in (0,2].
 \end{equation}
 We wish to point out that the above equation \eqref{1.1} represents both parabolic and hyperbolic type processes simultaneously since it models the classical wave equation when $\alpha=2 $ and diffusion (heat) equation when $\alpha=1.$
 In addition, we note that equation \eqref{1.1} describes the %classical diffusion when $\alpha=1$,
 process of anomalous sub-diffusion if $0<\alpha<1$ and super-diffusion when $1<\alpha<2$. The super-diffusion phenomenon studies intermediate properties between wave and diffusion behaviors of the system.
 An ample amount of research has been devoted to studying the popular class of convection-diffusion-reaction equations, which appear in all major areas like fluid dynamics, biological population, mathematical finance, ecology, the study of oceanic waves, etc.
 Researchers have investigated nonlinear convection-diffusion-reaction equations through qualitative, quantitative, and asymptotic methods.
  {In the literature, Cherniha et al. \cite{CRD2} have discussed the $(n+1)$-dimensional generalized convection-diffusion-reaction equation
  \begin{equation}
  	\label{cherniha}
  	 \dfrac{\partial u}{\partial t}= \bigtriangledown \cdot \left(F(u) \bigtriangledown u\right)+ K(u)\cdot \bigtriangledown u +R(u),
  \end{equation}
where $u=u(x_1,\dots,x_n,t),
 \bigtriangledown =\left(  \dfrac{\partial }{\partial x_1},\dots, \dfrac{\partial }{\partial x_n}\right) , t>0,x_r\in\mathbb{R},r=1,2,\dots,n,
 $ and the function $ F(u)>0$ denotes the diffusivity, the vector $K(u)=(K_1(u),\dots,K_n(u)) $ represents the convective transport of the medium which typically means the velocity and $R(u)$ is the reaction term. Also, they have studied the Lie symmetries of the equation \eqref{cherniha} with $n=2.$
 In 1983,  Dorodnitsyn et al. \cite{russian} have investigated the $(n+1)$-dimensional generalized diffusion-reaction equation
 \begin{equation}\label{dorodnitsyn}
 \dfrac{\partial u}{\partial t}= \sum_{r=1}^{n}\dfrac{\partial }{\partial x_r}\left[  F_r(u) \left(\dfrac{\partial u }{\partial x_r}\right) \right] + R(u), \quad F_r(u)>0,r=1,\dots,n,
 \end{equation}
where $u=u(x_1,\dots,x_n,t),x_r\in\mathbb{R},t>0$ and the functions $F_r(u),r=1,\dots,n,$ and $R(u)$ describe the diffusivity and kinetics of the processes, respectively. For $n=2$ and $n=3,$ they have presented the group properties of the equation \eqref{dorodnitsyn}. Also, we note that Polyanin and Zaitsev \cite{polyanin} have derived various types of exact solutions for the above equation \eqref{dorodnitsyn} with  $n=3$  using separation of the variable method with various transformations.
\\
In the present study, we consider the  most general form of the $(3+1)$-dimensional nonlinear time-fractional convection-diffusion-reaction (CDR)} equation in the form
\begin{equation}\label{rdeq}
	\dfrac{\partial^\alpha u}{\partial t^\alpha} =\sum\limits_{r=1}^{3} \dfrac{\partial}{\partial x_r}\left( F_{r}(u)\dfrac{\partial u}{\partial x_r}\right)+\sum\limits_{r=1}^{3}  K_{r}(u)\dfrac{\partial u}{\partial x_r}+R(u), \alpha \in (0,2]
\end{equation}
defined on the spatially bounded domain, $ \Omega \times [0,\infty) \subset \mathbb{R}^4 $
along with the appropriate initial and boundary conditions
\begin{eqnarray}
	&u(x_1,x_2,x_3,0)=\xi(x_1,x_2,x_3),\alpha\in(0,1],\label{a1}\\
	& u(x_1,x_2,x_3,0)=\xi(x_1,x_2,x_3) \, \& \,\dfrac{\partial u}{\partial t} \big{|}_{t=0}=\eta(x_1,x_2,x_3),\alpha\in(1,2],\label{a2}
\end{eqnarray}
{and}
\begin{eqnarray}\begin{aligned}
		u(x_1,x_2,x_3,t)\Big{|}_{\sum\limits_{i=1}^3\lambda_ix_i=0}=\left\{
		\begin{array}{ll}
			\delta_1(t),\text{if}\, \alpha\in(0,1]\label{bc},\\
			\delta_2(t),\text{if}\, \alpha\in(1,2],
		\end{array}  \right.\\
		\, \& \,
		u(x_1,x_2,x_3,t)\Big{|}_{\sum\limits_{i=1}^3\lambda_ix_i=l}=\left\{
		\begin{array}{ll}
			\delta_3(t),\text{if}\, \alpha\in(0,1],\\
			\delta_4(t),\text{if}\, \alpha\in(1,2],
		\end{array}  \right.
	\end{aligned}
\end{eqnarray}
where the fractional derivative of order $\alpha>0$ denoted by $ \dfrac{\partial^\alpha (\cdot)}{\partial t^\alpha}$ which is taken in the Caputo sense and defined \cite{pi,kai,kts} as
\begin{eqnarray}
	\begin{aligned}\label{c}
		\dfrac{\partial^\alpha u}{\partial t^\alpha}=
		\left\{ \begin{array}{ll}
			\dfrac{1}{\Gamma(p-\alpha)}\int\limits_{0}^{t}\dfrac{\partial ^{p} u(x_1,x_2,x_3,\tau)}{\partial \tau^{p} }(t-\tau)^{p-\alpha-1}d\tau, & p-1<\alpha< p ,\ p\in\mathbb{N}; \\
			\dfrac{\partial^ p u}{\partial t^p}, & \alpha=p \in\mathbb{N};
		\end{array}
		\right.
	\end{aligned}
\end{eqnarray}
and $\Omega=\big\{(x_1,x_2,x_3)\in \mathbb{R}^3 \big{|} 0\leq \sum\limits_{r=1}^3\lambda_rx_r\leq l,\ x_r,\lambda_r,l\in \mathbb{R},r=1,2,3\big\},$
the functions $F_{r}(u),K_r(u)$ and $R(u) $ represent diffusion process, convection process and reaction kinetics of the physical system under consideration, respectively for $u=u(x_1,x_2,x_3,t),t>0$ { and the functions $\delta_1(t),\delta_3(t)\in C^{1}([0,\infty))$, while $\delta_2(t),\delta_4(t)\in C^2([0,\infty)).$
Recently, Prakash et al. \cite{ppl} have discussed the exact solutions for the initial value problems of \eqref{rdeq} with $R(u)=0$ through direct approach of the invariant subspace method without any variable transformation. Here we wish to point out that when $\alpha\in(1,2],$ the above equation \eqref{rdeq} is known as the time-fractional convection-diffusion-reaction (CDR) wave equation, which helps to study the intermediate process between diffusion and wave phenomena simultaneously.
\\
We believe that to the  best of our knowledge, no one has discussed the invariant subspace method associated with the variable transformation for deriving
 	the exact solutions of the initial and boundary value problems of the given equation \eqref{rdeq}.} In this work, one of our aims is to find the various dimensional invariant subspaces for the given equation \eqref{rdeq} through the invariant subspace method associated with variable transformation. Using the invariant subspaces obtained, our second aim is to derive the following type of exact solutions for the initial and boundary value problems of the equation \eqref{rdeq}:
\begin{equation}
 	u(x_1,x_2,x_3,t)=\sum\limits_{r=1}^{n}F_{r}(t)G_r(\lambda_1x_1+\lambda_2x_2+\lambda_3x_3),\lambda_i\in\mathbb{R},i=1,2,3.
 	\label{3dsol}
 \end{equation}
The paper is structured as follows: Section 2 describes how to apply the invariant subspace method associated with variable transformation for the $(k+1)$-dimensional nonlinear time-fractional PDEs.
 Additionally, we provide a detailed study for constructing the exact solutions using the invariant subspaces.
 In section 3, we provide the description for the estimation of the invariant subspaces with different dimensions for the $(3+1)$-dimensional nonlinear time-fractional {CDR} equation.
  Section 4 explains the derivation of the exact solutions for the initial and boundary value problems of the $(3+1)$-dimensional nonlinear time-fractional {CDR} equation using the invariant subspaces.
   In section 5, the extension of the invariant subspace method associated with variable transformation for the $(k+1)$-dimensional nonlinear time-fractional PDEs with several time delays is investigated, and also the exact solutions of the $(3+1)$-dimensional delay nonlinear time-fractional {CDR} equation are derived. Then, in section 6, we provide a brief account of the applications of the theory developed in the earlier sections. Finally, section 7 provides a brief discussion and concluding remarks.
\section{Invariant subspace method associated with $(k+1)$-dimensional nonlinear time-fractional PDEs}
In this section, we present how to find the invariant subspaces associated with the following $(k+1)$-dimensional nonlinear time-fractional PDE
\begin{equation}\label{keq}
\dfrac{\partial ^\alpha u}{\partial t^\alpha}=\textbf{\textit{F}}[u], \alpha>0,t\geq0,
\end{equation}
 where   $ \textbf{\textit{F}}[u] $ is a sufficiently smooth $m$-th order nonlinear partial differential operator involving $ k $ independent space variables such that
 \begin{eqnarray}\label{kop}
\textbf{\textit{F}}[u]=\textbf{\textit{F}}\big(x_{1},x_2,\dots ,x_k,u,u^{(1)}_1,\dots,u^{(1)}_{k},u^{(2)}_{11},\dots,u^{(2)}_{r_1r_2},\dots,u^{(m)}_{r_1r_2\cdots r_m}\big).
\end{eqnarray}
Here $ u=u(x_1,x_2,\dots,x_k,t),$
$u^{(1)}_r=\dfrac{\partial u}{\partial x_r},u^{(2)}_{r_1r_2}=\dfrac{\partial^2 u}{\partial x_{r_1}\partial x_{r_2}},\dots,
u^{(m)}_{r_1r_2\cdots r_m}=\dfrac{\partial^m u}{\partial x_{r_1}\partial x_{r_2}\cdots\partial x_{r_m}}$,
$x_{r_s}\in \mathbb{R},r,r_s=1,2,\dots,k,s=1,2,\dots,m,$ and $k,m\in\mathbb{N}.$\\
 Now,  {we look for particular solutions of the form
 	% on the spatial variables $x_r,r=1,2,\dots,k$ as follows:
 \begin{equation}\label{wt}
 u(x_1,x_2,\dots,x_k,t) = w(z, t), z = \lambda_1 x_1 + \lambda_2 x_2 +\dots+ \lambda_k x_k, \end{equation}}
where $\lambda_r\in\mathbb{R},r=1,2,\dots,k, k\in\mathbb{N}.$
 {Using the ansatz \eqref{wt}, }{along with the invariant subspace method, the working procedure for finding exact solutions of \eqref{keq} can be summarized in the following steps:
	\begin{itemize}
		\item[Step-1 :]  {Using the ansatz \eqref{wt}, } the given $(k+1)$-dimensional equation \eqref{keq} reduces to the following $ (1+1) $-dimensional nonlinear time-fractional PDE,
	\begin{equation}\label{tpde}
		\dfrac{\partial ^\alpha w}{\partial t^\alpha}=\textbf{\textit{F}}_T[w],\alpha>0,
	\end{equation}
	where $ \textbf{\textit{F}}_T[w] $ is the transformed ordinary differential operator given by
	\begin{equation}\label{top}
		\textbf{\textit{F}}_T[w]=\textbf{\textit{F}}_T\big(z,w,w^{(1)}_z,\dots,
		w^{(m)}_z\big),w^{(j)}_z=\dfrac{\partial^j w}{\partial z^j}, j=1,2,\dots,m.
	\end{equation}
	\item[Step-2 :]
		Apply the invariant subspace method to the above reduced $ (1+1) $-dimensional nonlinear time-fractional PDE \eqref{tpde} for finding the invariant subspaces.
	\item[Step-3 :]
	Using the obtained invariant subspaces, the given $ (1+1) $-dimensional nonlinear time-fractional PDE \eqref{tpde} reduces to a system of fractional ODEs.
		\item[Step-4 :] Next, one can apply the well-known analytical methods to the obtained system of fractional ODEs. From this, we can look for particular form of the exact solution as follows:
	 $$ u(x_1,x_2,	\dots,x_k,t)=w(z,t)=\sum\limits_{s=1}^n C_s(t)\phi_s(z),$$
	 where $z=\lambda_1x_1+\lambda_2x_2+\cdots+\lambda_kx_k,n\in\mathbb{N}.$
\end{itemize}
Now,  {using the invariant subspace method as introduced by Galaktionov and Svirshchevskii  \cite{gs},}
let us explain how to find the invariant subspaces for the equation \eqref{tpde}.}
%%%%%%%%%%%%%%%%%%%%%%%%%%%%%%%%%%%%%%%%%%%%%%%%%%%%%%%%%%%%%%%%%%%%%%%%%%%%%%%%%%%%%%%%%%%%%%%%%%%%%%%%%%%%%%%%%%%%%%%%%%%%%%%%%%%%%%%%%%%%%%%%%%%%%%%%%%%%%%%%%%%%%%%%%%%%%%%%%%%%%%%%%%%%%%%%%%%%%%
First, we consider the $n$-{th} order linear homogeneous ODE
\begin{equation}\label{kode}
\dfrac{d^n y}{ d z^n}+p_{n-1}(z)\dfrac{d^{n-1} y}{ d z^{n-1}}+\cdots+p_0(z)y=0,
\end{equation}
where the functions $p_s(z),s=0,1,2,\dots,n-1,$ are continuous functions of z. Then there exist $n$-linearly independent solutions for the above  ODE  \eqref{kode}, say $\phi_s(z),s=1,2,\dots,n$.
 Let ${\mathbf{V}}_n$ be the linear space spanned by the above-mentioned linearly independent functions $ \phi_s(z), s = 1,2,..., n,$ which can be simply written as
 \begin{equation}\label{vs}
 \mathbf{V}_n =\text{Span}\{\phi_s(z)|s=1,2,\dots,n\}.
 \end{equation}
The above linear space (solution space) $\mathbf{V}_n$ is said to be invariant under the nonlinear ordinary differential operator $\textbf{\textit{F}}_T[w]$ given in \eqref{top} if
$w\in\mathbf{V}_n$ implies  $\textbf{\textit{F}}_T[w] \in \mathbf{V}_n. $ In other words, we can write as follows:
\begin{equation}\label{tinc}
w=\sum\limits_{s=1}^{n} C_s \phi_s(z)\, \text{ implies that } \,\textbf{\textit{F}}_T[w]=\sum\limits_{s=1}^{n} \Omega_s(C_1,C_2,\dots,C_n) \phi_s(z),
\end{equation}
where $  C_s\in\mathbb{R}, $ and the functions $ \Omega_s(C_1,C_2,\dots,C_n),s=1,2,\dots,n, $ denote the coefficient of expansion with respect to the basis set $ \{\phi_s(z)|s=1,2,\dots,n\}. $
\\
 Additionally, we wish to point out that the linear space $\mathbf{V}_n$ given in \eqref{vs} is invariant under the nonlinear partial differential operator  $ \textbf{\textit{F}}[u] $ given in \eqref{kop}   if
$w\in\mathbf{V}_n$ implies  $\textbf{\textit{F}}_T[w] \in \mathbf{V}_n. $
This means that if the transformed nonlinear ordinary differential operator
  $\textbf{\textit{F}}_T[w]$ admits the linear space $\mathbf{V}_n,$ then the  linear space $\mathbf{V}_n$ is also invariant under the given nonlinear partial differential operator
  $ \textbf{\textit{F}}[u] $ { and vice-versa.}

The following theorem helps to find the maximal dimension of the linear space for the given nonlinear differential operator \eqref{top}.
\begin{thm}\label{thm1}\cite{gs}
If the $n$-dimensional linear (vector) space is invariant under the given $ m $-th order nonlinear ordinary differential operator $ \textbf{{F}}_T[w]=\textbf{{F}}_T\big(z,w,w^{(1)}_z,\dots,
w^{(m)}_z\big), $ then $n\leq2m+1.$
\end{thm}
More precisely, by the virtue of maximal dimension theorem \ref{thm1}, the possible dimensions of linear space for the given $ m $-th order nonlinear
ordinary differential operator $ \textbf{\textit{F}}_T[w] $  are $ 1,2,\dots,2m+1. $
\\
 Now, we show how to find the exact solutions of the equation \eqref{keq} using the invariant subspace $ \mathbf{V}_n$ for the given differential operator $ \textbf{\textit{F}}_T[w] $ that is discussed below.
\begin{thm}\label{thm2}
Let % $u(x_1,x_2,\dots,x_k,t)=w(z,t),
$ \mathbf{V}_n =\text{Span}\{\phi_s(z)|s=1,2,\dots,n\} $ be an n-dimensional linear space. Suppose that $ \mathbf{V}_n $ is invariant under  the ordinary differential operator $ \textbf{\textit{F}}_T[w] $ given in \eqref{top}.
Then the  $(1+1)$-dimensional nonlinear time-fractional PDE
\begin{equation}\label{1pde}
	\dfrac{\partial^\alpha w}{\partial t^\alpha}= \textbf{\textit{F}}_T[w],\alpha>0,
\end{equation}
admits an exact solution in the finite separable form
  \begin{equation}\label{1pdesol}
 	w(z,t)=\sum\limits_{s=1}^n C_s(t)\phi_s(z),
 \end{equation}
if and only if  the $(k+1)$-dimensional nonlinear time-fractional PDE
\begin{equation}\label{rkeq}
	\dfrac{\partial ^\alpha u}{\partial t^\alpha}=\textbf{\textit{F}}[u], \alpha>0,
\end{equation}
admits an exact solution in the finite separable form
\begin{equation}\label{u2}
	u(x_1,x_2,\dots,x_k,t)=w(z,t)= \sum\limits_{s=1}^{n} C_{s}(t)\phi_s(\lambda_1x_1+\lambda_2x_2+\cdots+\lambda_kx_k),
\end{equation}
where $z=\lambda_1x_1+\lambda_2x_2+\cdots+\lambda_kx_k.$
Moreover, the functions $  C_{s}(t),s=1,2,\dots,n,  $ satisfy the following system of fractional ODEs
\begin{equation}\label{sys}
\dfrac{d ^\alpha C_{s}(t) }{d t^\alpha}=\Omega_s(C_1(t),C_2(t),\dots,C_n(t)),s=1,2,\dots,n.
\end{equation}
\end{thm}
Next, we present a specific study for finding the invariant subspaces and exact solutions of the $(3+1)$-dimensional generalized nonlinear time-fractional {CDR} equation \eqref{rdeq}.
%%%%%%%%%%%%%%%%%%%%%%%%%%%%%%%%%%%%%%%%%%%%%%%%%%%%%%%%%%%%%%%%%%%%%%%%%%%%%%%%%%%%%%%%%%%%%%%%%%%%%%%%%%%%%%%%%%%%%%%%%%%%%%%%%%%%%%%%%%%%%%%%%%%%%%
\section{Invariant subspaces associated with $(3+1)$-dimensional generalized nonlinear time-fractional {CDR} equation \eqref{rdeq}}
In this section, we present a detailed algorithmic study for finding the invariant subspaces of the $(3+1)$-dimensional generalized nonlinear time-fractional {CDR} equation \eqref{rdeq}. Thus, the nonlinear partial differential operator reads as follows:
 \begin{equation}\label{rd}
 \textbf{F}[u]= \sum\limits_{r=1}^{3} \dfrac{\partial}{\partial x_r}\left( F_{r}(u)\dfrac{\partial u}{\partial x_r}\right)+\sum\limits_{r=1}^{3}  K_{r}(u)\dfrac{\partial u}{\partial x_r}+R(u).
 \end{equation}
Now,  {using the ansatz \eqref{wt} with $k=3,$ } we can reduce the given partial differential operator \eqref{rd} with three independent space variables into the ordinary differential operator with one new independent space variable. Hence,  {we consider the particular solution of the equation \eqref{rdeq} as} $u(x_1,x_2,x_3,t)=w(z,t),z=\lambda_1x_1+\lambda_2x_2+\lambda_3x_3$.
Thus, the given $(3+1)$-dimensional generalized nonlinear time-fractional {CDR} equation \eqref{rdeq} is transformed into the $(1+1)$-dimensional generalized nonlinear time-fractional {CDR} equation in the form
{ \begin{equation}\label{trd}
 \dfrac{\partial^\alpha w}{\partial t^\alpha} = {\textbf{\textit{{F}}}}_T[w] \equiv   \dfrac{\partial}{\partial z}\left(F(w)\dfrac{\partial w}{\partial z}\right)+K(w)\dfrac{\partial w}{\partial z}+R(w), \alpha \in (0,2],
 \end{equation}
}
  and the corresponding initial and boundary conditions \eqref{a1}-\eqref{bc}  can be considered as follows,
  \begin{eqnarray}
 &w(z,0)=\xi(z),&\alpha\in(0,1],\label{ta1}\\
& w(z,0)=\xi(z) \,\,\, \& \,\, \,\dfrac{\partial w}{\partial t} \big{|}_{t=0}=\eta(z),& \alpha\in(1,2],\label{ta2}
\end{eqnarray}
{and}
\begin{eqnarray}
w(z,t)\Big{|}_{z=0}=\left\{
\begin{array}{ll}
\delta_1(t),\text{if}\, \alpha\in(0,1]\label{tbc},\\
\delta_2(t),\text{if}\, \alpha\in(1,2],
\end{array}  \right.
\, \& \quad
w(z,t)\Big{|}_{z=l}=\left\{
\begin{array}{ll}
\delta_3(t),\text{if}\, \alpha\in(0,1],\\
\delta_4(t),\text{if}\, \alpha\in(1,2],
\end{array}  \right.
\end{eqnarray}
{where the functions $F(w)=  \sum\limits_{r=1}^{3}\lambda_r^2  F_{r}(w)$ and $K(w)=\sum\limits_{r=1}^{3}\lambda_rK_{r}(w).$}
 Next, we present a detailed algorithmic study for finding finite-dimensional invariant subspaces for the nonlinear differential operator ${\textbf{\textit{{F}}}}_T[w] $ given in \eqref{trd}.
Now, we introduce the $n$-dimensional linear space
\begin{equation}\label{crdvs}
\mathbf{V}_n =\text{Span}\{ \phi_s(z)| s=1,2,\dots,n\},
\end{equation}
where the functions $ \phi_s(z), s=1,2,\dots,n,$ are linearly independent solutions of the following ODE:
\begin{equation}\label{ode}
\mathit{D}_n[y]\equiv\left[\dfrac{d^n }{ d z^n}+p_{n-1}(z)\dfrac{d^{n-1} }{ d z^{n-1}}+\cdots+p_0(z)\right]y=0.
\end{equation}
Here the functions { $p_s(z),s=0,1,2,\dots,n-1,$}
are continuous functions of z.
The solution space $ \mathbf{V}_n $ is invariant under the following nonlinear differential operator
{
\begin{equation}\label{trdop}
{\textbf{\textit{{F}}}}_T[w] =  \dfrac{\partial}{\partial z}\left(F(w)\dfrac{\partial w}{\partial z}\right)+K(w)\dfrac{\partial w}{\partial z}+R(w),
\end{equation}
 if $\textbf{\textit{F}}_T[w]\in \mathbf{V}_n
$ for all
$w \in \mathbf{V}_n.
$}
Thus, we get the following invariance conditions
\begin{eqnarray}
\label{inv1}
\mathit{D}_n\big[{\textbf{\textit{{F}}}}_T[w] \big]\equiv&\left[\dfrac{d^n }{ d z^n}+p_{n-1}(z)\dfrac{d ^{n-1} }{ d  z^{n-1}}+\cdots+p_0(z)\right]{\textbf{\textit{{F}}}}_T[w] =0
\\\nonumber
\text{whenever} \quad \mathit{D}_n[w]\equiv&\left[\dfrac{d ^n }{ d z^n}+p_{n-1}(z)\dfrac{d ^{n-1} }{d z^{n-1}}+\cdots+p_0(z)\right]w=0.
\end{eqnarray}
 By the maximal dimension theorem \ref{thm1}, the possible dimensions of the linear space  $ \mathbf{V}_n $  for the given differential operator $ {\textbf{\textit{{F}}}}_T[w] $   are $n=1,2,3,4,5,$  because dim$(\mathbf{V}_n)\leq 5.$

In order to determine the unknown functions { $
F_r(w),K_r(w),R(w), r=1,2,3,$ and $p_s(z), $  $s=0,1,2,\dots,n-1,$} and their corresponding invariant subspaces with different dimensions, we substitute
\begin{eqnarray}
&\dfrac{d ^n w}{ d z^n}=-\left[p_{n-1}(z)\dfrac{d ^{n-1} }{d z^{n-1}}+\cdots+p_0(z)\right]w
\end{eqnarray}
and the operator \eqref{trdop} into the equation \eqref{inv1},
we obtain an over-determined system of equations. Solving the obtained system, we obtain the differential operators with their corresponding linear spaces.
\\
 Now,  we present how to determine the two-dimensional invariant linear space with their corresponding differential operators.
 {\begin{thm}
		Consider the two-dimensional linear space $\mathbf{V}_2$ defined as the solution space of the second order linear homogeneous ODE
		\begin{eqnarray}\label{2ode}
			&\dfrac{d ^2 y}{ d z^2}+p_{1}(z)\dfrac{dy  }{d z}+p_0(z)y=0,
		\end{eqnarray}
		then there exists nonlinear differential operator ${\textbf{\textit{{F}}}}_T[w] $  that preserves the linear space defined by the ODE \eqref{2ode} when $p_1(z)p_0(z)=0.$
		\\
		The full description of the  two-dimensional invariant linear spaces and their corresponding differential operators ${\textbf{\textit{{F}}}}_T[w] $ are listed in Table \ref{crd1}.
	\end{thm}
	\begin{proof}	Let  $\mathbf{V}_2=\text{Span}\{\phi_1(z),\phi_2(z)\}$. Thus, $\text{dim}( \mathbf{V}_n)=n=2.$
		\\
		For $n=2$, the invariance condition \eqref{inv1} can be read as follows:
  	\begin{eqnarray}
\label{inv01}
\mathit{D}_2\big[{\textbf{\textit{{F}}}}_T[w] \big]\equiv\left[\dfrac{d^2 }{ d z^2}+p_{1}(z)\dfrac{d }{ d  z}+p_0(z)\right]{\textbf{\textit{{F}}}}_T[w] =0,\\
\nonumber\text{whenever}\, \dfrac{d ^2 w}{ d z^2}=-p_{1}(z)\dfrac{d w  }{d z}-p_0(z)w.
\end{eqnarray}
%%%%%%%%%%%%%%%%%%%%%%%%%%%%%%%%%%%%%%%%%%%%%%%%%%%%%%%%%%%%%%%%%%%%%%%%%%%%%%%%%%%%%%%%%%%%%%%%%%%%%%%%%%%%%%%%%%%%%%%%%%%%%%%%%%%%%%%%%%%%%%%%%%	
  Simplifying the above invariant condition \eqref{inv01} and equating the coefficients of $w_z, (w_z)^2$, etc., independently equal to zero, we obtain the following system of equations:
 \begin{eqnarray}\label{overD}
 	\begin{aligned}\label{odsys}
 		&\textbf{D}^{(3)}(F(w))=0,
 		-5\textbf{D}^{(2)}(F(w))p_1(z)+\textbf{D}^{(2)}(K(w))=0,
 	- 6\textbf{D}^{(2)}(F(w))p_0(z)w-	\\
 		&
 		 \textbf{D}(F(w))\left(4\dfrac{d}{d z} p_1(z)+3p_0(z)-4p_1^2(z)\right)-2p_1(z)\textbf{D}(K(w))
 		+\textbf{D}^{(2)}(R(w))=0,
 		\\
 		&-\textbf{D}(F(w))\left(4w\dfrac{d}{d z}p_0(z)-7wp_0(z)p_1(z)\right)
 		-3wp_0(z)\textbf{D}(K(w))
 		-F(w)\left(2\dfrac{d}{d z} p_0(z)
 		\right.\\&\left.
 		-2p_1(z)\dfrac{d}{d z} p_1(z)
 		+\dfrac{d^2}{d z^2} p_1(z)\right)
 		-K(w)\dfrac{d}{d z} p_1(z)=0,
 		3\textbf{D}(F(w))w^2p_0^2(z)-K(w)w\dfrac{d}{d z} p_0(z)
 	\\&	+F(w)\left( -w \dfrac{d^2}{d z^2} p_0(z)+2p_0(z)w\dfrac{d}{d z} p_1(z)\right) 
 		+\textbf{D}(R(w))wp_0(z)+R(w)p_0(z)=0,
 	\end{aligned}
 \end{eqnarray}
 where $ \textbf{D}^{(i)}(\cdot) $ denotes the $i$-th order total derivative of the
 corresponding functions, $i=1,2,$  the functions $F(w)=  \sum\limits_{r=1}^{3}\lambda_r^2  F_{r}(w),$ and $K(w)=\sum\limits_{r=1}^{3}\lambda_rK_{r}(w).$ Substituting $F(w)$ and $K(w)$ in \eqref{overD}, we obtain the following equations:
 \begin{eqnarray}
 	\begin{aligned}\label{odsys} &\sum\limits_{r=1}^{3}\lambda_r^2\textbf{D}^{(3)}F_{r}(w)=0,
  		-5\left( \sum\limits_{r=1}^{3}\lambda_r^2\textbf{D}^{(2)}F_{r}(w)\right) p_1(z)+ \sum\limits_{r=1}^{3}\lambda_r\textbf{D}^{(2)}K_{r}(w)=0,
 		\\
 		&
 		- \left( \sum\limits_{r=1}^{3}\lambda_r^2\textbf{D}F_{r}(w)\right)\left(4\dfrac{d}{d z} p_1(z)+3p_0(z)-4p_1^2(z)\right)- 6\left( \sum\limits_{r=1}^{3}\lambda_r^2\textbf{D}^{(2)}F_{r}(w)\right)p_0(z)w
 		\\
 		&	+\textbf{D}^{(2)}(R(w))-2p_1(z)\left( \sum\limits_{r=1}^{3}\lambda_r\textbf{D}K_{r}(w)\right)
 	=0,-\left( \sum\limits_{r=1}^{3}\lambda_r^2\textbf{D}F_{r}(w)\right)\Big(4w\dfrac{d}{d z}p_0(z)
 	\\&	-7wp_0(z)p_1(z)\Big)
 		-3wp_0(z)\left( \sum\limits_{r=1}^{3}\lambda_r\textbf{D}K_{r}(w)\right)-\left( \sum\limits_{r=1}^{3}\lambda_rK_{r}(w)\right)\dfrac{d}{d z} p_1(z)
 	\\&	-\left( \sum\limits_{r=1}^{3}\lambda_r^2F_{r}(w)\right)
 		\left(2\dfrac{d}{d z} p_0(z)
 		 			-2p_1(z)\dfrac{d}{d z} p_1(z)	+\dfrac{d^2}{d z^2} p_1(z)\right)
 		=0,
 		3\left( \sum\limits_{r=1}^{3}\lambda_r^2\textbf{D}F_{r}(w)\right)w^2p_0^2(z)
  \\&	+\left( \sum\limits_{r=1}^{3}\lambda_r^2F_{r}(w)\right)
 		\left( - \dfrac{d^2}{d z^2} p_0(z)+2p_0(z)\dfrac{d}{d z} p_1(z)\right)w -\left( \sum\limits_{r=1}^{3}\lambda_rK_{r}(w)\right)w\dfrac{d}{d z} p_0(z)
 	\\
 	&	+\textbf{D}(R(w))wp_0(z)+R(w)p_0(z)=0,
 	 	\end{aligned}
 \end{eqnarray}
 where $ \textbf{D}^{(i)}(\cdot) $ denotes the $i$-th order total derivative of
 the corresponding functions, $i=1,2.$
%%%%%%%%%%%%%%%%%%%%%%%%%%%%%%%%%%%%%%%%%%%%%%%%%%%%%%%%%%%%%%%%%%%%%%%%%%%%%%%%%%%%%%%%%%%%%%%%%%%%%%%%%%%%%%%%%%%%%%%%%%%%%%%%%%%%%%%%%%%%%%%%%%%%%%
The above system \eqref{odsys} is not solvable in general.
Thus we take  either $p_1(z)=0$ or $p_0(z)=0.$
Let us first consider the case $p_1(z)=0.$  Thus, we obtain one two-dimensional linear space
$
\mathbf{V}_2=\text{Span}\{ sin(\sqrt{\rho_0}z), cos(\sqrt{\rho_0}z)\}
$ with their corresponding differential operator ${\textbf{\textit{{F}}}}_T[w] =\lambda_1^2\dfrac{\partial}{\partial z}\left((\frac{\mu_1w^2}{3\lambda_1^2\rho_0}+\mu_4)\dfrac{\partial w}{\partial z} \right) +\lambda_1\mu_3\dfrac{\partial w}{\partial z} +\mu_1w^3+\mu_2w.$ Next, we consider $p_0(z)=0.$ In this case, we obtain the four different two-dimensional linear spaces with their corresponding distinct differential operators $ {\textbf{\textit{{F}}}}_T[w].$ For these two cases,  we have listed the obtained two-dimensional invariant subspaces with their corresponding differential operators,
%  {where  $\mu_1,\mu_2,\dots,$ are arbitrary constants,}
 in Table \ref{crd1}.
\end{proof}
 In a similar way, we can find the various other dimensional linear spaces with their corresponding differential operators $ {\textbf{\textit{{F}}}}_T[w].$ For example, we consider the dimensions of linear spaces to be 1,3 and 4. For these cases, the obtained linear spaces with their corresponding differential operators $ {\textbf{\textit{{F}}}}_T[w]$,
 % {where  $\mu_1,\mu_2,\dots,$ denote arbitrary constants,}
  are listed in Tables \ref{crd2} and \ref{crd3}.
%%%%%%%%%%%%%%%%%%%%%%%%%%%%%%%%%%%%%%%%%%%%%%%%%%%%%%%%%%%%%%%
\\
Additionally, we can consider the particular forms of $F_r(w),K_r(w),r=1,2,3,$ and $ R(w)$ as given below:
\begin{eqnarray}\label{cubic}
	\begin{aligned}
		&	F_1(w)=\beta_2w^2+\beta_1w+\beta_0,\\
		&	F_2(w)=\kappa_2w^2+\kappa_1w+\kappa_0,\\
		&	F_3(w)=\zeta_2w^2+\zeta_1w+\zeta_0,
		\\
		&	K_1(w)=f_2w^2+f_1w+f_0,\\
		&	K_2(w)=g_2w^2+g_1w+g_0,\\
		&	K_3(w)=h_2w^2+h_1w+h_0,
		\\
		&	R(w)=r_3w^3+r_2w^2+r_1w+r_0,
	\end{aligned}
\end{eqnarray}
where $\beta_i,\kappa_i,\zeta_i,f_i,g_i,h_i,i=0,1,2, $ and $r_j,j=0,1,2,3,$ are real arbitrary constants.
For this case,  the given differential operator $ {\textbf{\textit{{F}}}}_T[w]$
is obtained as
\begin{eqnarray}
	\begin{aligned}\label{plop}
	{\textbf{\textit{{F}}}}_T[w] &=\dfrac{\partial }{\partial z}\left[  \Big({\lambda_1^2}(\beta_2w^2+\beta_1w+\beta_0)+
{\lambda_2^2}(\kappa_2w^2+\kappa_1w+\kappa_0)+
{\lambda_3^2}(\zeta_2w^2+\zeta_1w+\zeta_0)\Big)\dfrac{\partial w}{\partial z}\right]
\\&+\left[ \lambda_1(f_2w^2+f_1w+f_0)
+
\lambda_2(g_2w^2+g_1w+g_0)
+
\lambda_3(h_2w^2+h_1w+h_0)\right] \dfrac{\partial w}{\partial z}
\\&
+r_3w^3+r_2w^2+r_1w+r_0.
	\end{aligned}
	\end{eqnarray}
Now let us consider dim($\mathbf{V}_n$)=2 and dim($\mathbf{V}_n$)=3. For these cases, we obtain various types of two and three-dimensional linear spaces  $\mathbf{V}_n$ with their corresponding quadratic and cubic nonlinear differential operators that are listed in Tables \ref{crd4} and \ref{crd5}.
\\
Here we would like to mention that in a similar way, we can find the various other dimensional linear spaces corresponding to the differential operators ${\textbf{\textit{{F}}}}_T[w]$ with different nonlinearity.
%\section{Tables}
\begin{sidewaystable}[htp]
	\caption{{Two-dimensional invariant subspaces for the differential operator $ {\textbf{\textit{{F}}}}_T[w] $ given in \eqref{trdop}.}}
	\begin{center}
		\begin{tabular}{cllc}%{p{1.25in}dddd}
			\toprule
			Cases& { \qquad \qquad Coefficients of the operator \eqref{trdop}}
			%& $\mathit{D}_n[w]\equiv\dfrac{d^n w}{d z^n}+\dots+\rho_0(z)w=0$
			& Invariant Subspaces
			\\
	%		&{$F(w)=\sum\limits_{r=1}^3\lambda_r^2F_r(w),K(w)=\sum\limits_{r=1}^3\lambda_rK_r(w),  \, \& \, R(w)$}
			%& $n=1,2,3$			&${ \mathbf{V}_2 =\text{Span}\{ \phi_1(z),\phi_2(z)\}} $ \\
			\midrule
			
			1.&
			{$\begin{array}{ll}
				F(w)={\lambda_1^2}\left( \dfrac{1}{2}\mu_5 w^2+\mu_6 w+\mu_7\right),
				K(w)={\lambda_1}\left( \mu_3w+\mu_4\right)
				,
				R(w)=\mu_1 w+\mu_2
				\\
			\end{array}
			$}
					& $
			\mathbf{V}_2 =\text{Span}\{ 1,z\}
			$
			\\
			\hline\\		
			2.&
			{
				$\left.\begin{array}{ll}
				F(w)={\lambda_1^2}\left(
				-\frac{1}{12}\big(\frac{-6\mu_5\rho_1\lambda_1w-\mu_1w^2+3\mu_2w}{\rho_1^2\lambda_1^2}\big)+\mu_7\right)
				\\
				K(w)={\lambda_1}\left( \frac{5}{12}\frac{\mu_1 w^2}{\rho_1 \lambda_1}+\mu_5w+\mu_6\right)
				,
					R(w)=\frac{1}{6}\mu_1w^3+\frac{1}{2}\mu_2 w^2+\mu_3 w+\mu_4\qquad\qquad
				\\
			\end{array}
			\right.$
		}
						& $
			\mathbf{V}_2 =\text{Span}\{ 1,e^{-\rho_1 z}\}
			$\\
			\hline
			\\
		
			3.&
			{
				$\begin{array}{ll}
				F(w)=\lambda_1^2(\frac{\mu_1w^2}{3\lambda_1^2\rho_0}+\mu_4),
				K(w)={\lambda_1}\mu_3,
					R(w)=\mu_1w^2+\mu_2
				\\
			\end{array}
			$
		}
					& $
			\mathbf{V}_2=\text{Span}\{ sin(\sqrt{\rho_0}z), cos(\sqrt{\rho_0}z)\}
			$\\
			\hline \\
			4.&
			{
				$\begin{array}{ll}
				F(w)={\lambda_1^2}\left( \mu_4w+\mu_5\right)
				,
				K(w)=0
				,
				R(w)=\mu_2w+\mu_3\\
			\end{array}
			$
		}
						& $
			\mathbf{V}_2=\text{Span}\{1,(z-\mu_1)^2\}
			$\\
			\hline
			\\
			
			5.&
			{
				$\begin{array}{ll}
				F(w)={\lambda_1^2}\left( \frac{\mu_1^2\mu_3}{4\lambda_1^2}w+\mu_6\right)
				,
							K(w)=0
				,
								R(w)=\frac{1}{2}\mu_3 w^2+\mu_4w+\mu_5\qquad\quad\quad\qquad\qquad
				\\
			\end{array}
			$
		}
			%&$\mathit{D}_1[w]\equiv\dfrac{d w}{d z}+\rho_0w=0$
			& $
			\mathbf{V}_2=\text{Span}\{1,sin(\frac{z+\mu_2}{\mu_1})\}
			$
	\\
			\bottomrule
		& 	Here, we consider $ \mu_i\in\mathbb{R},i=1,2,\dots,7.\qquad\qquad\qquad \qquad\qquad\qquad\qquad\quad\qquad\qquad$
		\end{tabular}
	\end{center}
	\label{crd1}
%\end{sidewaystable}\begin{sidewaystable}[htp]
\caption{
	{
		Different dimensional invariant subspaces for the differential operator $ {\textbf{\textit{{F}}}}_T[w] $ given in \eqref{trdop} .
	}
}
\begin{center}
\begin{tabular}{clll}%{p{1.25in}dddd}
\toprule
{Cases}& {\qquad\qquad Coefficients of the operator \eqref{trdop}}
%& $\mathit{D}_n[w]\equiv\dfrac{d^n w}{d z^n}+\dots+\rho_0(z)w=0$
& Invariant Subspaces
\\
%& {$F(w)=\sum\limits_{r=1}^3\lambda_r^2F_r(w),K(w)=\sum\limits_{r=1}^3\lambda_rK_r(w),  \, \& \, R(w)$}
%& $n=1,2,3$&$ \mathbf{V}_n =\text{Span}\{ \phi_s(z)| s=1,\dots,n\} $ \\
\midrule
6.&
{
	$\begin{array}{ll}
	F(w)={\lambda_1^2}\left( \frac{\mu_1 w}{2\lambda_1 \rho_0}+\mu_3\right)
,
	K(w)={\lambda_1}\left( \mu_1w+\mu_2\right)
,
R(w)=\mu_0w\qquad\qquad\qquad\qquad\quad
	\\
\end{array}
$
}
%&$\mathit{D}_1[w]\equiv\dfrac{d w}{d z}+\rho_0w=0$
& $
\mathbf{V}_1 =\text{Span}\{ e^{-\rho_0z}\}
$
\\

 7.&
 {
 	$\begin{array}{ll}
F(w)={\lambda_1^2}\left( {\mu_4w}+\mu_5\right)
,
K(w)={\lambda_1}\mu_3,
R(w)=\mu_1w+\mu_2\qquad\qquad\qquad\qquad\qquad\quad\quad
\\
\end{array}
$
}
%&$\mathit{D}_1[w]\equiv\dfrac{d w}{d z}+\rho_0w=0$
& $
 \mathbf{V}_3 =\text{Span}\{ 1,z,\frac{1}{2}z^2\}
 $\\

 8.&
 {
 	$\begin{array}{ll}
F(w)={\lambda_1^2}\left( \frac{1}{4}\frac{\mu_1w}{\lambda_1^2 \rho_1}+\mu_5\right)
,
K(w)={\lambda_1}\mu_4,
R(w)=\frac{1}{2}\mu_1w^2+\mu_2w+\mu_3\qquad
\qquad\qquad\qquad
\\
\end{array}
$}
& $
 \mathbf{V}_3 =\text{Span}\{ 1,sin(\sqrt{\rho_1}z),cos(\sqrt{\rho_1}z)\}
 $\\
 9. &
 {
 	$\begin{array}{ll}
F(w)={\lambda_1^2}\mu_5
,
K(w)=0,
R(w)=\mu_3w+\mu_4\qquad\qquad
\qquad\qquad\qquad\qquad\quad\quad\quad\quad\quad\quad
\\
\end{array}
$}
& $\begin{array}{ll}
 \mathbf{V}_3 =\text{Span}\{ 1,z,-\mu_1^2cos(\frac{\mu_2+z}{\mu_1})\}
\end{array} $\\

10. &
{
	$\begin{array}{ll}
F(w)={\lambda_1^2}\left( \frac{\mu_1w}{6\lambda_1^2\rho_1}+\mu_4\right) ,
K(w)={\lambda_1}\left( \frac{7\sqrt{2}}{12\lambda_1\sqrt{p_1}}\mu_1w+\mu_4\right)
,
R(w)=\frac{1}{2}\mu_1w^2+\mu_2w+\mu_3\quad
\\
\end{array}
$}
& $\left.\begin{array}{ll}
 \mathbf{V}_3 =\text{Span}\{ 1,e^{-\frac{1}{2}\sqrt{2p_1}z},e^{-\sqrt{2p_1}z}\}\\
  \end{array}
 \right.$\\

 11.&
 {
 	$\begin{array}{ll}
F(w)=\frac{\lambda_1\mu_2}{\rho_2},
K(w)=0,
R(w)=\mu_1w\qquad\qquad\qquad\qquad\qquad\qquad\qquad\qquad\quad\qquad\qquad
\\
\end{array}
$
}
& $\left.\begin{array}{ll}
 \mathbf{V}_4 =\text{Span}\{ 1,z,sin(\sqrt{\rho_2}z),cos(\sqrt{\rho_2}z)\}\\
  \end{array}
 \right.$\\
 \bottomrule
 	Here, & we consider $ \mu_i\in\mathbb{R},i=0,1,2,\dots,5.\qquad\qquad\qquad\qquad\quad\qquad\qquad\qquad\qquad\quad\qquad\qquad\quad$
\end{tabular}
\end{center}
\label{crd2}
\end{sidewaystable}

\begin{sidewaystable}[htp]
\caption{Different dimensional invariant subspaces for the differential operator  $ {\textbf{\textit{{F}}}}_T[w] $ given in \eqref{trdop}.}
\begin{center}
\begin{tabular}{cclc}%{p{1.25in}dddd}
\toprule
Cases&{ {Coefficients of the operator \eqref{trdop}}}
%& $\mathit{D}_n[w]\equiv\dfrac{d^n w}{d z^n}+\dots+\rho_0(z)w=0$
& Invariant Subspaces
\\
%&{	$F(w)=\sum\limits_{r=1}^3\lambda_r^2F_r(w),K(w)=\sum\limits_{r=1}^3\lambda_rK_r(w),  \, \& \, R(w)$}%& $n=1,2,3$&$ \mathbf{V}_n =\text{Span}\{ \phi_s(z)| s=1,\dots,n\} $ \\
\midrule
12. &
{
	$\left.\begin{array}{ll}
F(w)={\lambda_1^2}\mu_4
,
K(w)={\lambda_1}\mu_3
,R(w)=\mu_1w+\mu_2
\\
\end{array}
\right.$
}
& $\left.\begin{array}{ll}
\mathbf{V}_3 =\text{Span}\{ 1,e^{-\frac{1}{2}(\rho_1+\sqrt{p_1^2-4p_1})z},e^{-\frac{1}{2}(\rho_1-\sqrt{p_1^2-4p_1})z}\}\\
\mathbf{V}_3 =\text{Span}\{ 1,cosh(\sqrt{\rho_2}z),sinh(\sqrt{\rho_2}z)\}\\
 \mathbf{V}_3 =\text{Span}\{ 1,z,e^{-\rho_1z}\}\\
  \mathbf{V}_4 =\text{Span}\{ 1,z,cos(\sqrt{\rho_2}z),sin(\sqrt{\rho_2}z)\}\\
  \mathbf{V}_4 =\text{Span}\{ 1,z,z^2,e^{\rho_3z})\}\\
  \mathbf{V}_4 =\text{Span}\{ 1,z,z^2,z^3\}\\
 % \mathbf{V}_4 =\text{Span}\{ 1,z,-\rho_1^3sin(\frac{\rho_2+z}{\rho_1})\}
\end{array}
 \right.$\\
 \hline
 \\
13. &
{
	$\left.\begin{array}{ll}
F(w)={\lambda_1}\frac{\mu_3}{2\rho_3}
,
K(w)=-{\lambda_1}\mu_3
,R(w)=\mu_1w+\mu_2
\\
\end{array}
\right.$
}
& $\left.\begin{array}{ll}
 \mathbf{V}_3 =\text{Span}\Big\{ 1,z,4cosh\big[\frac{1}{2}(\frac{\rho_1}{\rho_2}+\frac{z}{\rho_1})\big]\big(cosh(\rho_3z)-sinh(\rho_3z)\big)\\ \qquad   +2sinh\big[\frac{1}{2}(\frac{\rho_1}{\rho_2}+\frac{z}{\rho_1})\big]\big(cosh(\rho_3z)-sinh(\rho_3z)\big)\Big\}\\
  \mathbf{V}_3 =\text{Span}\Big\{ 1, z,-\frac{2\rho_1^2}{(2\rho_1\rho_3+1)^2}sinh\big(\frac{(2\rho_1\rho_3+1)(z+\rho_2)}{2\rho_1}\big)
  \\
\qquad  -\frac{2\rho_1^2}{(2\rho_1\rho_3-1)^2}\Big[sinh\big(\frac{(2\rho_1\rho_3-1)(z+\rho_2)}{2\rho_1}\big)-cosh\big(\frac{(2\rho_1\rho_3-1)(z+\rho_2)}{2\rho_1}\big)\Big]
\\\qquad
+\frac{2\rho_1^2}{(2\rho_1\rho_3+1)^2}cosh\big(\frac{(2\rho_1\rho_3+1)(z+\rho_2)}{2\rho_1}\big)
  \Big\}\\

  \mathbf{V}_4 =\text{Span}\Big\{ 1,z,z^2, \frac{4\rho_1^3}{(2\rho_1\rho_3+1)^3}sinh\big(\frac{(2\rho_1\rho_3+1)(z+\rho_2)}{2\rho_1}\big)
  \\
\qquad  +\frac{4\rho_1^3}{(2\rho_1\rho_3-1)^3}\Big[\sinh\big(\frac{(2\rho_1\rho_3-1)(z+\rho_2)}{2\rho_1}\big)-cosh\big(\frac{(2\rho_1\rho_3+1)(z+\rho_2)}{2\rho_1}\big)\big]
\\\qquad
-\frac{4\rho_1^3}{(2\rho_1\rho_3-1)^3}cosh\big(\frac{(2\rho_1\rho_3-1)(z+\rho_2)}{2\rho_1}\big)
  \Big\}\\
\end{array}
 \right.$\\
\bottomrule
& Here, we consider 	$	 \mu_i\in\mathbb{R},i=1,2,\dots4.\qquad\qquad\qquad\qquad\quad$
\end{tabular}
\end{center}
\label{crd3}
\end{sidewaystable}

\begin{sidewaystable}[htp]
	\caption{{Different dimensional invariant subspaces for the differential operator  $ {\textbf{\textit{{F}}}}_T[w] $ given in \eqref{plop} with quadratic nonlinearity.}}
	\begin{center}
		\begin{tabular}{cclc}%{p{1.25in}dddd}
			\toprule
			Cases& {{Coefficients of the operator \eqref{plop}}}
			%& $\mathit{D}_n[w]\equiv\dfrac{d^n w}{d z^n}+\dots+\rho_0(z)w=0$
			& Invariant Subspaces
			\\
		%	& {$F(w)=\sum\limits_{r=1}^3\lambda_r^2F_r(w),K(w)=\sum\limits_{r=1}^3\lambda_rK_r(w),  \, \& \, R(w)$}
			%& $n=1,2,3$			&$ \mathbf{V}_n =\text{Span}\{ \phi_s(z)| s=1,\dots,n\} $ \\
			\midrule
			1.&
			{
				$\left.
				\begin{array}{ll}\left.
					\begin{array}{ll}
						F(w)=\lambda_1^2(\beta_1w+\beta_0)+
						\lambda_2^2(\kappa_1w+\kappa_0)+
						\lambda_3^2(\zeta_1w+\zeta_0)
						\qquad\qquad\qquad\qquad\quad\quad\\
						K(w)=\lambda_{1}(f_1w+f_0)+
						\lambda_2(
						g_1w+g_0)+
						\lambda_3
						(h_1w+h_0)
						,R(w)=r_1w+r_0
						\\
					\end{array}
					\right.\\
					\left.
					\begin{array}{ll}
						F(w)=\lambda_1^2(\beta_1w+\beta_0)+
						\lambda_2^2(\kappa_1w+\kappa_0)+
						\lambda_3^2(\zeta_1w+\zeta_0)\\
						K(w)=\lambda_{1}f_0
						+\lambda_2g_0
						+\lambda_3h_0
						,R(w)=0
						\\
					\end{array}
					\right.
				\end{array}
				\right.$
			}
			& $
			\mathbf{V}_2 =\text{Span}\{ 1,z\}
			$\\
			\hline \\
			
			2.&
			{
				$\left.\begin{array}{ll}
					\left.
					\begin{array}{ll}
						F(w)=\lambda_1^2(\beta_1w+\beta_0)+
						\lambda_2^2(\kappa_1w+\kappa_0)+
						\lambda_3^2(\zeta_1w+\zeta_0)\\
						K(w)=\lambda_{1}(f_1w+f_0)+
						\lambda_2(
						g_1w+g_0)+
						\lambda_3
						(h_1w+h_0)
						\\
						R(w)=\big[-2(\beta_1\lambda_1^2+\kappa_1\lambda_2^2+\zeta_1\lambda_3^2)\rho_1^2
						+\rho_1(f_1\lambda_1+g_1\lambda_2+h_1\lambda_3)\big]w^2+r_1w+r_0
					\end{array}
					\right.\\
					\left.
					\begin{array}{ll}
						F(w)=\lambda_1^2(\dfrac{\beta_1}{\lambda_1^2}w+\beta_0)+
						\lambda_2^2(\kappa_1w+\kappa_0)+
						\lambda_3^2(\zeta_1w+\zeta_0)\\
						K(w)=\lambda_{1}(\dfrac{2\beta_1\rho_1}{\lambda_1}w+f_0)+
						\lambda_2
						g_0+
						\lambda_3
						h_0
						,
						R(w)=0
					\end{array}
					\right.
				\end{array}\right.$
			}
			&
			$\mathbf{V}_2 =\text{Span}\{ 1,e^{-\rho_1z}\}
			$
			\\
			\hline
			\\					
			3.&{
				$\left.
				\begin{array}{ll}
					F(w)=\lambda_1^2(\beta_1w+\beta_0)+
					\lambda_2^2(\kappa_1w+\kappa_0)+
					\lambda_3^2(\zeta_1w+\zeta_0)\\
					K(w)=\lambda_{1}\big[\big(\frac{7(\beta_1\lambda_1^2+\kappa_1\lambda_2^2+\zeta_1\lambda_3^2)-3(g_1\lambda_2+h_1\lambda_3)}{3\lambda_1}\big)w+f_0\big]+
					\lambda_2(
					g_1w+g_0)+
					\lambda_3
					(h_1w+h_0)
					\\	
					R(w)=\big[\frac{2}{3}(\beta_1\lambda_1^2+\kappa_1\lambda_2^2+\zeta_1\lambda_3^2)\rho_1^2 \big] w^2+r_1w
					\\
				\end{array}
				\right.$
			}
			& $
			\mathbf{V}_2 =\text{Span}\{ e^{-\frac{1}{3}\rho_1z}, e^{-\frac{2}{3}\rho_1z}\}
			$\\
			\hline \\
			
			4.&
			{$ \left.
				\begin{array}{ll}
					F(w)=\lambda_1^2(\beta_1w+\beta_0)+
					\lambda_2^2(\kappa_1w+\kappa_0)+
					\lambda_3^2(\zeta_1w+\zeta_0)
					\qquad\qquad\qquad\qquad\qquad\qquad\\
					K(w)=\lambda_{1}f_0
					+\lambda_2g_0
					+\lambda_3h_0
					\\R(w)=2(\beta_1\lambda_1^2+\kappa_1\lambda_2^2+\zeta_1\lambda_3^2)\rho_1w^2+ r_1w+r_0
					\\
				\end{array}
				\right.$}
			& $
			\mathbf{V}_3 =\text{Span}\{ 1,sin(\sqrt{\rho_1}z),cos(\sqrt{\rho_1}z)\}
			$\\
			
			\bottomrule
		\end{tabular}
	\end{center}
	\label{crd4}
\end{sidewaystable}

\begin{sidewaystable}[htp]
\caption{{Different dimensional invariant subspaces for the cubic nonlinear differential operator $ {\textbf{\textit{{F}}}}_T[w] $ given in \eqref{plop}.}}
\begin{center}
\begin{tabular}{cccc}%{p{1.25in}dddd}
\toprule
Cases &{ {Coefficients of the operator \eqref{plop}}}
& Invariant Subspaces
\\
%& {$F(w)=\sum\limits_{r=1}^3\lambda_r^2F_r(w),K(w)=\sum\limits_{r=1}^3\lambda_rK_r(w),  \, \& \, R(w)$}
%& $n=1,2,3$&$ \mathbf{V}_n =\text{Span}\{ \phi_s(z)| s=1,\dots,n\} $ \\
\midrule

 5.&
 {$
 	\left.\begin{array}{ll}\left.\begin{array}{ll}
F(w)={\lambda_1^2}(\beta_2w^2+\beta_1w+\beta_0)+
{\lambda_2^2}(\kappa_2w^2+\kappa_1w+\kappa_0)+
{\lambda_3^2}(\zeta_2w^2+\zeta_1w+\zeta_0)
\\
K(w)=		\lambda_1\left[\big(\frac{5(\beta_2\lambda_1^2+\kappa_2\lambda_2^2+\zeta_2\lambda_3^2)\rho_1-(g_1\lambda_2+h_1\lambda_3)}{\lambda_1}\big
)w^2+f_1w+f_0\right]\\ \qquad
+
\lambda_2(g_2w^2+g_1w+g_0)
+
\lambda_3(h_2w^2+h_1w+h_0)
\\
R(w)=\big[2(\beta_2\lambda_1^2+\kappa_2\lambda_2^2+\zeta_2\lambda_3^2)\rho_1^2\big]w^3+
\big[-2(\beta_1\lambda_1^2+\kappa_1\lambda_2^2+\zeta_1\lambda_3^2)\rho_1^2
\\\qquad +(f_1\lambda_1+g_1\lambda_2+h_1\lambda_3)\rho_1\big]w^2+r_1w
\\
\end{array}
\right.
\\
\left.\begin{array}{ll}
F(w)=\lambda_1^2\Big(\big[\frac{r_3}{2\lambda_1^2\rho_1^2}\big]w^2+\big[\frac{(\lambda_1f_1+\lambda_2g_1+\lambda_3h_1)-r_2}{2\lambda_1^2\rho_1}\big]w+\beta_0\Big)
+\lambda_2^2\kappa_0
+\lambda_3^2\zeta_0
\\
K(w)=\lambda_1\Big[\big(\frac{5r_3}{2\lambda_1\rho_1}\big)w^2+f_1w+f_0\Big]
+\lambda_2(g_1w+g_0)
+\lambda_3(h_1w+h_0)
\\
R(w)=r_3w^3+r_2w^2+r_1w+r_0
\\
\end{array}
\right.
\end{array}
\right.
$
}
&
 $\mathbf{V}_2 =\text{Span}\{ 1,e^{-\rho_1z}\}
 $

\\\hline\\
6.&
{$
	\left.\begin{array}{ll}
F(w)={\lambda_1^2}(\beta_2w^2+\beta_1w+\beta_0)+
{\lambda_2^2}(\kappa_2w^2+\kappa_1w+\kappa_0)+
{\lambda_3^2}(\zeta_2w^2+\zeta_1w+\zeta_0)\qquad
\\
K(w)=
\lambda_1(f_1w+f_0)
+
\lambda_2(g_1w+g_0)
+
\lambda_3(h_1w+h_0)
\\
R(w)=r_1w+r_0
\\
\end{array}
\right.$
}
&
 $\mathbf{V}_2 =\text{Span}\{ 1,z\}
 $
\\
\bottomrule
\end{tabular}
\end{center}
\label{crd5}
\end{sidewaystable}
\section{Exact solutions of the equation \eqref{rdeq}}
This section presents how to construct the different types of exact solutions for the given $(3+1)$-dimensional generalized nonlinear time-fractional {CDR} equation \eqref{rdeq} using the obtained invariant linear spaces that are listed in Tables \ref{crd1}-\ref{crd5}.
\subsection{{Various types of} exact solutions for nonlinear time-fractional {CDR} equations}
In this subsection, let us first construct the exact solutions for the nonlinear time-fractional {CDR} equation \eqref{rdeq} using the obtained invariant linear spaces which are discussed below in detail.
\begin{example}
	%%%%%%%%%%%%%%%%%%%%%%%%%%%%%%%%%%%%%%%%%%%%%%%%%%%%%%%%%%%%%%%%%%%%%%%%%%%%%%%%%%%%%%%%%%%%%%%%%%%
	{Let us first consider the $(3+1)$-dimensional nonlinear time-fractional CDR equation \eqref{rdeq} in the form
		\begin{eqnarray}
			\begin{aligned}\label{ocrdf1}
				\dfrac{\partial^\alpha u}{\partial t^\alpha}=&\dfrac{\partial}{\partial x_1}\left[
				\left(-\dfrac{\lambda_2^2F_2(u)+\lambda_3^2F_3(u)}{\lambda_1^2}+\dfrac{\mu_1}{2\lambda_1 \rho_0} u+ \mu_3\right)\dfrac{\partial u}{\partial x_1}\right]
				\\&+ \sum_{r=2}^{3}
				\dfrac{\partial}{\partial x_r}
				\left(F_r (u)\dfrac{\partial u}{\partial x_r}\right) 			
				+	\left(-\dfrac{\lambda_2K_2(u)+\lambda_3K_3(u)}{\lambda_1}+\mu_1 u+ \mu_2\right)\left( \dfrac{\partial u}{\partial x_1}\right)
				\\&+ \sum_{r=2}^{3}K_r (u)\left( \dfrac{\partial u}{\partial x_r}\right) +
				\mu_0u, \alpha \in (0,2],
			\end{aligned}
		\end{eqnarray}
		along with the appropriate initial and boundary conditions \eqref{a1}-\eqref{bc}.
		Here $F_r(u)$ and $K_r(u),r=2,3$ are arbitrary functions of $u$  { and $ \mu_i\in\mathbb{R},i=0,1,2,3. $}
	%%%%%%%%%%%%%%%%%%%%%%%%%%%%%%%%%%%%%%%%%%%%%%%%%%%%%%%%%%%%%%%%%%%%%%%%%%%%%%%%%%%%%%%%%%%%%%%%%%%
	%%%%%%%%%%%%%%%%%%%%%%%%%%%%%%%%%%%%%%%%%%%%%%%%%%%%%%%%%%%%%%%%%%%%%%%%%%%%%%%%%%%%%%%%%%%%%%%%%%%
	For this case,}
	the {transformed} nonlinear time-fractional {CDR} equation \eqref{trd} reads as follows:
	\begin{eqnarray}\begin{aligned}\label{crdf1}
			\dfrac{\partial^\alpha w}{\partial t^\alpha}=\textbf{\textit{F}}_T[w]\equiv&\lambda_1^2\dfrac{\partial}{\partial z}\left[
			\left(\dfrac{\mu_1}{2\lambda_1 \rho_0} w+ \mu_3\right)\dfrac{\partial w}{\partial z}\right] + \lambda_1
			\left(\mu_1 w+ \mu_2\right)\dfrac{\partial w}{\partial z}+
			\mu_0w, \alpha \in (0,2],
	\end{aligned} \end{eqnarray}
	along with the following initial and boundary conditions:
	\begin{eqnarray}
		&w(z,0)=\xi(z),&\alpha\in(0,1],\label{ta1c1}\\
		& w(z,0)=\xi(z) \,\,\, \& \,\, \,\dfrac{\partial w}{\partial t} \big{|}_{t=0}=\eta(z),& \alpha\in(1,2],\label{ta2c1}
	\end{eqnarray}
	{and}
	\begin{eqnarray}
		w(z,t)\Big{|}_{z=0}=\left\{
		\begin{array}{ll}
			\delta_1(t),\text{if}\, \alpha\in(0,1]\label{tbc1},\\
			\delta_2(t),\text{if}\, \alpha\in(1,2],
		\end{array}  \right.
		\, \& \quad
		w(z,t)\Big{|}_{z=l}=\left\{
		\begin{array}{ll}
			\delta_3(t),\text{if}\, \alpha\in(0,1],\\
			\delta_4(t),\text{if}\, \alpha\in(1,2],
		\end{array}  \right.
	\end{eqnarray}
	which preserves the one-dimensional linear space
	$
	\mathbf{V}_1=\text{Span}\{e^{-\rho_0z }\} 	.
	$  {Note that this case is listed in Table \ref{crd2} of  case 6.}
	Thus, for  $  w=Ce^{-\rho_0z  }\in \mathbf{V}_1,$
	we obtain
	$$
	\textbf{\textit{F}}_T[Ce^{-\rho_0z}]= C\big(\rho_0^2\lambda_{1}^{2}\mu_3-\lambda_1\mu_2\rho_0+\mu_0\big)e^{-\rho_0z} \in \mathbf{V}_1.
	$$
	Then, there exist an exact solution of \eqref{crdf1} as follows
	\begin{eqnarray}\label{s1}
		w(z,t)=C(t)e^{-\rho_0z},
	\end{eqnarray}
	where
	\begin{equation}\label{sys1}
		\dfrac{d^\alpha C(t)}{d t^\alpha} =\gamma C(t), \gamma=\rho_0^2\lambda_{1}^{2}\mu_3-\lambda_1\mu_2\rho_0+\mu_0.
	\end{equation}
	For the integer orders $\alpha=1$ and $ \alpha=2, $ we get the exact solution of \eqref{crdf1} as
	%%%%%%%%%%%%%%%%%%%%%%%%%%%%%%%%%%%%%%%%%%%%%%%%%%%%%%%%%%%%%%%%%%%%%%%%%%%%%%%%%%%%%%%%%%%%%%%%%%%
	{ 	\begin{eqnarray}\label{ios1}
			w(z,t)=	\left\{
			\begin{array}{ll}
				A_0e^{-\rho_0z+\gamma t}, \text{if} \,\alpha=1,\\
				\big[A_0 cosh\big(\sqrt{\gamma} t\big)+\dfrac{A_1}{\sqrt{\gamma}}sinh\big(\sqrt{\gamma} t\big)\big]e^{-\rho_0z}, \text{if}\,\alpha=2.
			\end{array}
			\right.
	\end{eqnarray}}
	%%%%%%%%%%%%%%%%%%%%%%%%%%%%%%%%%%%%%%%%%%%%%%%%%%%%%%%%%%%%%%%%%%%%%%%%%%%%%%%%%%%%%%%%%%%%%%%%%%%
	Next, we explain how to derive the exact solution of \eqref{crdf1} for arbitrary order $\alpha,\alpha\in(0,2].$
	In this connection,
	we know that the Laplace transformation of the $\alpha$-th order Caputo fractional derivative \cite{mathai} as follows:
	$$\mathcal{L}\{ \dfrac{d^\alpha \tau(t)}{d t^\alpha};s\} = s^\alpha \mathcal{L}\{\tau(t);s\} -\sum\limits_{k=0}^{q-1}s^{\alpha-k-1} \dfrac{d^{k} \tau(t)}{d t^k}\Big|_{t=0},q-1\leq\alpha<q,q\in \mathbb{N}. $$
	First, let $\alpha\in(0,1].$
	Thus, applying the Laplace and inverse Laplace transformations on \eqref{sys1}, we get
	\begin{equation}\label{c1}
		C(t)= C(0)E_{\alpha,1}\big(\gamma t^{\alpha}\big),
	\end{equation}
	where $ E_{\alpha_1,\alpha_2}\big(\xi\big)$ is the $2$-parameter Mittag-Leffler function \cite{mathai}, defined as
	$ E_{\alpha_1,\alpha_2}\big(\xi\big)=\sum\limits_{s=0}^{\infty}\dfrac{\xi^{s}}{\Gamma(\alpha_1s+\alpha_2)},  \mathcal{R}(\alpha_i)>0,$ $i=1,2.$\\
	Next, for $ \alpha\in(1,2], $ on applying the Laplace transformation for the equation \eqref{sys1}, we get
	\begin{eqnarray}\begin{aligned}
			% &\mathcal{L}\{ \dfrac{d^\alpha}{d t^\alpha}C(t);s\} =\gamma \mathcal{L}\{C(t);s\},\\\label{c2}
			&\mathcal{L}\{C(t);s\}=\dfrac{s^{\alpha-1}}{s^{\alpha}-\gamma}C(0)+\dfrac{s^{\alpha-2}}{s^{\alpha}-\gamma}C'(0).\,
		\end{aligned}
	\end{eqnarray}
	Taking inverse Laplace transformation of the above equation, one obtains
	\begin{equation}\label{c2}
		C(t)= C(0)E_{\alpha,1}\big(\gamma t^{\alpha}\big)+ C'(0)tE_{\alpha,2}\big(\gamma t^{\alpha}\big).
	\end{equation}
	%along with the appropriate initial and boundary conditions given in \eqref{5.2}-\eqref{tbc1}.
	Substituting \eqref{c1} and \eqref{c2} in \eqref{s1}, we obtain the exact solution of \eqref{crdf1} in the form
	%%%%%%%%%%%%%%%%%%%%%%%%%%%%%%%%%%%%%%%%%%%%%%%%%%%%%%%%%%%%%%%%%%%%%%%%%%%%%%%%%%%%%%%%%%%%%%%%%%%
	{ 	\begin{eqnarray}
			w(z,t)=
			\left\{\begin{array}{ll}
				A_0 E_{\alpha,1}\big(\gamma t^{\alpha}\big)e^{-\rho_0z}, \alpha\in(0,1],
				\\
				\label{aos1}
				\big[A_0 E_{\alpha,1}\big(\gamma t^{\alpha}\big)+A_1tE_{\alpha,2}\big(\gamma t^{\alpha}\big)\big]e^{-\rho_0z}, \alpha\in(1,2],
			\end{array}
			\right.
		\end{eqnarray}
	}
	%%%%%%%%%%%%%%%%%%%%%%%%%%%%%%%%%%%%%%%%%%%%%%%%%%%%%%%%%%%%%%%%%%%%%%%%%%%%%%%%%%%%%%%%%%%%%%%%%%%
	where $ A_0=C(0),A_1=C'(0), $ and $\gamma=\rho_0^2\lambda_{1}^{2}\mu_3-\lambda_1\mu_2\rho_0+\mu_0. $
	Since $ E_{\alpha,1}\big(0\big)=1,
	\dfrac{d}{dt}[ E_{\alpha,1}\big(\gamma t^{\alpha}\big)]=t^{\alpha-1}E_{\alpha,\alpha}\big(\gamma t^{\alpha}\big),$
	$\dfrac{d}{dt}[t E_{\alpha,2}\big(\gamma t^{\alpha}\big)]=E_{\alpha,1}\big(\gamma t^{\alpha}\big) $ and
	$\dfrac{d}{dt} [t^{\alpha}E_{\alpha,\alpha+1}\big(\gamma t^{\alpha}\big)]=t^{\alpha-1}E_{\alpha,\alpha}\big(\gamma t^{\alpha}\big) $ which are given in \cite{mathai}.
	In addition, the exact solutions \eqref{aos1} satisfy the given initial and boundary conditions \eqref{ta1c1}-\eqref{tbc1} along with
	%%%%%%%%%%%%%%%%%%%%%%%%%%%%%%%%%%%%%%%%%%%%%%%%%%%%%%%%%%%%%%%%%%%%%%%%%%%%%%%%%%%%%%%%%%%%%%%%%%%
	{ \begin{eqnarray*}
			\begin{aligned}
				&	\xi(z)=A_0e^{-\rho_0z},\qquad\quad
				\eta(z)= A_1e^{-\rho_0z},\\
				& 	 \delta_1(t)=A_0 E_{\alpha,1}\big(\gamma t^{\alpha}\big),\quad
				\delta_2(t)=A_0E_{\alpha,1}\big(\gamma t^{\alpha}\big)+A_1tE_{\alpha,2}\big(\gamma t^{\alpha}\big),
				\\
				& \delta_3(t)=A_0e^{-\rho_0l} E_{\alpha,1}\big(\gamma t^{\alpha}\big),\, \text{and }
				\delta_4(t)=\big[A_0 E_{\alpha,1}\big(\gamma t^{\alpha}\big)+A_1tE_{\alpha,2}\big(\gamma t^{\alpha}\big)\big]e^{-\rho_0l}.
			\end{aligned}
		\end{eqnarray*}
	}
	%%%%%%%%%%%%%%%%%%%%%%%%%%%%%%%%%%%%%%%%%%%%%%%%%%%%%%%%%%%%%%%%%%%%%%%%%%%%%%%%%%%%%%%%%%%%%%%%%%%
	Additionally, we wish to point out that the obtained integer-order solutions \eqref{ios1} coincide with the fractional-order exact solutions \eqref{aos1} when $\alpha=1$ and $\alpha=2,$ respectively.
	%%%%%%%%%%%%%%%%%%%%%%%%%%%%%%%%%%%%%%%%%%%%%%%%%%%%%%%%%%%%%%%%%%%%%%%%%%%%%%%%%%%%%%%%%%%%%%%%%
	
	{In addition, we would like to point out that Prakash et al.\cite{ppl} have derived the exact solution for the equation \eqref{ocrdf1} along with $F_r(u)=c_{r1}u+c_{r0},$ $K_r(u)=l_{r1}u+l_{r0},c_{ri},l_{ri}\in\mathbb{R},i=0,1,r=2,3,$  and  $\mu_0=0 $ using the one-dimensional linear subspace $\mathbf{V}_1=\text{Span}\{e^{-(a_1x_1+a_2x_2+a_3x_3)}\},$ which is obtained from the direct approach of the invariant subspace method. The obtained solution \eqref{aos1} is similar to the obtained solution in \cite{ppl}. However, the invariant subspace method with variable transformation is easy to apply for finding the exact solution of nonlinear time-fractional PDEs. }
	\end{example}

%%%%%%%%%%%%%%%%%%%%%%%%%%%%%%%%%%%%%%%%%%%%%%%%%%%%%%%%%%%%%%%%%%%%%%%%%%%%%%%%%%%%%%%%%%%%%%%%%%%%%%%%%%%%%%%%%%%%%%%%%%%%%%%%%%%%%%%%%%%%%%%%%%%%%%%%	%%%%%%%%%%%%%%%%%%%%%%%%%%%%%%%%%%%%%%%%%%%%%%%%%%%%%%%%%%%%%%%%%%%%%%%%%%%%%%%%%%%%%%%%%%%%%%%%%%%
\begin{example}
	Now, we consider the following quadratic
	nonlinear time-fractional {CDR} equation
	%%%%%%%%%%%%%%%%%%%%%%%%%%%%%%%%%%%%%%%%%%%%%%%%%%%%%%%%%%%%%%%%%%%%%%%%%%%%%%%%%%%%%%%%%%%%%%%%%%%
	{\begin{eqnarray}\begin{aligned}\label{ocrdf2}
				\dfrac{\partial^\alpha u}{\partial t^\alpha}=&\dfrac{\partial}{\partial x_1}\left[
				\left(\dfrac{\beta_1}{\lambda_1^2}u
				+ \beta_0\right)\dfrac{\partial u}{\partial x_1}\right]
				+\kappa_0\dfrac{\partial^2 u}{\partial x_2^2}
				+\zeta_0\dfrac{\partial^2 u}{\partial x_3^2}
				\\
				&	+\left(\dfrac{2\beta_1\rho_1}{\lambda_1}u+f_0\right)
				\left(\dfrac{\partial u}{\partial x_1}\right)
				+g_0
				\left(\dfrac{\partial u}{\partial x_2}\right)
				+h_0
				\left(\dfrac{\partial u}{\partial x_3}\right),\alpha\in(0,2],
		\end{aligned} \end{eqnarray}
		along with the appropriate initial and boundary conditions \eqref{a1}-\eqref{bc}.
		The above equation \eqref{ocrdf2} gets transformed into}
	%%%%%%%%%%%%%%%%%%%%%%%%%%%%%%%%%%%%%%%%%%%%%%%%%%%%%%%%%%%%%%%%%%%%%%%%%%%%%%%%%%%%%%%%%%%%%%%%%%%
	\begin{eqnarray}\begin{aligned}\label{crdf2}
			\dfrac{\partial^\alpha w}{\partial t^\alpha}=\textbf{\textit{F}}_T[w]\equiv&\dfrac{\partial}{\partial z}\left[
			\left( {\beta_1} w+ {\lambda_1^2}\beta_0 +\lambda_2^2 \kappa_0+\lambda_3^2\zeta_0 \right)\dfrac{\partial w}{\partial z}\right]
			\\
			& +
			\left({2\beta_1 \rho_1} w+\lambda_1 f_0+\lambda_2g_0+\lambda_3h_0\right)\dfrac{\partial w}{\partial z},\alpha\in(0,2],
	\end{aligned} \end{eqnarray}
	along with the given initial and boundary conditions \eqref{ta1}-\eqref{tbc}.
	\\
	Here, the differential operator $ \textbf{\textit{F}}_T[w] $ admits the two-dimensional exponential linear space  $ \mathbf{V}_2=\text{Span}\{1,e^{-\rho_1z}\}, $ which is listed in {case 2 of Table \ref{crd4}}.
	For $\alpha\in(0,2],$ we find the exact solutions of the equation \eqref{crdf2} as follows:
	{
		\begin{equation}
			w(z,t)=\left\{
			\begin{array}{ll}
				b_0+a_0E_{\alpha,1}(\gamma t^\alpha)e^{-\rho_1z},\,\, \text{if} \, \quad\alpha\in(0,1],\\
				b_1t+b_0+ {C_1(t)}e^{-\rho_1z}, \, \,\text{if} \quad \alpha\in(1,2],\end{array}\right.\label{fos}
	\end{equation}}
	where $ \gamma=\rho_1^2(\beta_0\lambda_1^2+\kappa_0\lambda_2^2+\zeta_0\lambda_3^2-\beta_1b_0)-\rho_1(f_0\lambda_1+g_0\lambda_2+h_0\lambda_3) $ and the function
	$C_1(t) $ satisfies the following fractional-order ODE
	$$\dfrac{d^\alpha C_1(t)}{dt^\alpha}=-\rho_1^2\beta_1(b_1t+b_0)C_1(t)+\rho_1C_1(t)
	\big[\rho_1(\lambda_1^2\beta_0+\lambda_2^2\kappa_0+\lambda_3^2\zeta_0)-(f_0\lambda_1+g_0\lambda_2+h_0\lambda_3)\big],$$
	which may not be solvable in general.
	If we choose $b_1=0$, we get $C_1(t)=a_0E_{\alpha,1}(\gamma_2t^\alpha)+a_1tE_{\alpha,2}(\gamma t^\alpha)$ when $\alpha\in(1,2].$
	For this case, we obtain the exact solution of \eqref{crdf2} as follows:
	$$w(z,t)=b_0+\big[a_0E_{\alpha,1}(\gamma t^\alpha)+a_1tE_{\alpha,2}(\gamma t^\alpha)\big]e^{-\rho_1z}, \text{if} \, \alpha\in(1,2].$$
	Note that when $ \alpha=1 $ and $\alpha=2,$ the obtained exact solution of \eqref{crdf2} is as follows:
	%%%%%%%%%%%%%%%%%%%%%%%%%%%%%%%%%%%%%%%%%%%%%%%%%%%%%%%%%%%%%%%%%%%%%%%%%%%%%%%%%%%%%%%%%%%%%%%%%%%
	{ \begin{equation}
			w(z,t)=
			\left\{ \begin{array}{ll}
				b_0+e^{\gamma t-\rho_1z},\,\text{if}\,\alpha=1,\\
				b_0+\big[a_0cosh(\sqrt{\gamma }t)+\frac{a_1}{\sqrt{\gamma}}sinh(\sqrt{\gamma}t)\big]e^{-\rho_1z}, \,\text{if}\,\alpha=2.
			\end{array} \right.    \label{ios2}
	\end{equation}}
	%%%%%%%%%%%%%%%%%%%%%%%%%%%%%%%%%%%%%%%%%%%%%%%%%%%%%%%%%%%%%%%%%%%%%%%%%%%%%%%%%%%%%%%%%%%%%%%%%%%
	In addition, we note that the
	fractional-order exact solution \eqref{fos} coincides with integer-order exact solution \eqref{ios2} if $\alpha=1$ and $\alpha=2. $
	It is also observed that the obtained solutions \eqref{fos} satisfy the given initial and boundary conditions \eqref{ta1}-\eqref{tbc} along with
	%%%%%%%%%%%%%%%%%%%%%%%%%%%%%%%%%%%%%%%%%%%%%%%%%%%%%%%%%%%%%%%%%%%%%%%%%%%%%%%%%%%%%%%%%%%%%%%%%%%
	{
		\begin{eqnarray*}
			\begin{aligned}
				&\xi(z)=b_0+a_0e^{-\rho_1z},
				\qquad 	\eta(z)=a_1e^{-\rho_1z},
				\\
				& \delta_1(t)=
				b_0+a_0E_{\alpha,1}(\gamma t^\alpha),
				\,	\delta_2(t)=
				b_0+a_0E_{\alpha,1}(\gamma t^\alpha)+a_1tE_{\alpha,2}(\gamma t^\alpha),
				\\
				& \delta_3(t)=
				b_0+a_0e^{-\rho_1l}E_{\alpha,1}(\gamma t^\alpha),
				\,\text{and } 	\delta_4(t)=
				b_0+\big[a_0E_{\alpha,1}(\gamma t^\alpha)+a_1tE_{\alpha,2}(\gamma t^\alpha)\big]e^{-\rho_1l}.
			\end{aligned}
		\end{eqnarray*}
	}	
	%%%%%%%%%%%%%%%%%%%%%%%%%%%%%%%%%%%%%%%%%%%%%%%%%%%%%%%%%%%%%%%%%%%%%%%%%%%%%%%%%%%%%%%%%%%%%%%%%%%%%%%
	{Additionally, we observe that in \cite{ppl}, exact solution of the above equation \eqref{ocrdf2} with initial conditions was derived using the 4-dimensional linear space $\mathbf{V}_4=\text{Span}\{e^{-2\rho_1\lambda_1x_1}, $ $e^{-(2\rho_1\lambda_1x_1+a_2x_2)}, e^{-(2\rho_1\lambda_1x_1+a_3x_3)}, e^{-(2\rho_1\lambda_1x_1+a_2x_2+a_3x_3)}\},$ which is obtained from the direct approach of the invariant subspace method. }
\end{example}
%%%%%%%%%%%%%%%%%%%%%%%%%%%%%%%%%%%%%%%%%%%%%%%%%%%%%%%%%%%%%%%%%%%%%%%%%%%%%%%%%%%%%%%%%%%%%%%%%%%%%%%%%%%%%%%%%%%%%%%%%%%%%%%%%%%%%%%%%%%%%%%%%%%%%%%%
%\textbf{Example-3 : Exact solutions using polynomial invariant subspace}
\begin{example}
	Consider the nonlinear time-fractional convection-diffusion equation of the form
	%%%%%%%%%%%%%%%%%%%%%%%%%%%%%%%%%%%%%%%%%%%%%%%%%%%%%%%%%%%%%%%%%%%%%%%%%%%%%%%%%%%%%%%%%%%%%%%%%%%
	{\begin{eqnarray}\begin{aligned}\label{ocrdf3}
				\dfrac{\partial^\alpha u}{\partial t^\alpha}=&\dfrac{\partial}{\partial x_1}\left[
				\left(\beta_1u
				+ \beta_0\right)\dfrac{\partial u}{\partial x_1}\right]
				+\dfrac{\partial}{\partial x_2}\left[(\kappa_1u+\kappa_0) \dfrac{\partial u}{\partial x_2}\right]
				\\
				&+\dfrac{\partial}{\partial x_3}\left[\left(\zeta_1u+\zeta_0\right)\dfrac{\partial u}{\partial x_3}\right]
				+f_0
				\left(\dfrac{\partial u}{\partial x_1}\right)
				+g_0
				\left(\dfrac{\partial u}{\partial x_2}\right)
				+h_0
				\left(\dfrac{\partial u}{\partial x_3}\right),\alpha\in(0,2],
		\end{aligned} \end{eqnarray}
		along with the appropriate initial and boundary conditions \eqref{a1}-\eqref{bc}.
		Under the transformation \eqref{wt}, the above equation \eqref{ocrdf3} is transformed into the following form }
	%%%%%%%%%%%%%%%%%%%%%%%%%%%%%%%%%%%%%%%%%%%%%%%%%%%%%%%%%%%%%%%%%%%%%%%%%%%%%%%%%%%%%%%%%%%%%%%%%%%
	\begin{eqnarray}\begin{aligned}\label{crdf3}
			\dfrac{\partial^\alpha w}{\partial t^\alpha}\equiv\textbf{\textit{F}}_T[w]=&\dfrac{\partial}{\partial z}\left[
			\left(( \lambda_1^2\beta_1+\lambda_2^2\kappa_1+\lambda_3^2\zeta_1)w
			+ \lambda_1^2\beta_0+ \kappa_0\lambda_2^2+\lambda_3^2 \zeta_0\right)\dfrac{\partial w}{\partial z}\right]
			%+\lambda_2^2 \dfrac{\partial}{\partial z}\left[ \dfrac{\partial w}{\partial z}\right]
			\\
			&%+\lambda_3^2 \dfrac{\partial}{\partial z}\left[\left(\right)\dfrac{\partial w}{\partial z}\right]
			+\left( \lambda_1f_0+\lambda_2g_0+\lambda_3h_0\right)
			\left(\dfrac{\partial w}{\partial z}\right),\alpha\in(0,2],
	\end{aligned} \end{eqnarray}
	along with the following initial and boundary conditions:
	\begin{eqnarray}
		&w(z,0)=\xi(z),&\alpha\in(0,1],\label{ta1c2}\\
		& w(z,0)=\xi(z) \,\,\, \& \,\, \,\dfrac{\partial w}{\partial t} \big{|}_{t=0}=\eta(z),& \alpha\in(1,2],\label{ta2c2}
	\end{eqnarray}
	{and}
	\begin{eqnarray}
		w(z,t)\Big{|}_{z=0}=\left\{
		\begin{array}{ll}
			\delta_1(t),\text{if}\, \alpha\in(0,1]\label{tbc2},\\
			\delta_2(t),\text{if}\, \alpha\in(1,2],
		\end{array}  \right.
		\, \& \quad
		w(z,t)\Big{|}_{z=l}=\left\{
		\begin{array}{ll}
			\delta_3(t),\text{if}\, \alpha\in(0,1],\\
			\delta_4(t),\text{if}\, \alpha\in(1,2].
		\end{array}  \right.
	\end{eqnarray}
	The differential operator $ \textbf{\textit{F}}_T[w] $ given in \eqref{crdf3}  admits a 2-dimensional linear space $\mathbf{V}_2=\text{Span}\{1,z\},$ {which is listed in case 1 of Table \ref{crd4}.} For the integer values $\alpha=1$ and $\alpha=2,$ we obtain exact solutions of \eqref{crdf3} as follows,
	{\begin{equation}
			w(z,t)=\left\{
			\begin{array}{ll}
				b_0(\gamma_1b_0+\gamma_2)t+a_0+b_0z,  \text{if}\, \alpha=1,\\
				%	\left.	\begin{array}{ll}
					\frac{1}{12}t^4b_1^2\gamma_1+t^3(\frac{1}{3}b_1b_0\gamma_1
					+\frac{1}{6}b_1\gamma_2)
					\\+\frac{1}{2}b_0(b_0\gamma_1+\gamma_2)t^2+a_1t+a_0+(b_1t+b_0)z, \text{if}\, \alpha=2,
					%	\end{array}\right.
			\end{array}\right.\label{iso3}
	\end{equation}}
	where $\gamma_1=\lambda_1^2\beta_1+\lambda_2^2\kappa_1+\lambda_3^2\zeta_1$ and $\gamma_2=\lambda_1f_0+
	\lambda_2g_0+\lambda_3h_0.$\\
	For $\alpha\in(0,2], $ the exact solutions of the equation \eqref{crdf3} are obtained as follows,
	{\begin{eqnarray}
			w(z,t)=	\left\{
			\begin{array}{ll}
				b_0(\gamma_1b_0+\gamma_2)\dfrac{t^\alpha}{\Gamma(\alpha+1)}+a_0+b_0z,  \,\text{if}\, \alpha\in(0,1],\\
				%	\left\{	\begin{array}{ll}
					\dfrac{t^{\alpha+2}}{\Gamma(\alpha+3)}b_1^2\gamma_1
					+
					\dfrac{t^{\alpha+1}}{\Gamma(\alpha+2)}(2b_1b_0\gamma_1+b_1\gamma_2)\\
					+b_0(b_0\gamma_1+\gamma_2)\dfrac{t^\alpha}{\Gamma(\alpha+1)}
					+a_1t+a_0+(b_1t+b_0)z, \,\text{if}\, \alpha\in(1,2].\label{as3}
					%	\end{array}\right.
			\end{array}\right.
		\end{eqnarray}
	}
	The above solutions \eqref{as3} coincide with the integer-order solutions \eqref{iso3} if $\alpha=1$ and $ \alpha=2 $. Also, we note that the exact solutions \eqref{as3} satisfy the given initial and boundary conditions  \eqref{ta1c2}-\eqref{tbc2}
	with
	{
		\begin{eqnarray*}
			\begin{aligned}
				&	\xi(z)=a_0+b_0z,\qquad \eta(z)=a_1+b_1z ,\\
				& \delta_1(t)=b_0(\gamma_1b_0+\gamma_2)\dfrac{t^\alpha}{\Gamma(\alpha+1)}+a_0,
				\\
				& 	\delta_2(t)=\dfrac{t^{\alpha+2}}{\Gamma(\alpha+3)}b_1^2\gamma_1
				+
				\dfrac{t^{\alpha+1}}{\Gamma(\alpha+2)}(2b_1b_0\gamma_1+b_1\gamma_2)
				+b_0(b_0\gamma_1+\gamma_2)\dfrac{t^\alpha}{\Gamma(\alpha+1)}
				+a_1t+a_0,
				\\
				& \delta_3(t)=b_0(\gamma_1b_0+\gamma_2)\dfrac{t^\alpha}{\Gamma(\alpha+1)}+a_0+b_0l, \text{ and }
				\\
				& \delta_4(t)=
				\dfrac{t^{\alpha+2}}{\Gamma(\alpha+3)}b_1^2\gamma_1
				+
				\dfrac{t^{\alpha+1}}{\Gamma(\alpha+2)}(2b_1b_0\gamma_1+b_1\gamma_2)
				+b_0(b_0\gamma_1+\gamma_2)\dfrac{t^\alpha}{\Gamma(\alpha+1)}
				\\& \qquad\quad
				+a_1t+a_0+(b_1t+b_0)l.
			\end{aligned}
		\end{eqnarray*}
	}
	Also, we have shown the two-dimensional (2D) and three-dimensional (3D) graphical representations of the arbitrary-order exact solutions \eqref{as3} for different values of $\alpha,\ \alpha\in(0,2],$ with parameter values $\gamma_1=\gamma_2=-10,a_0=a_1=10,b_0=200,b_1=-100,$ in Figures \ref{fig1a} and \ref{fig1b}.
	\begin{figure}[h!]
		\begin{subfigure}{0.5\textwidth}
			\includegraphics[width=7.5cm, height=6cm]{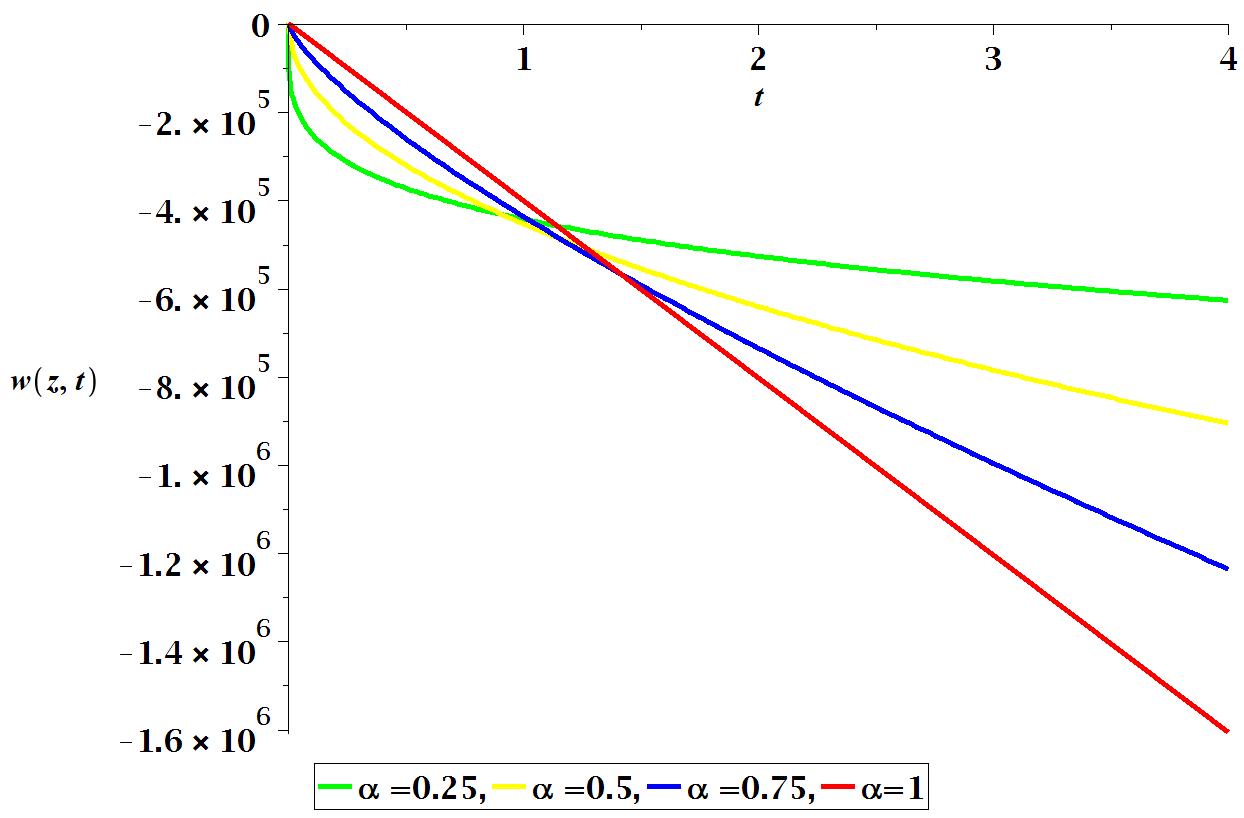}
			\caption{}
		\end{subfigure}
		\begin{subfigure}{0.5\textwidth}
			\includegraphics[width=8.5cm, height=6cm]{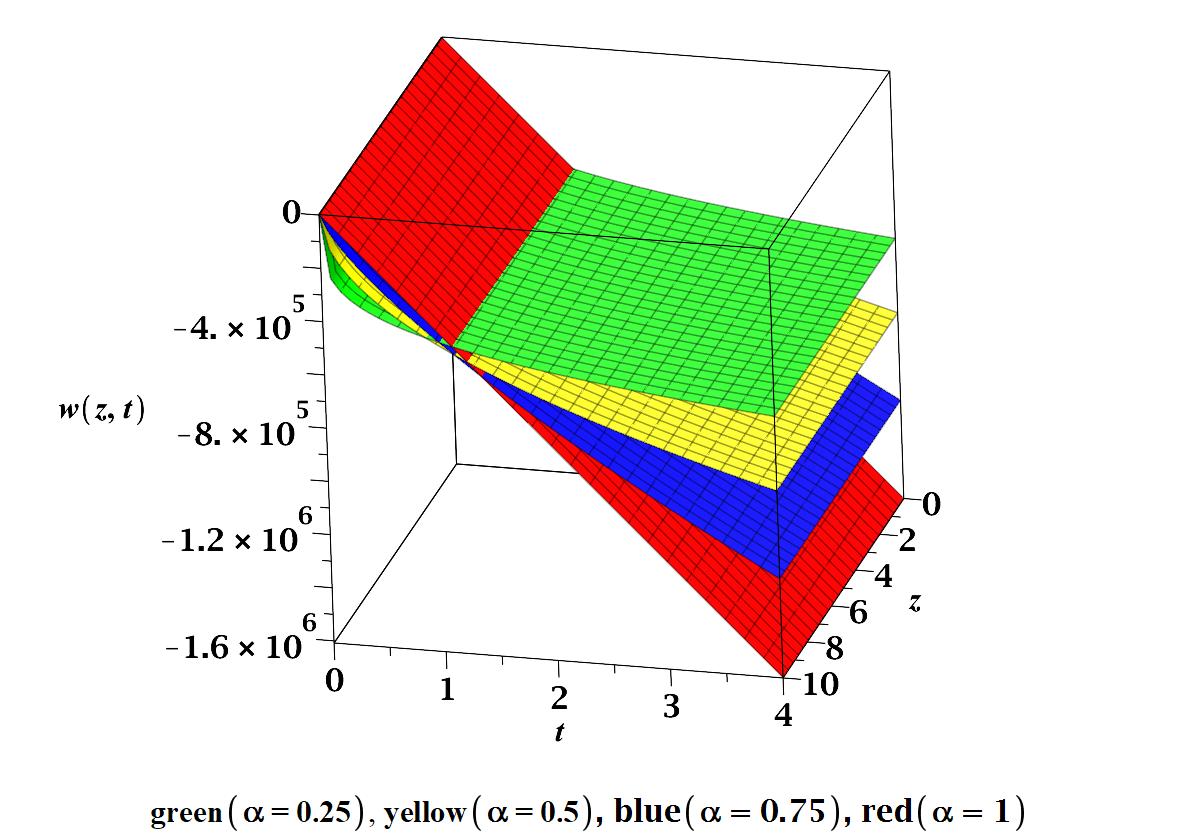}
			\caption{}
		\end{subfigure}
		\caption{ {{ (a) 2D and (b) 3D graphical representations of the arbitrary-order exact solutions \eqref{as3} for various values of $\alpha$, $\alpha\in(0,1]$.}}}
		\label{fig1a}
	\end{figure}
	\begin{figure}[h!]
		\begin{subfigure}{0.5\textwidth}
			\includegraphics[width=7.5cm, height=6cm]{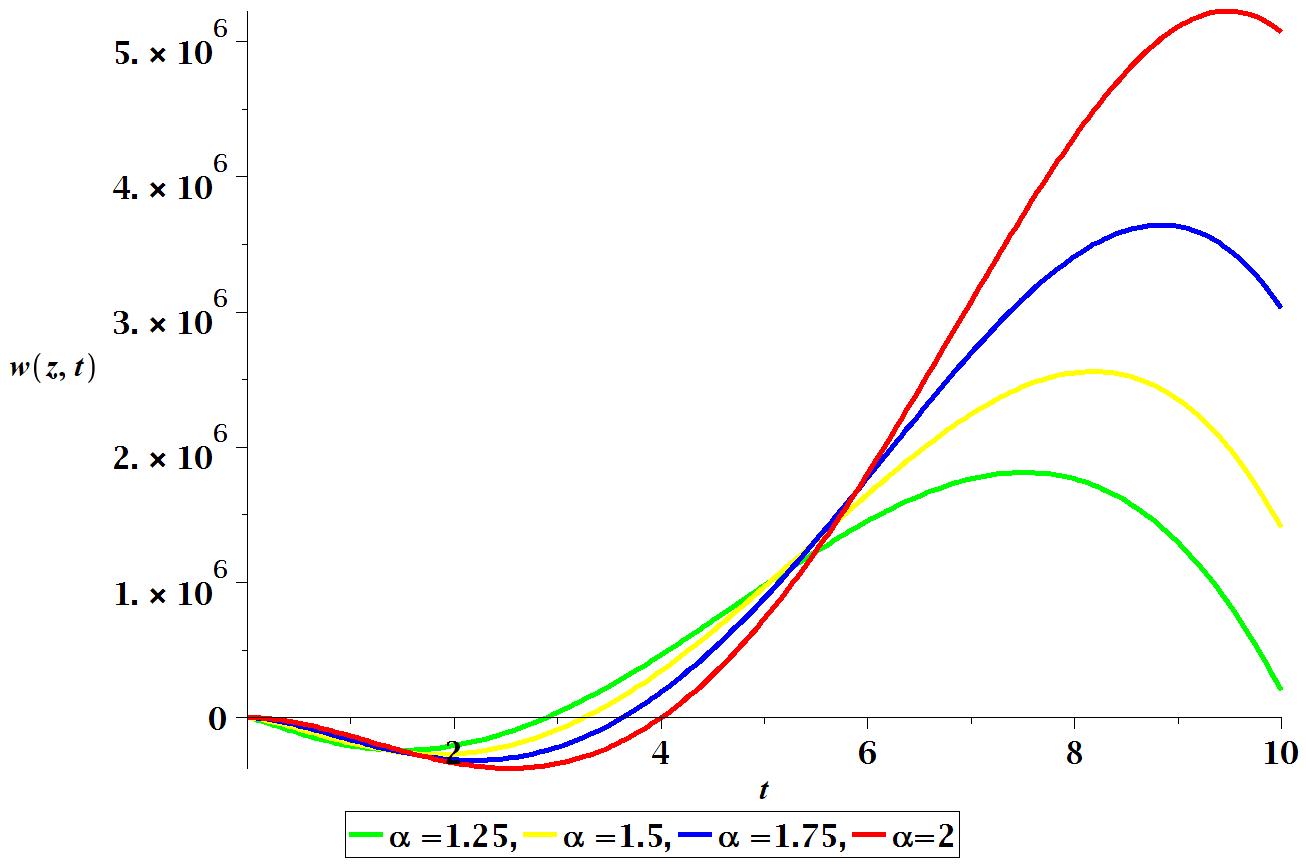}
			\caption{}
		\end{subfigure}
		\begin{subfigure}{0.5\textwidth}
			\includegraphics[width=8.5cm, height=6cm]{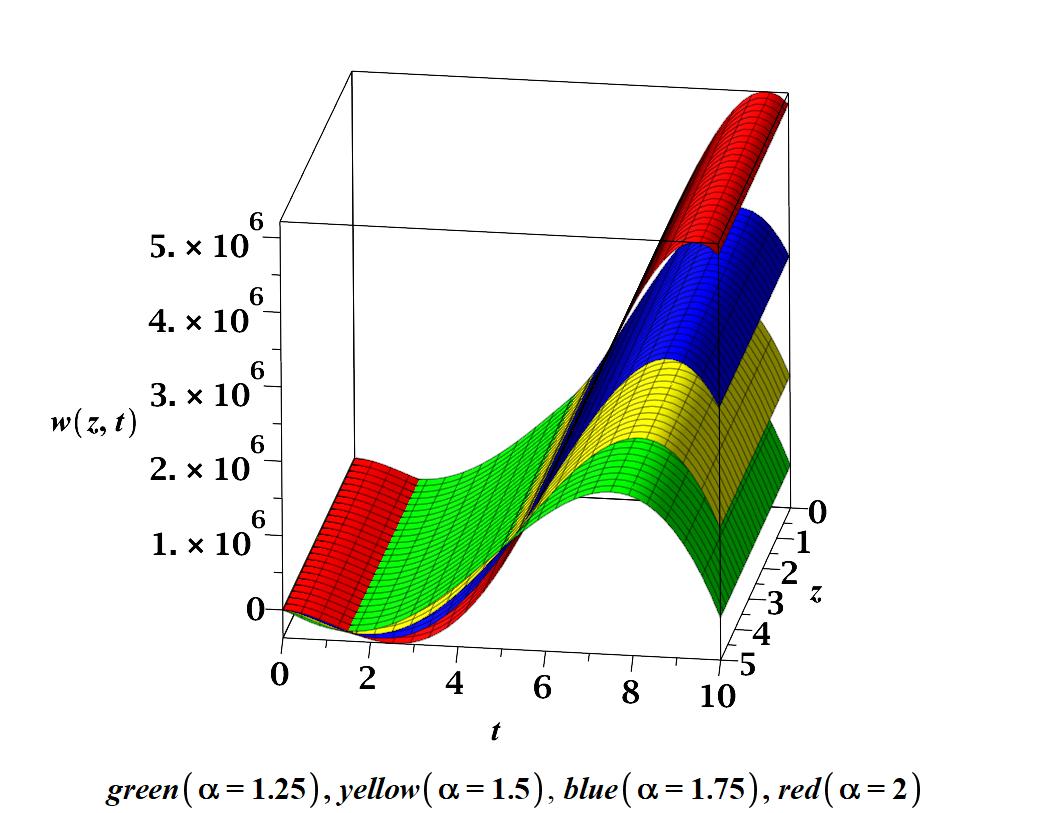}
			\caption{}
		\end{subfigure}
		\caption{ {{(a) 2D and (b) 3D graphical representations of the arbitrary-order exact solutions \eqref{as3} for various values of $\alpha$, $\alpha\in(1,2]$.}}}
		\label{fig1b}
	\end{figure}
	%%%%%%%%%%%%%%%%%%%%%%%%%%%%%%%%%%%%%%%%%%%%%%%%%%%%%%%%%%%%%%%%%%%%%%%%%%%%%%%%%%%%%%%%%%%%%%%%%%%%%%%%%%%%%%%%%%%%%%%%%%%%%%%%%%%%%%%%%%%%%%%%%%%%%%%%%%%%%%%%%%%%%%%%%%%%%%%%%%%%%%%%%%%%%%%%%%%%%%%%%%%%%%
	{Additionally, we note that the exact solution of \eqref{ocrdf3} was derived in \cite{ppl} using the four-dimensional invariant subspace $\mathbf{V}_4=\text{Span}\{1,x_1,x_2,x_3\}$, which was discussed through the direct approach of the invariant subspace method. From this, we can observe that the invariant subspace method associated with the variable transformation technique is easy to apply for finding the exact solutions of the higher-dimensional nonlinear time-fractional PDEs because the variable transformation allows one to reduce higher-dimensional nonlinear time-fractional PDEs into $(1+1)$-dimensional nonlinear time-fractional PDEs
		\begin{nt} The graphical depictions demonstrate the physical relevance of the acquired exact solutions to the considered time-fractional convection-reaction-diffusion equation. We have mainly observed that the obtained exact solutions behave differently for various ranges of $\alpha\in(0,2],$ as $\alpha$ changes. Additionally, we note that Figures 1 and 2 show the slow and fast diffusion processes, respectively.
		\end{nt}
	}
\end{example}

%%%%%%%%%%%%%%%%%%%%%%%%%%%%%%%%%%%%%%%%%%%%%%%%%%%%%%%%%%%%%%%%%%%%%%%%%%%%%%%%%%%%%%%%%%%%%%%%%%%%%%%%%%%%%%%%%%%%%%%%%%%%%%%%%%%%%%%%%%%%%%%%%%%%%%%%
\begin{example}
	Let us consider the following transformed nonlinear time-fractional diffusion-reaction equation
	\begin{eqnarray}\begin{aligned}\label{tri}
			\dfrac{\partial^\alpha w}{\partial t^\alpha} =\textbf{\textit{F}}_T[w]\equiv& \lambda_1^2\dfrac{\partial}{\partial z}\left[\big(\dfrac{1}{4}\dfrac{\mu_1}{\lambda_1^2 \rho_1}w+
			\mu_5\big)\dfrac{\partial w}{\partial z}\right]+
			\dfrac{1}{2}\mu_1w^2+\mu_2w+\mu_3,\alpha\in(0,2],
	\end{aligned} \end{eqnarray}
	 { where  $ \mu_i\in\mathbb{R},i=1,\dots,5, $} along with the following initial and boundary conditions:
	\begin{eqnarray}
		&w(z,0)=\xi(z),&\alpha\in(0,1],\label{ta1c3}\\
		& w(z,0)=\xi(z) \,\,\, \& \,\, \,\dfrac{\partial w}{\partial t} \big{|}_{t=0}=\eta(z),& \alpha\in(1,2],\label{ta2c3}
	\end{eqnarray}
	{and}
	\begin{eqnarray}
		w(z,t)\Big{|}_{z=0}=\left\{
		\begin{array}{ll}
			\delta_1(t),\text{if}\, \alpha\in(0,1]\label{tbc3},\\
			\delta_2(t),\text{if}\, \alpha\in(1,2],
		\end{array}  \right.
		\, \& \quad
		w(z,t)\Big{|}_{z=l}=\left\{
		\begin{array}{ll}
			\delta_3(t),\text{if}\, \alpha\in(0,1],\\
			\delta_4(t),\text{if}\, \alpha\in(1,2].
		\end{array}  \right.
	\end{eqnarray}
	The nonlinear differential operator $ \textbf{\textit{F}}_T[w] $ preserves the 3-dimensional trigonometric linear space
	$\mathbf{V}_3=\text{Span}\{1,sin(\sqrt{\rho_1}z),cos(\sqrt{\rho_1}z)\}$ as given {in case 8 of Table \ref{crd2}},
	which helps to reduce the given equation \eqref{tri} into
	\begin{eqnarray*}
		\left.
		\begin{array}{ll}
			\dfrac{d^\alpha C_1(t)}{d t^\alpha} =\dfrac{1}{4}\mu_1(C_1(t)^2+C_2(t)^2+C_3(t)^2)+\mu_2C_1(t)+\mu_3,
			\\
			\dfrac{d^\alpha C_r(t)}{d t^\alpha} =-\lambda_1^2\mu_5\rho_1C_r(t)+\dfrac{3}{4}\mu_1C_1(t)C_r(t)+\mu_2C_r(t),r=2,3.
		\end{array}
		\right.
	\end{eqnarray*}
	The above system of fractional ODEs may not be solvable in general. If we take $\mu_1=0,$ we obtain the exact solutions of \eqref{tri} as follows,
	{ \begin{eqnarray}
			w(z,t)=
			\left\{\begin{array}{ll}\label{ls1}
				% \left\{\begin{array}{ll}\label{ls1}
					a_1E_{\alpha,1}(\mu_2t^\alpha)+
					\mu_3t^\alpha E_{\alpha,\alpha+1}(\mu_2t^\alpha)
					+E_{\alpha,1}(\gamma t^\alpha)\big[A_1sin(\sqrt{\rho_1}z)\\
					+ A_2cos(\sqrt{\rho_1}z)\big]
					,\text{if}\,
					\alpha\in(0,1],
					% \end{array}\right.
				\\
				%\left\{\begin{array}{ll}
					E_{\alpha,1}(\gamma t^\alpha)\big[A_1sin(\sqrt{\rho_1}z)
					+ A_2cos(\sqrt{\rho_1}z)\big]
					\\
					+tE_{\alpha,2}(\gamma t^\alpha)\big[A_3sin(\sqrt{\rho_1}z)
					+ A_4cos(\sqrt{\rho_1}z)\big]
					\\
					+\mu_3t^\alpha E_{\alpha,\alpha+1}(\mu_2t^\alpha)
					+ a_1E_{\alpha,1}(\mu_2t^\alpha)
					+a_2tE_{\alpha,2}(\mu_2t^\alpha),
					\text{if}\, \alpha\in(1,2],
					% \end{array}  \right.
			\end{array}
			\right.
		\end{eqnarray}
	}
	where $a_1,a_2,A_i\in\mathbb{R},i=1,2,3,4,$  and $\gamma=\mu_2-\lambda_1^2\mu_5\rho_1.$ The above exact solutions \eqref{ls1} satisfy the given initial and boundary conditions \eqref{ta1c3}-\eqref{tbc3} along with
	{
		\begin{eqnarray*}
			\begin{aligned}
				&\xi(z)=a_1+A_1sin(\sqrt{\rho_1}z)+A_2cos(\sqrt{\rho_1}z),
				\\& \eta(z)=a_2+A_3sin(\sqrt{\rho_1}z)+A_4cos(\sqrt{\rho_1}z),
				\\&  \delta_1(t)=a_1E_{\alpha,1}(\mu_2t^\alpha)+
				\mu_3t^\alpha E_{\alpha,\alpha+1}(\mu_2t^\alpha)
				+E_{\alpha,1}(\gamma t^\alpha) A_2 ,
				\\&  \delta_2(t)=A_2 E_{\alpha,1}(\gamma t^\alpha)
				+tE_{\alpha,2}(\gamma t^\alpha)A_4
				+\mu_3t^\alpha E_{\alpha,\alpha+1}(\mu_2t^\alpha)
				+ a_1E_{\alpha,1}(\mu_2t^\alpha)
				+a_2tE_{\alpha,2}(\mu_2t^\alpha),
				\\&  \delta_3(t)=
				a_1E_{\alpha,1}(\mu_2t^\alpha)+
				\mu_3t^\alpha E_{\alpha,\alpha+1}(\mu_2t^\alpha)
				+AE_{\alpha,1}(\gamma t^\alpha), \text{ and }
				\\& \delta_4(t)=
				A E_{\alpha,1}(\gamma t^\alpha)
				+BtE_{\alpha,2}(\gamma t^\alpha)
				+\mu_3t^\alpha E_{\alpha,\alpha+1}(\mu_2t^\alpha)
				+ a_1E_{\alpha,1}(\mu_2t^\alpha)
				+a_2tE_{\alpha,2}(\mu_2t^\alpha),
			\end{aligned}
		\end{eqnarray*}
	}
	where $ A=A_1sin(\sqrt{\rho_1}l)
	+ A_2cos(\sqrt{\rho_1}l),$ and $B=A_3sin(\sqrt{\rho_1}l)
	+ A_4cos(\sqrt{\rho_1}l). $
	Here, we have also shown the 2D and 3D graphical representations of the arbitrary-order exact solutions \eqref{ls1} for different values of $\alpha, \alpha\in(0,2],$ in Figures \ref{fig2} and \ref{fig3}.
	\begin{figure}[h!]
		\begin{subfigure}{0.5\textwidth}
			\includegraphics[width=7.25cm, height=5.5cm]{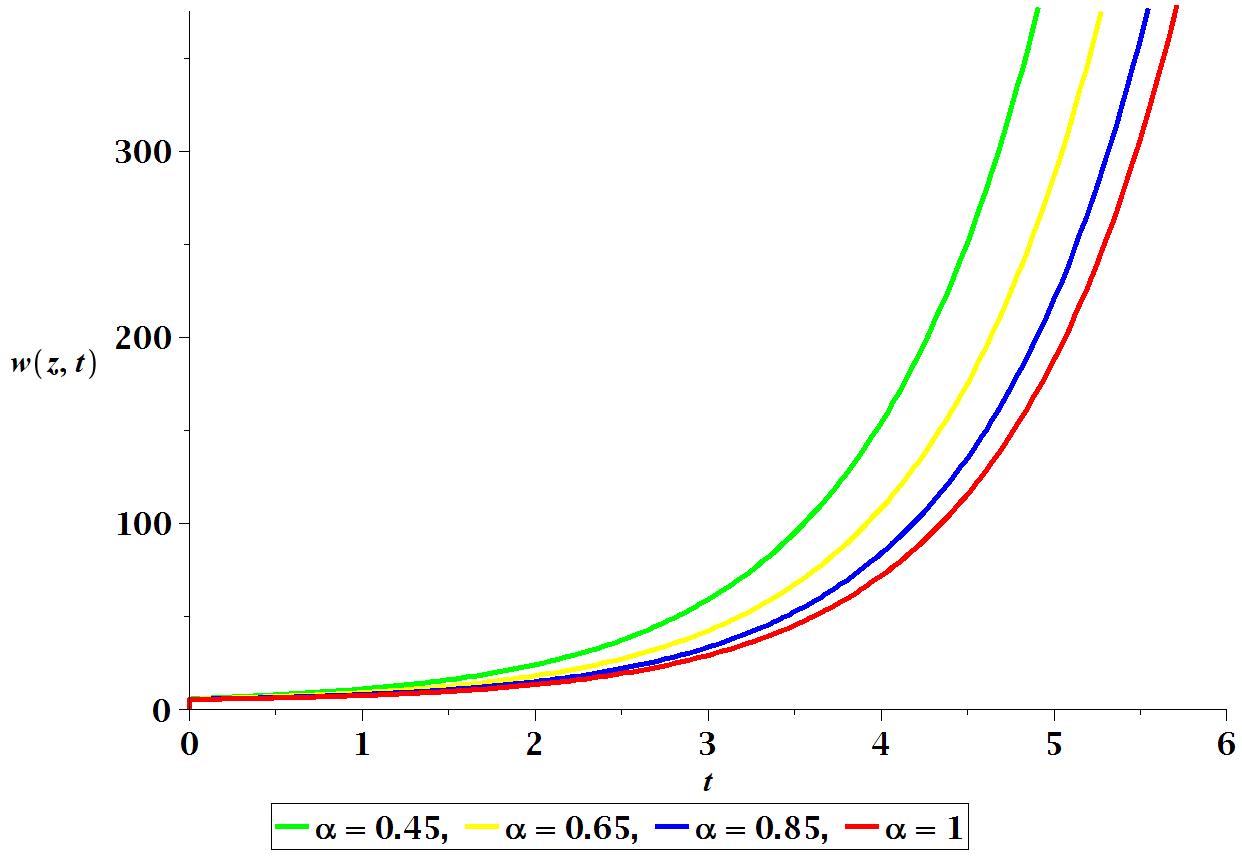}
			\caption{}
		\end{subfigure}
		\begin{subfigure}{0.5\textwidth}
			\includegraphics[width=8.5cm, height=5.5cm]{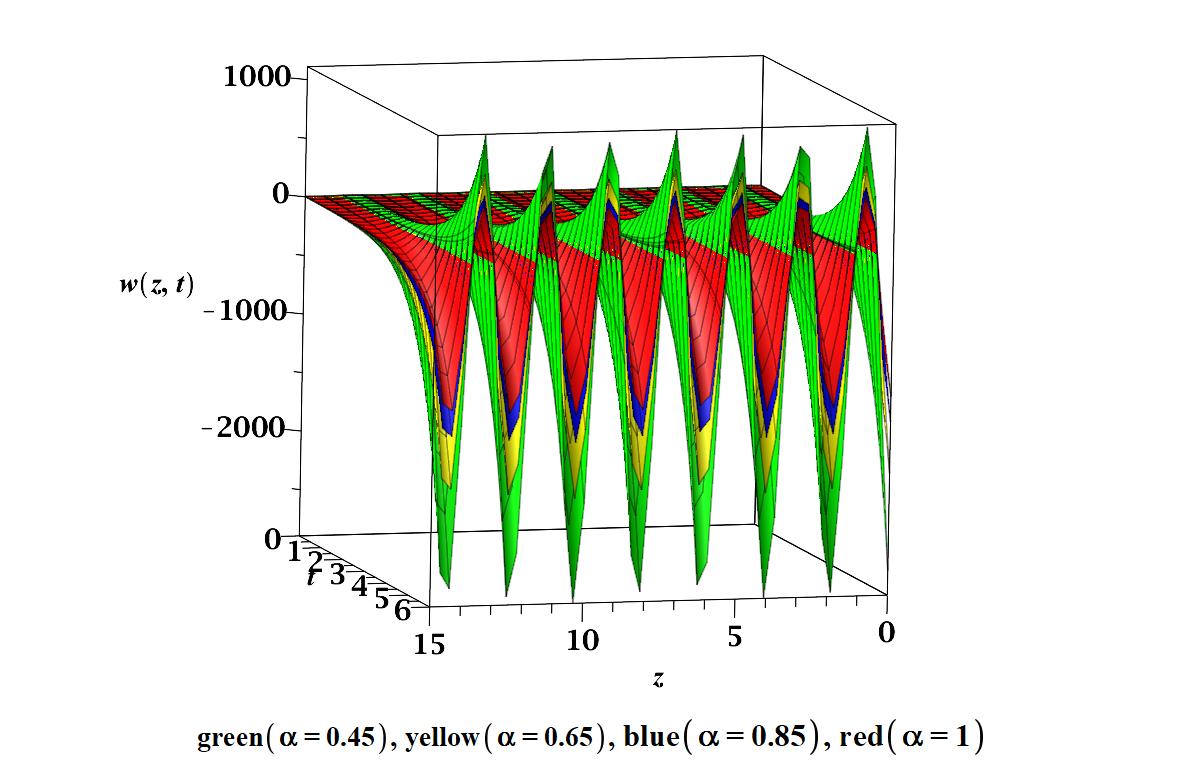}\caption{}
		\end{subfigure}
		\caption{ {(a) 2D and (b) 3D graphical representations of the arbitrary-order exact solutions \eqref{ls1} with parameter values $A_1=1,A_2=-2,a_1=3,\gamma=\mu_2=1,\mu_3=-4,\rho_1=9,$  for various values of $\alpha$, $\alpha\in(0,1].$}}\label{fig2}
	\end{figure}
	\begin{figure}[h!]
		\begin{subfigure}{0.5\textwidth}
			\includegraphics[width=8cm, height=6.5cm]{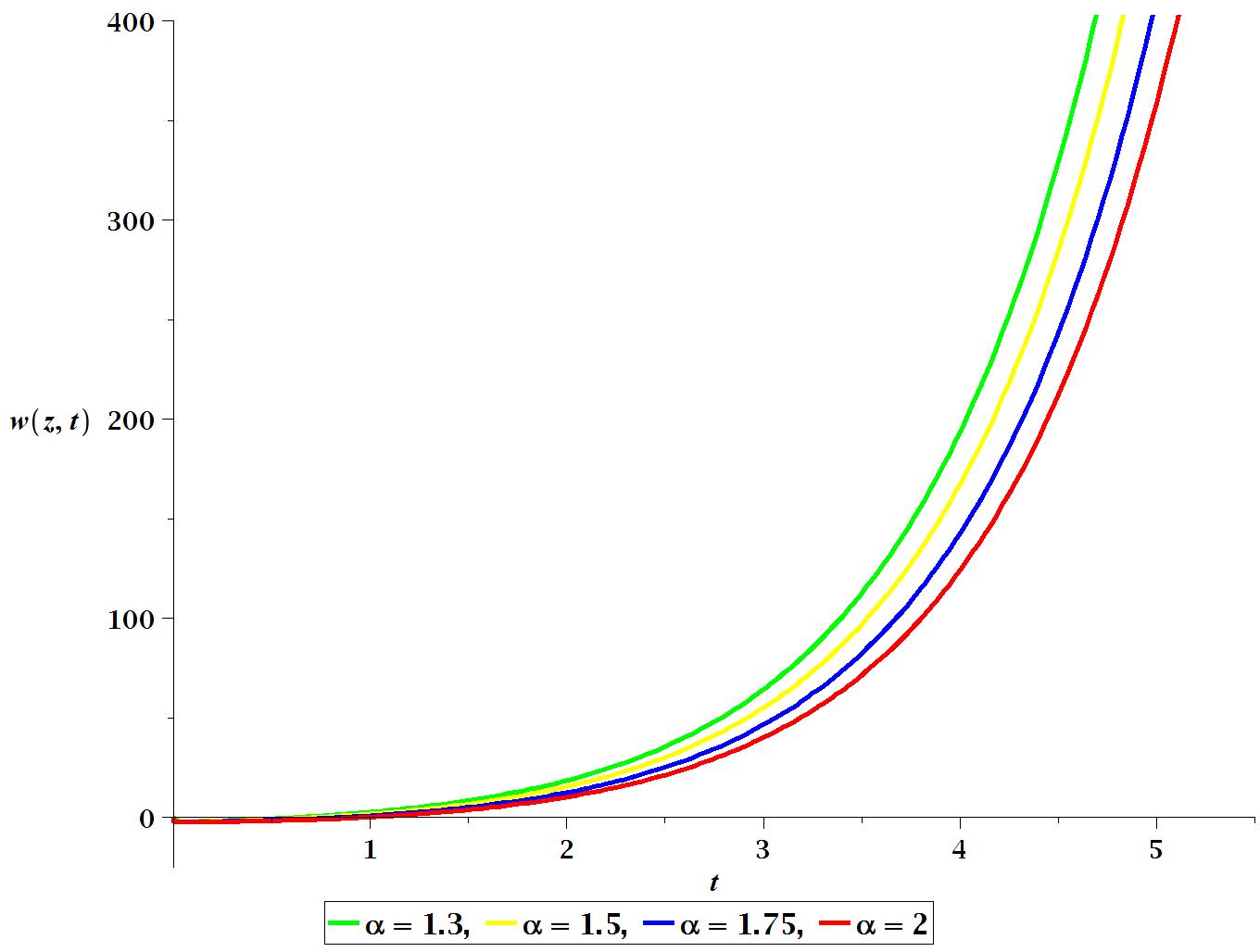}\caption{}
		\end{subfigure}
		\begin{subfigure}{0.5\textwidth}
			\includegraphics[width=9cm, height=6.5cm]{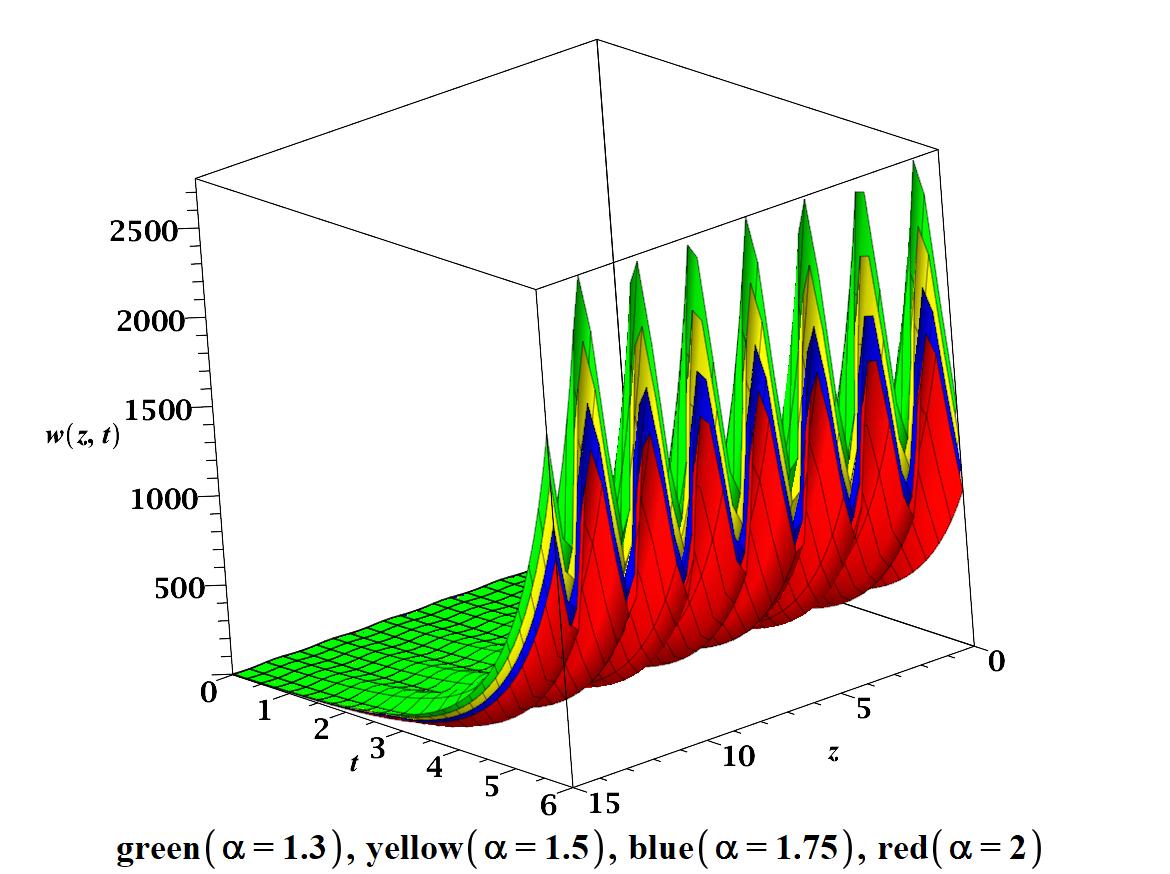}\caption{}
		\end{subfigure}
		\caption{ {{(a) 2D and (b) 3D graphical representations of the arbitrary-order exact solutions \eqref{ls1} with parameter values $A_1=A_3=\gamma=\mu_2=1,A_2=-2,A_4=-1,a_2=-3,a_1=3,\mu_3=4,\rho_1=9,$ for various values of $\alpha$, $\alpha\in(1,2].$}}}
		\label{fig3}
	\end{figure}
\end{example}
\subsection{{Various types of }exact solutions for some linear cases of \eqref{rdeq}}
%\begin{flushleft}	{Construction of exact solution of \eqref{rdeq} using exponential invariant subspaces}\end{flushleft}
\begin{example}
	Consider the following transformed linear time-fractional {CDR equation
		\begin{eqnarray}\begin{aligned}\label{lc1}
				\dfrac{\partial^\alpha w}{\partial t^\alpha} =\textbf{\textit{F}}_T[w]\equiv& \lambda_1^2\mu_4\left(
				\dfrac{\partial ^2w}{\partial z^2}\right)+
				\lambda_1\mu_3\left( \dfrac{\partial w}{\partial z}\right) +\mu_1w+\mu_2,\alpha\in(0,2],
	\end{aligned} \end{eqnarray}}
	 { where $ \mu_1,\mu_2,\mu_3,\mu_4\in\mathbb{R}, $}   along with the following initial and boundary conditions:
	\begin{eqnarray}
		&w(z,0)=\xi(z),&\alpha\in(0,1],\label{ta1l1}\\
		& w(z,0)=\xi(z) \,\,\, \& \,\, \,\dfrac{\partial w}{\partial t} \big{|}_{t=0}=\eta(z),& \alpha\in(1,2],\label{ta2l1}
	\end{eqnarray}
	{and}
	\begin{eqnarray}
		w(z,t)\Big{|}_{z=0}=\left\{
		\begin{array}{ll}
			\delta_1(t),\text{if}\, \alpha\in(0,1]\label{tbcl1},\\
			\delta_2(t),\text{if}\, \alpha\in(1,2],
		\end{array}  \right.
		\, \& \quad
		w(z,t)\Big{|}_{z=l}=\left\{
		\begin{array}{ll}
			\delta_3(t),\text{if}\, \alpha\in(0,1],\\
			\delta_4(t),\text{if}\, \alpha\in(1,2].
		\end{array}  \right.
	\end{eqnarray}
	The given equation \eqref{lc1} preserves the $(n+1) $-dimensional exponential invariant subspace $  \mathbf{V}_{n+1}=\text{Span}\{1,e^{-\rho_1z},\ldots,e^{-\rho_nz} \}. $
	For $\alpha=1$ and $\alpha=2$, the given equation \eqref{lc1} possesses the following exact solutions,
	{\begin{eqnarray}
			w(z,t)=\left\{\begin{array}{ll}
				a_1e^{\mu_1t}+\mu_2(e^{\mu_1 t}-1)+\sum\limits_{s=1}^nA_se^{\gamma_s t-\rho_sz}, \,  \text{if} \, \alpha=1,\\
				%\left\{\begin{array}{ll}
					a_1cosh(\sqrt{\mu_1}t)+\frac{a_2}{\sqrt{\mu_1}}sinh(\sqrt{\mu_1}t)+\frac{\mu_2}{\mu_1}cosh(\sqrt{\mu_1}t)
					\\+\sum\limits_{s=1}^n\big(A_s cosh(\sqrt{\gamma_s}t)+\frac{B_s}{\sqrt{\gamma_s}}sinh(\sqrt{\gamma_s}t)\big)e^{-\rho_sz}, \,  \text{if}\, \alpha=2,\label{ils1}
					%\end{array}\right.
				\end{array}\right.
			\end{eqnarray}
		}
		where $a_1,a_2,A_s,B_s,\in\mathbb{R},s=1,2\dots,n,$  and $\gamma_s=\lambda_1^2\rho_s^2\mu_4-\lambda_1\mu_3\rho_s+\mu_1,s=1,2,\dots,n,n\in\mathbb{N}.$
		Thus, we obtain the exact solutions of the equation \eqref{lc1} as follows,
		{\begin{eqnarray}\label{les1}
				w(z,t)=\left\{\begin{array}{ll}a_1E_{\alpha,1}(\mu_1t^\alpha)+
					\mu_2t^\alpha E_{\alpha,\alpha+1}(\mu_1t^\alpha)
					+\sum\limits_{s=1}^{n}A_sE_{\alpha,1}(\gamma_s t^\alpha)e^{-\rho_sz}
					,\text{if}\, \alpha\in(0,1],\\
					%\left\{\begin{array}{ll}
						\sum\limits_{s=1}^{n}\big(A_sE_{\alpha,1}(\gamma_s t^\alpha)
						+B_stE_{\alpha,2}(\gamma_s t^\alpha)\big)e^{-\rho_sz}
						+\mu_2t^\alpha E_{\alpha,\alpha+1}(\mu_1t^\alpha)
						\\+ a_1E_{\alpha,1}(\mu_1t^\alpha)
						+a_2tE_{\alpha,2}(\mu_1t^\alpha),\text{if}\, \alpha\in(1,2].
						%\end{array}\right.
					\end{array}\right.
			\end{eqnarray}}
			Here also it should be noted that the obtained exact solutions \eqref{les1} satisfy the given initial and boundary conditions  \eqref{ta1l1}-\eqref{tbcl1} such that
			{
				$$ \xi(z)=a_1+\sum\limits_{s=1}^nA_se^{-\rho_sz},
				\qquad \eta(z)=a_2+\sum\limits_{s=1}^nB_se^{-\rho_sz}, $$
				\begin{eqnarray*}
					\begin{aligned}
&\delta_1(t)=a_1E_{\alpha,1}(\mu_1t^\alpha)+
						\mu_2t^\alpha E_{\alpha,\alpha+1}(\mu_1t^\alpha)
						+\sum\limits_{s=1}^n{A_s}E_{\alpha,1}(\gamma_s t^\alpha),\\ &\delta_2(t)=a_1E_{\alpha,1}(\mu_1t^\alpha)
						+a_2tE_{\alpha,2}(\mu_1t^\alpha)
						+\mu_2t^\alpha E_{\alpha,\alpha+1}(\mu_1t^\alpha)
						\\& \qquad\quad +\sum\limits_{s=1}^n\big({A_s}E_{\alpha,1}(\gamma_s t^\alpha)
						+{B_s}tE_{\alpha,2}(\gamma_s t^\alpha)\big),
						\\
						&
						\delta_3(t)=a_1E_{\alpha,1}(\mu_1t^\alpha)+
						\mu_2t^\alpha E_{\alpha,\alpha+1}(\mu_1t^\alpha)
						+\sum\limits_{s=1}^n\hat{A_s}E_{\alpha,1}(\gamma_s t^\alpha),
						\\&
						\delta_4(t)= a_1E_{\alpha,1}(\mu_1t^\alpha)
						+a_2tE_{\alpha,2}(\mu_1t^\alpha)
						+\mu_2t^\alpha E_{\alpha,\alpha+1}(\mu_1t^\alpha)
						\\& \qquad\quad  +\sum\limits_{s=1}^n\big(\hat{A_s}E_{\alpha,1}(\gamma_s t^\alpha)
						+\hat{B_s}tE_{\alpha,2}(\gamma_s t^\alpha)\big) ,
					\end{aligned}
			\end{eqnarray*}}
			where $\hat{A_s}=A_se^{-\rho_sl},$ and $\hat{B_s}=B_se^{-\rho_sl},s=1,2,\dots,n.$
			Additionally, it is also observed that the non-integer order solutions \eqref{les1} coincide with the integer-order solutions \eqref{ils1} if $\alpha=1$ and $\alpha=2.$
			
		\end{example}
		%%%%%%%%%%%%%%%%%%%%%%%%%%%%%%%%%%%%%%%%%%%%%%%%%%%%%%%%%%%%%%%%%%%%%%%%%%%%%%%%%%%%%%%%%%%%%%%%%%%%%%%%%%%%%%%%%%%%%%%%%%%
		%\subsubsection{Construction of exact solution using hyperbolic invariant subspaces}
		\begin{example}
			Consider the following linear time-fractional diffusion-reaction equation
			{
				\begin{eqnarray}\begin{aligned}\label{lc2}
						\dfrac{\partial^\alpha w}{\partial t^\alpha} =& \lambda_1^2\mu_3\left(
						\dfrac{\partial^2 w}{\partial z^2}\right)+
						\mu_1w+\mu_2,\alpha\in(0,2],
			\end{aligned} \end{eqnarray}}
			 { where $ \mu_1,\mu_2,\mu_3\in\mathbb{R}, $} along with the given initial and boundary conditions \eqref{ta1}-\eqref{tbc}.\\
			The given equation \eqref{lc2} admits the $(2n+1)  $-dimensional hyperbolic linear space
			$$  \mathbf{V}_{2n+1}=\text{Span}\{1,sinh(\rho_1z),cosh(\rho_1z),\ldots,sinh(\rho_nz),cosh(\rho_nz)\}. $$
			For $\alpha=1$ and $\alpha=2$, the given equation \eqref{lc2} yields the following  exact solutions,
			{
				\begin{eqnarray}
					w(z,t)=\left\{
					\begin{array}{ll}
						a_1e^{\mu_1t}+\mu_2(e^{\mu_1t}-1)+ \sum\limits_{s=1}^n e^{\gamma_s t}\big[A_{2s-1}sinh(\rho_s z)+A_{2s}cosh(\rho_s z)\big],   \text{if} \, \alpha=1,\\
						%\left\{\begin{array}{ll}
							a_1cosh(\sqrt{\mu_1}t)+\frac{a_2}{\sqrt{\mu_1}}sinh(\sqrt{\mu_1}t)+\frac{\mu_2}{\mu_1}cosh(\sqrt{\mu_1}t)
							\\
							+ \sum\limits_{s=1}^n\big[A_{2s-1}sinh(\rho_s z)
							+A_{2s}cosh(\rho_s z)\big]cosh(\sqrt{\gamma_s}t)
							\\+\frac{1}{\sqrt{\gamma_s}}sinh(\sqrt{\gamma_s}t)\big[B_{2s-1}sinh(\rho_s z)
							+B_{2s}cosh(\rho_s z)\big],   \text{if}\, \alpha=2,\label{ils2}
							%\end{array}\right.
						\end{array}\right.
				\end{eqnarray}}
				where $a_1,a_2,A_s,B_s \in\mathbb{R},s=1,2,\dots,2n,$ and $\gamma_s=\lambda_1^2\mu_3\rho_s^2+\mu_1,s=1,2,\dots,n$. \\
				Thus, we obtain the exact solutions of the equation \eqref{lc2} in the form
				{
					\begin{eqnarray}
						w(z,t)=\left\{
						\begin{array}{ll}
							%\left\{ 	\begin{array}{ll}
								\sum\limits_{s=1}^nE_{\alpha,1}(\gamma_s t^\alpha)\big[A_{2s-1}sinh(\rho_s z)+A_{2s}cosh(\rho_s z)\big]\\
								+a_1E_{\alpha,1}(\mu_1t^\alpha)+
								\mu_2t^\alpha E_{\alpha,\alpha+1}(\mu_1t^\alpha)
								, \text{if}\, \alpha\in(0,1],
								% \end{array}\right.
							\\
							% \left\{  \begin{array}{ll}
								\sum\limits_{s=1}^n E_{\alpha,1}(\gamma_s t^\alpha)\big[A_{2s-1}sinh(\rho_s z)+A_{2s}cosh(\rho_s z)\big]+  a_1E_{\alpha,1}(\mu_1t^\alpha)
								\\ +a_2tE_{\alpha,2}(\mu_1t^\alpha)
								+ \sum\limits_{s=1}^n tE_{\alpha,2}(\gamma_s t^\alpha)\big[B_{2s-1}sinh(\rho_s z)
								+B_{2s}cosh(\rho_s z)\big]
								\\     +\mu_2t^\alpha E_{\alpha,\alpha+1}(\mu_1t^\alpha) ,\, \text{if}\, \alpha\in(1,2].
								\label{ls2}
								%\end{array}\right.
							\end{array}\right.
					\end{eqnarray}}
					The above solutions \eqref{ls2} satisfy the given initial and boundary conditions \eqref{ta1}-\eqref{tbc} along with
					{
						\begin{eqnarray*}
							\begin{aligned}
								& \xi(z)=a_1
								+\sum\limits_{s=1}^n\big[A_{2s-1}sinh(\rho_s z)+A_{2s}cosh(\rho_s z)\big] ,
								\\&
								\eta(z)= a_2
								+\sum\limits_{s=1}^n\big[B_{2s-1}sinh(\rho_s z)
								+B_{2s}cosh(\rho_s z)\big],
								\\&
								\delta_1(t)= a_1E_{\alpha,1}(\mu_1t^\alpha)+
								\mu_2t^\alpha E_{\alpha,\alpha+1}(\mu_1t^\alpha)
								+\sum\limits_{s=1}^n E_{\alpha,1}(\gamma_s t^\alpha){A_{2s}} ,
								\\&
								\delta_2(t)=a_1E_{\alpha,1}(\mu_1t^\alpha)
								+a_2tE_{\alpha,2}(\mu_1t^\alpha)
								+\mu_2t^\alpha E_{\alpha,\alpha+1}(\mu_1t^\alpha)
								\\ & \qquad
								\quad+\sum\limits_{s=1}^n\left[ E_{\alpha,1}(\gamma_s t^\alpha){A_{2s}}+tE_{\alpha,2}(\gamma_s t^\alpha){B_{2s}}\right]
								,
							\end{aligned}
						\end{eqnarray*}
						\begin{eqnarray*}
							\begin{aligned}
								&\delta_3(t)= a_1E_{\alpha,1}(\mu_1t^\alpha)+
								\mu_2t^\alpha E_{\alpha,\alpha+1}(\mu_1t^\alpha)
								+\sum\limits_{s=1}^n E_{\alpha,1}(\gamma_s t^\alpha)\hat{A} ,\text{ and }
								\\&
								\delta_4(t)=
								a_1E_{\alpha,1}(\mu_1t^\alpha)
								+a_2tE_{\alpha,2}(\mu_1t^\alpha)
								+\mu_2t^\alpha E_{\alpha,\alpha+1}(\mu_1t^\alpha)
								\\ & \qquad
								\quad  +\sum\limits_{s=1}^n\left[ E_{\alpha,1}(\gamma_s t^\alpha)\hat{A}
								+tE_{\alpha,2}(\gamma_s t^\alpha)\hat{B}\right]
								,	\end{aligned}
					\end{eqnarray*}}
					where $\hat{A}=\sum\limits_{s=1}^n\big[A_{2s-1}sinh(\rho_sl)+A_{2s}cosh(\rho_s l)\big]$ and $ \hat{B}=\sum\limits_{s=1}^n\big[B_{2s-1}sinh(\rho_s l)+B_{2s}cosh(\rho_sl)\big]. $
					It should be noted that the exact solutions \eqref{ls2} coincide with the integer-order solutions \eqref{ils2} if $\alpha=1$ and $\alpha=2.$
					% \subsubsection{Construction of exact solutions using the combinations of trigonometric and exponential invariant subspaces}
				\end{example}
				%%%%%%%%%%%%%%%%%%%%%%%%%%%%%%%%%%%%%%%%%%%%%%%%%%%%%%%%%%%%%%%%%
				\begin{example}  Consider the following linear time-fractional diffusion-reaction equation
					{  \begin{eqnarray}\begin{aligned}\label{lc3}
								\dfrac{\partial^\alpha w}{\partial t^\alpha} =& -\mu_2\lambda_1^2 \left( \dfrac{\partial^2 w}{\partial z^2}\right)+
								\mu_1w,& \alpha\in(0,2],
					\end{aligned} \end{eqnarray}}
					 {where $ \mu_1,\mu_2\in\mathbb{R}, $}  along with the given initial and boundary conditions \eqref{ta1}-\eqref{tbc}.\\
					The given equation \eqref{lc3} admits the $ (3n+1)$-dimensional trigonometric and exponential invariant subspace
					$$ \mathbf{V}_{3n+1}=\text{Span}\{1,e^{\varrho_1z},\ldots,e^{\varrho_n z},sin(\sqrt{\rho_1}z),cos(\sqrt{\rho_1}z),\ldots,sin(\sqrt{\rho_n}z),cos(\sqrt{\rho_n}z)\}.  $$
					Thus, the obtained  exact solutions of \eqref{lc3} for $\alpha=1$ and $\alpha=2$ are  as follows,
					{\begin{eqnarray}
							w(z,t)=\left\{
							\begin{array}{ll}
								a_1e^{\mu_1t}+\sum\limits_{s=1}^n{B_s}e^{\beta_s t-\varrho_sz}+\sum\limits_{s=1}^ne^{\gamma_s t}\big[A_{2s-1}sin(\sqrt{\rho_s}z)+{A_{2s}}cos(\sqrt{\rho_s}z)\big],\, \text{if}\, \alpha=1,
								\\
								%\left\{\begin{array}{ll}
									a_1cosh(\sqrt{\mu_1}t)
									+a_2\dfrac{sinh(\sqrt{\mu_1}t)}{\sqrt{\mu_1}}
									+\sum\limits_{s=1}^ncosh(\sqrt{\gamma_s}t)\big[A_{2s-1}sin(\sqrt{\rho_s}z)
									\\+{A_{2s}}3cos(\sqrt{\rho_s}z)\big]
									+\sum\limits_{s=1}^n\dfrac{sinh(\sqrt{\gamma_s}t)}{\sqrt{\gamma_s}})\big[{M_{2s-1}}sin(\sqrt{\rho_s}z)
									\\
									+{M_{2s}}cos(\sqrt{\rho_s}z)\big]
									+\sum\limits_{s=1}^n\big[{B_s}cosh(\sqrt{\beta_s}t)+{N_s}\dfrac{sinh(\sqrt{\beta_s}t)}{\sqrt{\beta_s}})\big]e^{\varrho_sz},
									\text{if}\, \alpha=2,\label{ils3}
									%\end{array}\right.	
								\end{array}\right.
						\end{eqnarray}}
						where $a_1,a_2,A_s,B_s,M_s,N_s \in\mathbb{R},$ and $\gamma_s=\lambda_1^2\mu_2\rho_s+\mu_1,s=1,2,\dots,n.$\\
						Now, for $\alpha\in(0,2]$, we obtain the exact solutions of \eqref{lc3} as follows:
						{ \begin{eqnarray}
								w(z,t)=\left\{
								\begin{array}{ll}
									%\left\{ 	\begin{array}{ll}
										a_1E_{\alpha,1}({\mu_1t^\alpha})+\sum\limits_{s=1}^n{B_s}E_{\alpha,1}(\beta_s t^\alpha)e^{\varrho_sz}+\sum\limits_{s=1}^n E_{\alpha,1}(\gamma_s t^\alpha)\big[A_{2s-1}sin(\sqrt{\rho_s}z)
										\\+{A_{2s}}cos(\sqrt{\rho_s}z)\big]
										,  \text{if}\, \alpha\in(0,1],
										%\end{array}\right.
										\\
										% \left\{ \begin{array}{ll}
											a_1E_{\alpha,1}({\mu_1t^\alpha})+a_2tE_{\alpha,2}(\mu_1 t^\alpha)
											+\sum\limits_{s=1}^n\big[{B_s}E_{\alpha,1}(\beta_s t^\alpha)+{N_s}tE_{\alpha,2}(\beta_s t^\alpha)\big]e^{\varrho_sz}
											\\
											+\sum\limits_{s=1}^n E_{\alpha,1}(\gamma_s t^\alpha)\big[A_{2s-1}sin(\sqrt{\rho_s}z)
											+{A_{2s}}3cos(\sqrt{\rho_s}z)\big]
											\\+\sum\limits_{s=1}^n\big[{M_{2s-1}}sin(\sqrt{\rho_s}z)
											+{M_{2s}}cos(\sqrt{\rho_s}z)\big] tE_{\alpha,2}(\gamma_s t^\alpha),  \text{if}\, \alpha\in(1,2].
										\end{array}\right.\label{ls3}
										%\end{array}\right.
								\end{eqnarray}}
								% It is observed that the integer order solutions in \eqref{ils3} agree with the exact solutions obtained in \eqref{ls3} for the equation \eqref{lc3}.
								The obtained exact solutions \eqref{ls3} satisfy the given initial and boundary conditions  \eqref{ta1}-\eqref{tbc} along with
								{ \begin{eqnarray*}\begin{aligned}
											&	\xi(z)=a_1+\sum\limits_{s=1}^n\big({B_s}e^{\varrho_sz}+ {A_{2s-1}}sin(\sqrt{\rho_s}z)+{A_{2s}}cos(\sqrt{\rho_s}z)\big),
											\\
											& \eta(z)=a_2+\sum\limits_{s=1}^n\big({N_s}e^{\varrho_sz}+ {M_{2s-1}}sin(\sqrt{\rho_s}z)+{M_{2s}}cos(\sqrt{\rho_s}z)\big),\\
											& \delta_1(t)=
											{a_1}E_{\alpha,1}({\mu_1t^\alpha})+
											\sum\limits_{s=1}^n\big[{A_{2s}}E_{\alpha,1}(\gamma_s t^\alpha)+{B_s}E_{\alpha,1}(\beta_s t^\alpha)\big],
											\\&
											\delta_2(t)=
											a_1E_{\alpha,1}({\mu_1t^\alpha})+a_2tE_{\alpha,2}(\mu_1 t^\alpha)+\sum\limits_{s=1}^n\big[{A_{2s}}E_{\alpha,1}(\gamma_s t^\alpha)+{B_s}E_{\alpha,1}(\beta_s t^\alpha)\big]
											\\&\qquad\quad+
											\sum\limits_{s=1}^n\big[{M_{2s}}tE_{\alpha,2}(\gamma_s t^\alpha)+{N_s}tE_{\alpha,2}(\beta_s t^\alpha)\big]
											, \\
											%\end{aligned}\end{eqnarray*} \begin{eqnarray*}\begin{aligned}
											& 		  \delta_3(t)=
											a_1E_{\alpha,1}({\mu_1t^\alpha})+\sum\limits_{s=1}^n{B_s}E_{\alpha,1}(\beta_s t^\alpha)+A\sum\limits_{s=1}^n E_{\alpha,1}(\gamma_s t^\alpha),
											\\&
											\delta_4(t)=
											a_1E_{\alpha,1}({\mu_1t^\alpha})+a_2tE_{\alpha,2}(\mu_1 t^\alpha)+B\sum\limits_{s=1}^n{B_s}E_{\alpha,1}(\beta_s t^\alpha)
											\\&\qquad\quad
											+A_1\sum\limits_{s=1}^n E_{\alpha,1}(\gamma_s t^\alpha)+B\sum\limits_{s=1}^n{N_s}tE_{\alpha,2}(\beta_s t^\alpha)+A_2\sum\limits_{s=1}^nt E_{\alpha,2}(\gamma_s t^\alpha),
										\end{aligned}
								\end{eqnarray*}}
								where $B=\sum\limits_{s=1}^n e^{\varrho_sl},A=\sum\limits_{s=1}^nA_{2s},A_1=\sum\limits_{s=1}^n\big[{A_{2s-1}}sin(\sqrt{\rho_s}l)+{A_{2s}}cos(\sqrt{\rho_s}l)\big],$ and $A_2=\sum\limits_{s=1}^n\big[{M_{2s-1}}sin(\sqrt{\rho_s}l)+{M_{2s}}cos(\sqrt{\rho_s}l)\big]. $
							\end{example}
							%%%%%%%%%%%%%%%%%%%%%%%%%%%%%%%%%%%%%%%%%%%%%%%%%%%%%%%%%%%%%%%%%%%%%%%%%%%%%%%%%%%%%%%%%%%%%%%%%%%%%%%%%%%%%%%%%%%%%%%%%%%%%%%%%%
							% \subsubsection{Construction of exact solutions using polynomial linear space}
							\begin{example}  Finally, we consider the linear time-fractional diffusion-convection equation in the following form
								{  \begin{eqnarray}\begin{aligned}\label{lc4}
											\dfrac{\partial^\alpha w}{\partial t^\alpha} =&  \lambda_1^2\mu_2\left( \dfrac{\partial^2 w}{\partial z^2}\right)
											+
											\lambda_1 \mu_1\left( \dfrac{\partial w}{\partial z}\right) ,\alpha\in(0,2],
								\end{aligned} \end{eqnarray}}
								 { where $ \mu_1,\mu_2\in\mathbb{R}, $}  along with the given initial and boundary conditions \eqref{ta1}-\eqref{tbc}.\\
								The above equation \eqref{lc4} preserves the linear space $\mathbf{V}_n=\text{Span}\{1,z,z^2,z^3,\ldots,z^{n-1}\}.$ Let $n=5$.
								The integer-order solutions of equation \eqref{lc4} are then obtained as follows,
								{ \begin{eqnarray}
										w(z,t)=
										\left\{
										% \begin{array}{ll}\left\{
											\begin{array}{ll}
												\sum\limits_{r=1}^5A_rz^{5-r}
												+t(\gamma_0z^3 +\gamma_1z^2+\gamma_3z+\gamma_6)+\dfrac{t^2}{2}(\gamma_2z^2+\gamma_4z+\gamma_7)\\+\dfrac{t^3}{6}(\gamma_5z+\gamma_8)+\lambda_1\mu_1\gamma_5\dfrac{t^4}{24}, \text{if}\, \alpha=1,
												% \end{array}\right.
											\\
											%\left\{\begin{array}{ll}
												\sum\limits_{r=1}^5(A_r+B_rt)z^{5-r}+ \dfrac{t^2}{2!}(\gamma_0z^3 +\gamma_1z^2+\gamma_3z+\gamma_6)
												\\
												+\dfrac{t^3}{3!}(\beta_0z^3 +\beta_1z^2+\beta_3z+\beta_6)
												+\dfrac{t^4}{4!}(\gamma_2z^2+\gamma_4z+\gamma_7)+\dfrac{t^5}{5!}(\beta_2z^2
												\\
												+\beta_4z+\beta_7)
												+\dfrac{t^6}{6!}(\gamma_5z+\gamma_8)+\dfrac{t^7}{7!}(\beta_5z+\beta_8)
												+
												\lambda_1\mu_1\gamma_5\dfrac{t^8}{8!}
												\\
												+\lambda_1\mu_1\beta_5\dfrac{t^9}{9!}
												, \text{if}\, \alpha=2,\label{ils4}
											\end{array}\right.
											%\end{array}\right.
					\end{eqnarray}}
		where			{\begin{eqnarray*}					\begin{aligned}
			&	\gamma_8=2\lambda_1^2\mu_2\gamma_2+\lambda_1\mu_1\gamma_4,\\
			&\gamma_7=2\lambda_1^2\mu_2\gamma_1+\lambda_1\mu_1\gamma_3,\\
			& \gamma_6=2\lambda_1^2\mu_2A_3+\lambda_1\mu_1A_4+\mu_2,\\
			& \gamma_5=2\lambda_1\mu_1\gamma_2,\\
			&	\gamma_4=24\lambda_1^3\mu_2\mu_1A_1+2\lambda_1\mu_1\gamma_1,\, \\
					& \gamma_3=6\lambda_1^2\mu_2A_2+2\lambda_1\mu_1A_3, \\
			&\gamma_2=12\lambda_1^2\mu_1^2A_1,\\
					&\gamma_1=12\lambda_1^2\mu_2A_1+3\lambda_1\mu_1A_2,\\
				& \gamma_0=4\lambda_1\mu_1A_1 , 		    \end{aligned}
							\begin{aligned}
									& \beta_8= 2\lambda_1^2\mu_2\beta_2+\lambda_1\mu_1\beta_4,\\
									& \beta_7=2\lambda_1^2\mu_2\beta_1+\lambda_1\mu_1\beta_3,\\	&\beta_6=2\lambda_1^2\mu_2B_3+\lambda_1\mu_1B_4,\\
												&  		  \beta_5= 2\lambda_1\mu_1\beta_2,\\
												& \beta_4=24\lambda_1^3\mu_2\mu_1B_1+2\lambda_1\mu_1\beta_1,\\
												& \beta_3=6\lambda_1^2\mu_2B_2+2\lambda_1\mu_1B_3, \\
												& \beta_2=12\lambda_1^2\mu_1^2B_1, \\
												& \beta_1=12\lambda_1^2\mu_2B_1+3\lambda_1\mu_1B_2, \\
												&\beta_0= 4\lambda_1\mu_1B_1,
											\end{aligned}
									\end{eqnarray*} }
									and $A_r,B_r\in\mathbb{R},r=1,2,\dots,5.$
									\\
									For $\alpha\in(0,2],$ the obtained exact solutions of \eqref{lc4} are in the form
									{ \begin{eqnarray}
											\begin{aligned}
												w(z,t)=&
												%	\left\{			\begin{array}{ll}
													%	 \left\{\begin{array}{ll}
														\sum\limits_{r=1}^5A_rz^{5-r}
														+\dfrac{t^{\alpha}}{\Gamma(\alpha+1)}(\gamma_0z^3 +\gamma_1z^2+\gamma_3z+\gamma_6)
														\\&+\dfrac{t^{2\alpha}}{\Gamma(2\alpha+1)}(\gamma_2z^2+\gamma_4z+\gamma_7)
														+\dfrac{t^{3\alpha}}{\Gamma(3\alpha+1)}(\gamma_5z+\gamma_8)
														\\&
														+\lambda_1\mu_1\gamma_5\dfrac{t^{4\alpha}}{\Gamma(4\alpha+1)}, \text{if}\, \alpha\in(0,1],
														%\end{array}\right.
														\\
													\end{aligned}
												\end{eqnarray}
												\begin{eqnarray}
													\begin{aligned}
														% \left\{\begin{array}{ll}
															w(z,t)=&	\sum\limits_{r=1}^5(A_r+B_rt)z^{5-r}+ \dfrac{t^{\alpha}}{\Gamma(\alpha+1)}(\gamma_0z^3 +\gamma_1z^2+\gamma_3z+\gamma_6)
															\\&	+\dfrac{t^{\alpha+1}}{\Gamma(\alpha+2)}(\beta_0z^3
															+
															\beta_1z^2+\beta_3z+\beta_6)
															+\dfrac{t^{2\alpha}}{\Gamma(2\alpha+1)}(\gamma_2z^2+\gamma_4z+\gamma_7)
															\\&
															+\dfrac{t^{2\alpha+1}}{\Gamma(2\alpha+2)}(\beta_2z^2+\beta_4z+\beta_7)
															+\dfrac{t^{3\alpha}}{\Gamma(3\alpha+1)}(\gamma_5z+\gamma_8)
															\\& +\dfrac{t^{3\alpha+1}}{\Gamma(3\alpha+2)}(\beta_5z+\beta_8)
															+ \lambda_1\mu_1\gamma_5\dfrac{t^{4\alpha}}{\Gamma(4\alpha+1)}
															\\&\label{ls4b}
															+\lambda_1\mu_1\beta_5\dfrac{t^{4\alpha+1}}{\Gamma(4\alpha+2)}
															, \text{if}\, \alpha\in(1,2].\label{ls4}
															% \end{array}\right.
														%\end{array}\right.
													\end{aligned}
											\end{eqnarray}}
											The above fractional-order exact solutions \eqref{ls4}-\eqref{ls4b}  satisfy the given initial and boundary conditions \eqref{ta1}-\eqref{tbc} along with
											{ \begin{eqnarray*}
													\begin{aligned}
														&  \xi(z)=\sum\limits_{r=1}^5A_rz^{5-r},\qquad \eta(z)=\sum\limits_{r=1}^5B_rz^{5-r},\\
														& \delta_1(t)=
														A_5+\dfrac{t^{\alpha}}{\Gamma(\alpha+1)}\gamma_6+\dfrac{t^{2\alpha}}{\Gamma(2\alpha+1)}\gamma_7+\dfrac{t^{3\alpha}}{\Gamma(3\alpha+1)}\gamma_8+\lambda_1\mu_1\gamma_5\dfrac{t^{4\alpha}}{\Gamma(4\alpha+1)},\\
														& \delta_2(t)=
														A_5+B_5t+ \dfrac{t^{\alpha}}{\Gamma(\alpha+1)}\gamma_6+\dfrac{t^{\alpha+1}}{\Gamma(\alpha+2)}\beta_6
														+\dfrac{t^{2\alpha}}{\Gamma(2\alpha+1)}\gamma_7+\dfrac{t^{2\alpha+1}}{\Gamma(2\alpha+2)}\beta_7
														\\& \qquad\quad	+\dfrac{t^{3\alpha}}{\Gamma(3\alpha+1)}\gamma_8
														+\dfrac{t^{3\alpha+1}}{\Gamma(3\alpha+2)}\beta_8
														+ \lambda_1\mu_1\gamma_5\dfrac{t^{4\alpha}}{\Gamma(4\alpha+1)}
														+\lambda_1\mu_1\beta_5\dfrac{t^{4\alpha+1}}{\Gamma(4\alpha+2)}
														,
														\\
														&  \delta_3(t)=
														a_1+\dfrac{t^{\alpha}}{\Gamma(\alpha+1)}a_2+\dfrac{t^{2\alpha}}{\Gamma(2\alpha+1)}a_3+\dfrac{t^{3\alpha}}{\Gamma(3\alpha+1)}a_4+\lambda_1\mu_1\gamma_5\dfrac{t^{4\alpha}}{\Gamma(4\alpha+1)},\\
														& \delta_4(t)= a_1+b_1t+ \dfrac{t^{\alpha}}{\Gamma(\alpha+1)}a_2+\dfrac{t^{\alpha+1}}{\Gamma(\alpha+2)}b_2
														+\dfrac{t^{2\alpha}}{\Gamma(2\alpha+1)}a_3+\dfrac{t^{2\alpha+1}}{\Gamma(2\alpha+2)}b_3
														\\& \qquad\quad		+\dfrac{t^{3\alpha}}{\Gamma(3\alpha+1)}a_4
														+\dfrac{t^{3\alpha+1}}{\Gamma(3\alpha+2)}b_4
														+ \lambda_1\mu_1\gamma_5\dfrac{t^{4\alpha}}{\Gamma(4\alpha+1)}
														+\lambda_1\mu_1\beta_5\dfrac{t^{4\alpha+1}}{\Gamma(4\alpha+2)}
														, 		 	\end{aligned}
												\end{eqnarray*}
											}
											where $ a_1=  \sum\limits_{r=1}^5A_rl^{5-r},b_1=\sum\limits_{r=1}^5B_rl^{5-r},$
											$a_2=\gamma_0l^3 +\gamma_1l^2+\gamma_3l+\gamma_6,b_2=\beta_0l^3    +
											\beta_1l^2+\beta_3l+\beta_6  ,$ $a_3=\gamma_2l^2+\gamma_4l+\gamma_7,b_3=\beta_2l^2+\beta_4l+\beta_7,$ $a_4=\gamma_5l+\gamma_8,$ and $b_4=\beta_5l+\beta_8.$ \\
											Also, we observe that when $\alpha=1$ and $\alpha=2,$ the fractional-order exact solutions \eqref{ls4}-\eqref{ls4b} coincide with integer-order solutions \eqref{ils4}.
		\end{example}			\section{Extension of invariant subspace method associated with variable transformation to $(k+1)$-dimensional nonlinear time-fractional PDEs involving several linear time delays}
		This section presents how we can extend the invariant subspace method associated with variable transformation to $(k+1)$-dimensional nonlinear time-fractional PDEs with several linear time delays.
\subsection{Estimation of invariant subspaces associated with variable transformation for the $(k+1)$-dimensional nonlinear time-fractional PDE involving several linear time delays}
Let us consider the following generalized $(k+1)$-dimensional nonlinear time-fractional PDE with several linear time delays
% we describe schematic study for constructing the invariant subspaces associated with variable transformation for  time delay $(k+1)$-dimensional TFNPDE. Thus, w
\begin{eqnarray}\label{D1}
\begin{aligned}
& \frac{\partial^{\alpha}u}{\partial t^{\alpha}}=\tilde{\textbf{F}}[u,\tilde{{u_i}}]\equiv\textit{\textbf{F}}[u] + \sum_{i=1}^{N}\kappa_i\tilde{{u_i}},\ \alpha>0,\ t>0,\\
& u(x_1,\dots,x_k,t)= \nu(x_1,\dots,x_k,t) \quad \text{if} \; t \in [-\hat{\tau},0],
\end{aligned}
\end{eqnarray}
{where}   $\kappa_i>0, i=1,\dots,N,$ $N\in \mathbb{N},$ $\textit{\textbf{F}}[u]$ as given in \eqref{kop}, $u=u(x_1,\dots,x_k,t)$, {$\tilde{{u_i}}=u(x_1,\dots,x_k,t-\tau_i),$} $\tau_i>0 $  and $\hat{\tau}=max \{ \tau_i: i=1,\dots,N \}$, $x_i\in\mathbb{R},i=1,\dots,k.$
%Here, $\dfrac{\partial^{\alpha}}{\partial t^{\alpha}}(\cdot) $ denotes the Caputo fractional derivative \eqref{c} of order $\alpha$, and .
\\
{Here, the methodology of the invariant subspace method associated with the transformation is the same as discussed earlier in section 2.}
	%%%%%%%%%%%%%%%%%%%%%%%%%%%%5555555%%%%%%%%%%%%%%%%%%%%%%%%%%%%%%%%%%%%%%%%%%%%%%%%%%%%%%%%%%%%%%%%%%%%%Proceeding with the steps 1-4 mentioned as before, let us first we apply the substitution $u=w(z,t),, on the above nonlinear partial differential operator with time delay
  {Using the ansatz \eqref{wt},} the differential operator $\tilde{\textbf{F}}[u,{\tilde{{u_i}}}]$ gets transformed into
\begin{equation}\label{top2}
\tilde{\textbf{\textit{F}}}_T[w,\hat{w_i}]=\textbf{\textit{F}}_T\big(z,w,w^{(1)}_z,\dots,
w^{(m)}_z\big)+ \sum_{i=1}^{N}\kappa_i\hat{w_i},w^{(j)}_z=\dfrac{\partial^j w}{\partial z^j},
\end{equation}
where $ \hat{w_i}=w(z,t-\tau_i),\ \tau_i>0, \hat{\tau}=max \{ \tau_i: i=1,2,..,N \}, j=1,2,\dots,m. $
Thus,
the linear space $\mathbf{V}_n(n<\infty)$ given in \eqref{vs} is an invariant subspace of the nonlinear differential operator $\tilde{\textbf{F}}_T[w,\hat{w_i}]$ in \eqref{top2}  if {$\tilde{\textbf{F}}_T[\mathbf{V}_n] \subseteq \mathbf{V}_n.$} Thus, when $w=\sum\limits_{s=1}^{n}C_s\phi_s(z),$
%for every $ w\in \mathbf{V}_n$ implies .
$$
\tilde{\textbf{F}}_T\left[\sum_{s=1}^{n}C_s\phi_s(z), \sum_{s=1}^{n}\tilde{C}_s\phi_s(z)\right]=  \sum \limits^{n} _{s=1}\Omega_s(C_1,C_2,\dots,C_n)\phi_s(z) \label{D3} +\sum_{i=1}^{N}\sum_{s=1}^{n} \kappa_i\tilde{C}_s\phi_s(z),
$$
where $\tilde{C}_s,C_s \in \mathbb{R}$ and $\Omega_s $ denote the coefficients with respect to the basis set $ \Big\{\phi_s\Big(\sum\limits_{r=1}^k\lambda_rx_r\Big)\Big{|}\\
 s=1,2,\dots,n \Big\}.$
We would like to mention that the linear space $\mathbf{V}_n$ given in \eqref{vs} is invariant under the nonlinear ordinary differential operator $\tilde{\textbf{F}}_T[w,\hat{w_i}]$ given in \eqref{top2} if and only if  the nonlinear partial differential operator {$\tilde{\textbf{F}}[u,\tilde{{u_i}}]$} preserves the linear space  $\mathbf{V}_n$ given in \eqref{vs} along with variable transformation $ z=\sum\limits_{r=1}^k\lambda_rx_r $ and vice-versa.

 Next, we give a detailed study for deriving the exact solutions of the initial and boundary value problems for the time delay $(3+1)$-dimensional generalized nonlinear time-fractional {CDR} equation using the invariant subspace method associated with the variable transformation $ z=\lambda_1x_1+\lambda_2x_ 2+\lambda_3x_3$.
\subsection{Estimation of invariant subspaces associated with variable transformation for the time delay $(3+1)$-dimensional generalized nonlinear time-fractional {CDR} equation}
Consider the  $(3+1)$-dimensional generalized nonlinear time-fractional {CDR} equation with several linear time delays of the form,
{ \begin{eqnarray}\begin{aligned}\label{crdeqD}
 &\dfrac{\partial^\alpha u}{\partial t^\alpha} =\tilde{\textbf{F}}[u,\tilde{u}_i]\equiv \sum\limits_{r=1}^{3} \dfrac{\partial}{\partial x_r}\left( F_{r}(u)\dfrac{\partial u}{\partial x_r}\right)+\sum\limits_{r=1}^{3}  K_{r}(u)\dfrac{\partial u}{\partial x_r}+R(u)+\sum\limits_{i=1}^N \kappa_i\tilde{u}_i, \alpha \in (0,2],\\
 & u(x_1,x_2,x_3,t)= \nu(x_1,x_2,x_3,t) \quad \text{if} \; t \in [-\hat{\tau},0],\hat{\tau}=max \{ \tau_i: i=1,2,..,N \},
\end{aligned}
\end{eqnarray}
where $F_r(u),K_r(u),R(u)$ and $\tilde{u}_i=u(x_1,x_2,x_3,t-\tau_i),\tau_i>0,i=1,2,\dots,N,r=1,2,3,$} describe diffusion, convection, reaction term  involving time delay, respectively.\\
 {Using the ansatz \eqref{wt} with $ k=3,$} the given $(3+1)$-dimensional generalized time-fractional time delay {CDR} equation \eqref{crdeqD} is transformed to a $(1+1)$-dimensional generalized nonlinear time-fractional delay {CDR} equation as follows:
\begin{eqnarray}\begin{aligned}\label{TcrdeqD}
 &\dfrac{\partial^\alpha w}{\partial t^\alpha} ={\tilde{\textbf{F}}_T[w,\hat{w}_i]}\equiv  \dfrac{\partial}{\partial z}\left( F(w)\dfrac{\partial w}{\partial z}\right)+ K(w)\dfrac{\partial w}{\partial z}+R(w)+\sum\limits_{i=1}^N \kappa_i\hat{w}_i, \alpha \in (0,2],\\
 & w(z,t)= \nu(z,t) \quad \text{if} \; t \in [-\hat{\tau},0],\hat{\tau}=max \{ \tau_i: i=1,2,..,N \}.
\end{aligned}
\end{eqnarray}
Here $F(w)= \sum\limits_{r=1}^{3}\lambda_r^2F_r(w),K(w)=\sum\limits_{r=1}^{3}\lambda_rK_r(w), $ $ w(z,t)=u(x_1,x_2,x_3,t),$ and $\hat{w}_i=w(z,t-\tau_i)=u(x_1,x_2,x_3,t-\tau_i),$ where $z=\lambda_1x_1+\lambda_2x_ 2+\lambda_3x_3,\tau_i>0,i=1,2,\dots,N.$\\
%where $F(w),K(w),R(w)$ and $ describes diffusion, convection, reaction and the time delay terms in $(1+1)$-dimension, respectively.
Next, we discuss the efficacy and applicability of the invariant subspace method along with variable transformation for deriving the exact solutions to the initial and boundary value problem for the $(3 + 1)$-dimensional generalized nonlinear time-fractional {CDR} equation with linear time delay.
\subsection{Exact solution of initial and boundary value problem for the time delay nonlinear time-fractional {CDR} equation \eqref{TcrdeqD}}
Consider the time delay nonlinear time-fractional {CDR} equation in the following form
  \begin{eqnarray}\begin{aligned}\label{crdf1D}
 \dfrac{\partial^\alpha w}{\partial t^\alpha}={\tilde{\textbf{\textit{F}}}_T[w,\hat{w}]}\equiv&\lambda_1^2\dfrac{\partial}{\partial z}\left[
 \left(\dfrac{\mu_1}{2\lambda_1 \rho_0} w+ \mu_3\right)\dfrac{\partial w}{\partial z}\right] + \lambda_1
 \left(\mu_1 w+ \mu_2\right)\dfrac{\partial w}{\partial z}
 \\&+ \mu_0w+\kappa w(z,t-\tau),\\
& w(z,t)= \nu(z,t)=\delta_0(t)e^{-\rho_0z} \quad \text{if} \; t \in [-{\tau},0],\tau>0, \alpha \in (0,2],
\end{aligned} \end{eqnarray}
    { where $\kappa, \mu_i\in \mathbb{R}, i=0,1,2,3,$} along with the appropriate initial and boundary conditions
 \begin{eqnarray}
 &w(z,0)=\xi(z),&\alpha\in(0,1],\label{DIC1}\\
& w(z,0)=\xi(z) \,\,\, \& \,\, \,\dfrac{\partial w}{\partial t} \big{|}_{t=0}=\eta(z),& \alpha\in(1,2],\label{DIC2}
\end{eqnarray}
{and}
\begin{eqnarray}
w(z,t)\Big{|}_{z=0}=\left\{
\begin{array}{ll}
\delta_1(t),\text{if}\, \alpha\in(0,1]\label{DBC},\\
\delta_2(t),\text{if}\, \alpha\in(1,2],
\end{array}  \right.
\, \& \quad
w(z,t)\Big{|}_{z=l}=\left\{
\begin{array}{ll}
\delta_3(t),\text{if}\, \alpha\in(0,1],\\
\delta_4(t),\text{if}\, \alpha\in(1,2].
\end{array}  \right.
\end{eqnarray}
{It is easy to observe that the above-given operator $ \tilde{\textbf{\textit{F}}}_T[w,\hat{w}] $ preserves the linear space
$
\mathbf{V}_1=\text{Span}\{e^{-\rho_0z }\},
$
which is listed in  case 6 of  Table \ref{crd2} without delay term.}

Proceeding in a similar way as above, for $\alpha\in(0,2],$ the exact solutions of \eqref{crdf1D} are obtained as
\begin{eqnarray}
w(z,t)=\left\{
\begin{array}{ll}
%\left\{\begin{array}{ll}
	 e^{-\rho_0z}\left\{ A \sum\limits_{m=0}^{n}\kappa^m(t-m\tau)^{\alpha m} E_{\alpha,\alpha m+1}^{m+1}(\gamma(t-m\tau)^{\alpha}) \right.
	 \\
 +\left.    \left[     \sum\limits_{m=0}^{n}\kappa^{m+1}(t-m\tau)^{\alpha( m+1)-1} E_{\alpha,\alpha( m+1)}^{m+1}(\gamma(t-m\tau)^{\alpha})\right]
\right. \\
 \left.\ast[\delta_0(t-\tau)T(t-\tau)] \right\},
 \alpha\in(0,1],
%\end{array}\right.
\\
%\left\{\begin{array}{ll}
	e^{-\rho_0z}\left\{ A\left(E_{\alpha,1}^{1}(\gamma t^\alpha)+\sum\limits_{m=1}^{n}\kappa^m(t-m\tau)^{\alpha m} E_{\alpha,\alpha m+1}^{m+1}(\gamma(t-m\tau)^{\alpha})\right) \right.
 \\+
   \hat{A}\left(tE_{\alpha,2}^{1}(\gamma_2t^{\alpha})+\sum\limits_{m=1}^{n}\kappa^m(t-m\tau)^{\alpha m+1} E_{\alpha,\alpha m+2}^{m+1}(\gamma (t-m\tau)^{\alpha})\right)
\\ +   \int_{0}^{t}\Big( \sum\limits_{m=0}^{n}\kappa^{m+1}(y-m\tau)^{\alpha( m+1)-1} E_{\alpha,\alpha( m+1)}^{m+1}(\gamma_2(y-m\tau)^{\alpha})
\\
\times \left.\delta_0(t-\tau-y)T(t-\tau-y)\Big)dy \right\}, \alpha\in(1,2],\label{DSol}
%\end{array}\right.
\end{array}\right.
\end{eqnarray}
where $\gamma=\rho_0^2\lambda_{1}^{2}\mu_3-\lambda_1\mu_2\rho_0+\mu_0,$  $A,\hat{A}\in \mathbb{R}$,  $n-1<\dfrac{t}{\tau}\leq n,$ $0<\Big{|} \dfrac{\kappa e^{-\tau s}}{s^\alpha-\gamma}\Big{|}<1,$
 $T(t)=\left\{\begin{array}{
ll} 1 , \quad t<0,\\
0, \quad t\geq 0
\end{array}\right.$ and $E_{a,b}^c(y)=\sum\limits_{m=0}^\infty\dfrac{(c)_m y^m}{m!\Gamma(am+b)} $ is the generalized three parameter Mittag-Leffler function with $(c)_m=\dfrac{\Gamma(c+m)}{\Gamma(c)},m\in\mathbb{N}.$
Additionally, $\ast$ denotes the convolution of functions defined as $a(y)\ast b(y)=\int\limits_0^y {a(y-x)b(x)dx}.$
\\
The obtained exact solutions satisfy the initial and boundary conditions  \eqref{DIC1}-\eqref{DBC} with
$ \xi(z)=Ae^{-\rho_0z},\eta(z)=\hat{A}e^{-\rho_0z}, $
\begin{eqnarray*}
	\begin{aligned}
		&\delta_1(t)=
				A \sum\limits_{m=0}^{n}\kappa^m(t-m\tau)^{\alpha m} E_{\alpha,\alpha m+1}^{m+1}(\gamma(t-m\tau)^{\alpha})
			+    \left[     \sum\limits_{m=0}^{n}\kappa^{m+1}(t-m\tau)^{\alpha( m+1)-1}\right.
				\\
			 &\qquad\quad\left.E_{\alpha,\alpha( m+1)}^{m+1}(\gamma(t-m\tau)^{\alpha})\right]\ast[\delta_0(t-\tau)T(t-\tau)] ,
	\end{aligned}\end{eqnarray*}
\begin{eqnarray*}
\begin{aligned}		&	\delta_2(t)=
		A\left(E_{\alpha,1}^{1}(\gamma t^\alpha)+\sum\limits_{m=1}^{n}\kappa^m(t-m\tau)^{\alpha m} E_{\alpha,\alpha m+1}^{m+1}(\gamma(t-m\tau)^{\alpha})\right)
		\\& \qquad\quad+
		\hat{A}\left(tE_{\alpha,2}^{1}(\gamma_2t^{\alpha})+\sum\limits_{m=1}^{n}\kappa^m(t-m\tau)^{\alpha m+1} E_{\alpha,\alpha m+2}^{m+1}(\gamma (t-m\tau)^{\alpha})\right)
		+  \int\limits_{0}^{t}\Big(   \sum\limits_{m=0}^{n}\kappa^{m+1}
		\\&\qquad\quad
		(y-m\tau)^{\alpha( m+1)-1} E_{\alpha,\alpha( m+1)}^{m+1}(\gamma_2(y-m\tau)^{\alpha})\delta_0(t-\tau-y)T(t-\tau-y)\Big)dy,
	\\
%\begin{eqnarray*}\begin{aligned}	
		&\delta_3(t)=
			B \sum\limits_{m=0}^{n}\kappa^m(t-m\tau)^{\alpha m} E_{\alpha,\alpha m+1}^{m+1}(\gamma(t-m\tau)^{\alpha})
			+    \left[     \sum\limits_{m=0}^{n}\kappa^{m+1}(t-m\tau)^{\alpha( m+1)-1}\right.\\
			& \qquad\quad\left. E_{\alpha,\alpha( m+1)}^{m+1}(\gamma(t-m\tau)^{\alpha})\right]\ast[\delta_0(t-\tau)T(t-\tau)] ,
			\\
		&	\delta_4(t)=
		B\left(E_{\alpha,1}^{1}(\gamma t^\alpha)+\sum\limits_{m=1}^{n}\kappa^m(t-m\tau)^{\alpha m} E_{\alpha,\alpha m+1}^{m+1}(\gamma(t-m\tau)^{\alpha})\right)
				\\& \qquad\quad+
				\hat{B}\left(tE_{\alpha,2}^{1}(\gamma_2t^{\alpha})+\sum\limits_{m=1}^{n}\kappa^m(t-m\tau)^{\alpha m+1} E_{\alpha,\alpha m+2}^{m+1}(\gamma (t-m\tau)^{\alpha})\right)
				+  \int\limits_{0}^{t} \Big(  \sum\limits_{m=0}^{n}\kappa^{m+1} \\&\qquad\quad (y-m\tau)^{\alpha( m+1)-1}E_{\alpha,\alpha( m+1)}^{m+1}(\gamma_2(y-m\tau)^{\alpha})\delta_0(t-\tau-y)T(t-\tau-y)\Big)dy,
				\end{aligned}
\end{eqnarray*}
 where $B=Ae^{-\rho_0l}, $ and $\hat{B}=\hat{A}e^{-\rho_0l} $.
 \section{Applications}
The class of diffusion equations is one of most fundamental class of equations in physical sciences. Thus, the study of behavior and asymptotic nature of generalizations of this kind of equations has been an interesting area of study for scientists for ages. The process of diffusion has been developed by Fick from the fact that concentration gradient in an isotropic medium is proportional to the rate of transportation of diffusing particles across a unit area, which is mathematically viewed \cite{crank} as
$$\mathbf{F}=\mathit{D}\left( \dfrac{\partial u}{\partial x}\right) ,$$
where $ \mathbf{F} $ is the rate of transfer per unit area, $u$ is the concentration of diffusing particle with space variable $x$ and diffusion coefficient $\mathit{D}.$
When diffusion is assumed to be only along the $x$-direction, Fick formulated the one-dimensional diffusion equation as follows,
\begin{equation}\label{ap}
 \dfrac{\partial u}{\partial t} = D \left( \dfrac{\partial^2 u}{\partial x^2}\right) .
 \end{equation}
Direct analogy with equations of heat conduction initially formulated by Fourier was a remarkable observation during that period.
For one-dimensional vertical flow, one may consider a combination of the equation of continuity for conservation of water mass \cite{thesis}
\begin{equation}\label{ap1}
\dfrac{\partial u}{\partial t}=-\left( \dfrac{\partial R}{\partial x}\right)
\end{equation}
along with the Buckingham-Darcy law for unsaturated flow \cite{thesis}
\begin{equation}\label{ap2}
R=-F(u)\left( \dfrac{\partial u}{\partial x}\right) +Q(u).
\end{equation}
The above equations \eqref{ap1}-\eqref{ap2} lead to the (1+1)-dimensional convection-diffusion equation \cite{thesis}
\begin{eqnarray}\label{ap3}
\begin{aligned}
\dfrac{\partial u}{\partial t} = \dfrac{\partial }{\partial x}& \left(F(u)\dfrac{\partial u}{\partial x} \right)-&\left( \dfrac{d Q(u)}{d u}\right) \left( \dfrac{\partial u}{\partial x}\right) ,\\
 & \uparrow_\text{capillarity}  & \uparrow_\text{gravity}\quad
\end{aligned}
\end{eqnarray}
which can be written as
\begin{eqnarray}\label{ap4}
\begin{aligned}
\dfrac{\partial u}{\partial t} = \dfrac{\partial }{\partial x} \left(F(u)\dfrac{\partial u}{\partial x} \right)+K(u)\left( \dfrac{\partial u}{\partial x}\right) ,
\end{aligned}
\end{eqnarray}
where $K(u)=-\left( \dfrac{d Q(u)}{d u}\right) $, $u=u(x,t)$ is the volumetric water content at time $t$ in the depth $x$ below the soil surface, $F(u)$ is the concentration dependent soil-water diffusivity and $Q(u)$ is the concentration-dependent hydraulic conductivity.
 {Note that Liu \cite{liu} has discussed the various kinds of exact solutions of \eqref{ap3} using the invariant subspace method.}
The equation \eqref{ap4} is called a generalized $(1+1)$-dimensional nonlinear convection-diffusion equation.
{In science and engineering, most of the applications of the above type of equations are in modeling as in the form of diffusion, diffusion-convection (or advection) and diffusion-reaction (or absorption) equations. The general class of $(1+1)$-dimensional nonlinear convection-diffusion-reaction equation reads as follows \cite{GK,ps1},
\begin{equation}\label{1}
	u_t= (F(u)u_x)_{x}+K(u)u_x+R(u),\quad u=u(x,t),x\in\mathbb{R},t\geq0,
\end{equation}
where $u_t=\dfrac{\partial u}{\partial t},$ $u_x=\dfrac{\partial u}{\partial x},$  $F(u)$ is the diffusion coefficient, $K(u)$ is the convective term and $R(u)$ gives the kinetics of the system.
 {Equation \eqref{1} reduces to the following well-known nonlinear reaction-diffusion equations  \cite{GK}:}
 {\begin{itemize}
	\item[(1)] When $F(u)=1,$  $K(u)=0$, and $R(u)=u(1-u)$, equation \eqref{1} is referred to as the well-known Fisher equation or logistic equation \cite{GK}.
	\item[(2)] The above equation \eqref{1} becomes the Newell-Whitehead equation or amplitude equation  \cite{GK} if $F(u)=1,$  $K(u)=0$, and $R(u)=u(1-u^2)$.
	\item[(3)] If $F(u)=1,$  $K(u)=0$, and $R(u)=u^2(1-u)$, then the above equation \eqref{1} is called the Zeldovich equation  \cite{GK}.
	\item[(4)] Equation \eqref{1} is known as the Nagumo equation or bistable equation \cite{GK} if $F(u)=1,$  $K(u)=0$, and $R(u)=u(1-u)(u-\beta)$ with $\beta\in(0,1)$.
	\item[(5)] When $F(u)=1,$  $K(u)=0$, and $R(u)$-arbitrary, equation \eqref{1} can be viewed as the KPP equation  \cite{GK}, which is a generalization of the Fisher equation, the Newell-Whitehead equation, the Zeldovich equation and the Nagumo equation.
	\item[(6)] When $F(u)=mu^{m-1},$  $K(u)=0$, and $R(u)=\pm u^p$, $m,p>0$, the equation \eqref{1} is called the porous media equation \cite{GK} with absorption (source).
\end{itemize}}
Here we wish to point out some applications of the invariant subspace method to the above equations \eqref{ap4} and \eqref{1} as discussed below.
\begin{itemize}
	\item[(a)] When $F(u)=c_1$, equation \eqref{ap4}  reduces to the generalized Burgers' equation  \cite{GK}, which admits the two-dimensional linear space $\mathbf{V}_2=\text{Span}\left\{1,x\right\}$ if $K(u)=c_2u$.
	\item[(b)] If $F(u)=u^{-\frac{3}{2}}$ and $K(u)=0$, then the above equation \eqref{ap4} is known as the fast diffusion equation \cite{gs} which possesses an exact solution based on the linear space $\mathbf{V}_4=\text{Span}\left\{1,x,x^2,x^3\right\}$ associated with $v=u^{-\frac{3}{2}}.$
\item[(c)] When $F(u)=u^{-\frac{4}{3}},K(u)=0,$ and $R(u)=u^{\frac{7}{3}}$, equation \eqref{1} is referred to as the fast diffusion equation with reaction term \cite{gs}, which possesses exact solution based on the linear space $\mathbf{V}_{5}=\text{Span}\left\{1,x,x^2,x^3,x^4\right\}$ along with $v=u^{-\frac{4}{3}}.$
\item[(d)] If $F(u)=u^{-\frac{4}{3}},K(u)=0,$ and $R(u)=-u^{-\frac{1}{3}}$, then the equation \eqref{1} is called a fast diffusion equation with absorption term \cite{gs}, which possesses exact solution based on the linear space $\mathbf{V}_{5}=\text{Span}\left\{1,\cos(\frac{4}{\sqrt{3}}x),\sin(\frac{4}{\sqrt{3}}x),\cos(\frac{2}{\sqrt{3}}x),\sin(\frac{2}{\sqrt{3}}x)\right\}$ along with $v=u^{-\frac{4}{3}}$.
\item[(e)] For $ F(u)=u^{\sigma},K(u)=0, R(u)=-u^{1-\sigma},\sigma>0 $, the above equation \eqref{1} represents a porous medium equation with absorption \cite{gs} which possesses the exact solution based on the linear space $ \mathbf{V}_2 =\text{Span}\{1,x^2\}$ along with $v=u^{\sigma}.$
\end{itemize}
%%%%%%%%%%%%%%%%%%%%%%%%%%%%%%%%%%%%%%%%%%%%%%%%%%%%%%%%%%%%%%%%%%%%%%%%%%%%%%%%%%%%%%%%%%%%%%%%%%%%%%%%%%%%%%%%%%%%%%%%%%%%%%%%%%%%%%%%%%%%%%%%%%%%%
Recently, invariant subspaces and exact solutions of $(1+1)$-dimensional generalized nonlinear convection-diffusion-reaction equations \eqref{1} with power-law nonlinearities have been discussed in \cite{ps1}.
In the literature, some remarkable works can be seen for deriving the exact solutions of higher-dimensional nonlinear convection-diffusion-reaction equations that are given below.
\begin{itemize}
\item[$\bullet$] Exact solutions of the quadratic wave equation \cite{gs} with $(2+1)$-dimensions, $u_{tt}=\bigtriangledown\cdot(u\bigtriangledown u)+bu+a,$ have been derived based on the 3-dimensional invariant subspace $ \mathbf{V}_3=\text{Span}\{1,x_1^2,x_2^2\} $ in \cite{gs}, where $ u=u(x_1,x_2,t),x_i\in\mathbb{R},i=1,2,t\geq 0. $
\item[$\bullet$]
 The invariant subspace of the porous medium equation \cite{gs} with $(N+1)$-dimensions, $u_{t}=\bigtriangledown\cdot(u^\sigma\bigtriangledown u)+au^{1-\sigma}+bu,\sigma\neq0,$ is $ \mathbf{V}_{N+1}=\text{Span}\{1,x_1^2,x_2^2,\dots,x_N^2\},$  where $ u=u(x_1,x_2,\dots,x_N,t),x_i\in\mathbb{R},i=1,2,\dots,N,t\geq 0.$
\item[$\bullet$] The 9-dimensional linear subspace $\mathbf{V}_9= \text{Span}\{1,x_1,x_2,x_1^2,x_1x_2,x_2^2,x_1r,x_2r,r^2\},r=(x_1+x_2)^2$ can be used to exhibit the solutions of the $(2+1)$-dimensional fast diffusion equation \cite{gs}
 $u_{t}=\bigtriangledown\cdot(\frac{1}{u}\bigtriangledown u),u=u(x_1,x_2,t).$
\end{itemize}
}
However, in complex systems, it is observed that the diffusion process is non-predictable (anomalous) as it does not always follow the Gaussian statistics or Fick's law which are characteristics of the normal Brownian diffusion. It can be seen that anomalous diffusion is dominant in many complex systems of physics and biology \cite{book1,lenzi}. Metzler and Klafter \cite{ralf} have derived anomalous diffusion process mathematically by continuous-time random walk scheme in terms of fractional-order derivatives in the form
\begin{equation} \label{AD}
\dfrac{\partial^\alpha u}{\partial t^\alpha} -\dfrac{t^{-\alpha}}{\Gamma(1-\alpha)}u_0(x)=\mathcal{H}_\alpha\dfrac{\partial}{\partial x}\left(\dfrac{\partial^2 u}{\partial x^2}\right), \,
\end{equation}
where $u_0(x)$ is the initial value of diffusion and $\mathcal{H}_\alpha\, \text{ is the generalized diffusion coefficient.}$
The anomalous behavior is characterized by their nonlinear power-law  time dependence of the mean square displacement, that is
$$\big< x^2(t)\big>= \dfrac{2\mathcal{H}_\alpha}{\Gamma(1+\alpha)} t^\alpha.$$
Note that the diffusion exponent $\alpha\ (>0)$ divides the domain of anomalous diffusion \cite{ralf} into three types that are (i) If $\alpha\in(0,1),$ the process represents a sub-diffusion, (ii) when $\alpha\in(1,2]$, the process represents a ballistic (super) diffusion and (iii) when $\alpha=1$, the process represents the normal Brownian diffusion (time dependent and linear mean square displacement, i.e.,$\big< x^2(t)\big>\sim kt$).\\
Due to the rich applications of anomalous diffusion and its generalizations in all fields of science and engineering, many scientists devoted their time to investigating its structural and dynamical properties.
Different numerical methods were employed to establish the properties of the anomalous behavior of diffusion in complex systems like Monte-Carlo simulation,
 forward Euler difference formula \cite{yuste}, green's function method \cite{ahan2},
 to name a few.
Recently, exact solutions of time-fractional linear diffusion equation were studied through the Lie symmetries in \cite{bb1}.
 The Lie-symmetries of time-fractional linear diffusion equations with variable coefficient were studied by Sahadevan and Prakash \cite{pra5}.
In \cite{pra4}, Sahadevan and Prakash have studied the exact solutions of $(1+1)$-dimensional time-fractional reaction-diffusion equations and $(1+1)$-dimensional time-fractional convection-diffusion equations through the invariant subspace method.
Also, exact solutions of $(1+1)$-dimensional time-fractional generalized nonlinear reaction-diffusion with time delay equations have been derived through the invariant subspace method \cite{pp2}.
 Exact solutions of $(2+1)$-dimensional biological population model
$$
 \dfrac{\partial^\alpha u}{\partial t^\alpha}=\dfrac{\partial^2 (u^2)}{\partial x_1^2}+\dfrac{\partial^2 (u^2)}{\partial x_2^2}+\kappa u^a(1-\gamma u^a)
 $$
 were studied extensively for various parameter values of $\kappa, a,\gamma$ using  separation of the variable method \cite{rui}, invariant subspace method associated with variable transformation \cite{IVSM} and direct approach of invariant subspace method \cite{ppa}.

 %{Very recently, Prakash et al. \cite{ppa,ppl} have derived the finite separable exact solutions for the initial value problems of the $(2+1)$-dimensional nonlinear generalized time-fractional convection-diffusion-reaction equations} and $(3+1)$-dimensional nonlinear generalized time-fractional convection-diffusion equations through the direct approach of invariant subspace method.
% %Very recently,
%%Prakash et al. \cite{ppl} have investigated the exact solutions for initial value problems of the $(3+1)$-dimensional nonlinear generalized time-fractional convection-diffusion equation using the direct approach of the invariant subspace method.
%In this article, we have applied the invariant subspace method associated with variable transformation for deriving the exact solutions of the $(k+1)$-dimensional nonlinear time-fractional PDE \eqref{keq}. Also, we have derived the exact solutions of the form \eqref{3dsol} for the initial and time-dependent Dirichlet boundary value problems of the $(3+1)$-dimensional nonlinear time-fractional generalized {CDR} equation \eqref{rdeq} through the invariant subspace method associated with variable transformation.}
%% as$$u(x_1,x_2,x_3)=\sum\limits_{i=1}^3\psi_i(t)\phi_i(\lambda_1x_1+\lambda_2x_2+\lambda_3x_3),\lambda_i,x_i\in\mathbb{R},t>0.$$
\section{{Discussion and concluding remarks}}
In this paper, we have systematically investigated how to apply the invariant subspace method associated with variable transformation for deriving the exact solutions of the $(k+1)$-dimensional nonlinear time-fractional PDEs in detail. Also, this detailed specific study was used for finding the various types of exact solutions of the $(3+1)$-dimensional nonlinear time-fractional convection-diffusion-reaction equation explicitly, along with the appropriate initial and boundary conditions. Moreover, we note that the obtained exact solutions of the equation as mentioned above can be written in terms of polynomial, exponential, trigonometric, hyperbolic, and Mittag-Leffler functions. In addition, the discussed method was extended for the $(k+1)$-dimensional nonlinear time-fractional PDEs with several linear time delays, and also the exact solutions of the $(3+1)$-dimensional nonlinear time-fractional delay convection-diffusion-reaction equation were derived using the discussed method. It is well-known that the non-integer order derivatives have some unusual properties, such as violation of the standard form of the Leibniz rule, chain rule, and semigroup property. Due to these reasons, there are no well-defined analytical methods for nonlinear non-integer order PDEs.

{We wish to point out that in \cite{ppl}, Prakash et al. have investigated the invariant subspace method for finding the exact solutions of $(k+1)$-dimensional nonlinear time-fractional PDEs without any variable transformation. Also, they have derived various types of exact solutions for the $(3+1)$-dimensional nonlinear time-fractional convection-diffusion equation using the direct approach of the invariant subspace method. From this, we can look for exact solutions of the $(k+1)$-dimensional nonlinear time-fractional PDEs of the form \eqref{kda}.
%\begin{equation}	\label{da}	u(x_1,x_2,x_3,t) =\sum\limits_{i_1,i_2,i_3} \Psi_{i_1,i_2,i_3}(t)\prod_{r=1}^{3}\Phi^r_{rs}(x_r),i_r=1,2,\dots,n_r,r=1,2,3.\end{equation}
This work has been extended to derive exact solutions for a higher-dimensional time-fractional equation using the invariant subspace method with variable transformation.
%This work has studied to derive exact solutions for $(3+1)$-dimensional time-fractional convection-diffusion-reaction equation using the invariant subspace method along with variable transformation
%$$ u(x_1,x_2,x_3,t)=w(\lambda_1x_1+\lambda_2x_2+\lambda_3x_3,t).$$
Also, Abdel Kader et al. \cite{IVSM} have investigated exact solutions of a nonlinear time-fractional $(2+1)$-dimensional biological population model with variable
coefficients using the invariant subspace method associated with variable transformation.
So, we can expect the particular form of exact solutions \eqref{ttt} for (k+1)-dimensional time-fractional PDEs from these studies.
%\begin{equation}\label{vt}
%	u(x_1,x_2,\dots,x_3,t)=w(\lambda_1x_1+\dots+\lambda_kx_k,t)=\sum\limits_{s=1}^n C_s(t)\phi_s(\lambda_1x_1+\dots+\lambda_kx_k).
%\end{equation}
The derivation of the exact solution \eqref{kda}  may not be straightforward for higher-dimensional nonlinear time-fractional PDEs using the invariant subspace method without any variable transformation. However, it is easy to apply this study for the higher-dimensional case because it allows
one to reduce the $(k+1)$-dimensional equation to the $(1+1)$-dimensional equation. Hence these investigations show that the discussed method is a very important, efficient, and powerful analytical tool to derive the exact solutions to the initial and boundary value problems of the nonlinear time-fractional higher-dimensional PDEs in science and engineering.
The applicability of the method has been already discussed in the literature for time-space fractional nonlinear PDEs \cite{s1,pp1}.  The discussed method may be applied to systematically find the exact solutions of space fractional nonlinear PDEs with some additional assumptions, which will be studied in the future.}
\section*{Acknowledgements}
 The work of M. L. is supported by a DST-SERB National Science Chair.

\end{document}